\pdfoutput=1
\RequirePackage{ifpdf}
\ifpdf 
\documentclass[pdftex]{sigma}
\else
\documentclass{sigma}
\fi

\usepackage{mathtools}

\newcommand{\CI}{{\cal I}}

\newcommand{\CN}{{\cal N}}

\newcommand{\CW}{{\cal W}}

\newcommand{\res}{\operatorname{Res}}

\newcommand{\tcircled}[1]{{\text{\textcircled{#1}}}}

\DeclarePairedDelimiter\ceil{\lceil}{\rceil}
\DeclarePairedDelimiter\floor{\lfloor}{\rfloor}
\def\IN{{\mathbb N}}
\def\IZ{{\mathbb Z}}
\def\IR{{\mathbb R}}
\def\IC{{\mathbb C}}
\def\IP{{\mathbb P}}

\usepackage{tikz}
\newcommand*\circled[1]{\tikz[baseline=(char.base)]{
 \node[shape=circle,draw,inner sep=2pt] (char) {#1};}}

\newcommand{\re}{{\rm e}}
\newcommand{\ri}{{\rm i}}
\newcommand{\rd}{{\rm d}}

\newcommand{\Li}{\mathop{\rm Li}\nolimits}

\newcommand{\espilon}{{\epsilon}}
\newcommand{\I}{{\mathrm i}}
\newcommand{\E}{{\mathrm e}}

\newcommand{\be}{\begin{equation}}
\newcommand{\ee}{\end{equation}}
\newcommand{\ba}{\begin{aligned}}
\newcommand{\ea}{\end{aligned}}

\numberwithin{equation}{section}

\usepackage{subcaption}
\usepackage[labelsep=period,labelfont=bf]{caption}

\def\({\left(}
\def\){\right)}

\newdimen\tableauside\tableauside=1.0ex
\newdimen\tableaurule\tableaurule=0.4pt
\newdimen\tableaustep
\def\phantomhrule#1{\hbox{\vbox to0pt{\hrule height\tableaurule width#1\vss}}}
\def\phantomvrule#1{\vbox{\hbox to0pt{\vrule width\tableaurule height#1\hss}}}
\def\sqr{\vbox{%
 \phantomhrule\tableaustep
 \hbox{\phantomvrule\tableaustep\kern\tableaustep\phantomvrule\tableaustep}%
 \hbox{\vbox{\phantomhrule\tableauside}\kern-\tableaurule}}}
\def\squares#1{\hbox{\count0=#1\noindent\loop\sqr
 \advance\count0 by-1 \ifnum\count0>0\repeat}}
\def\tableau#1{\vcenter{\offinterlineskip
 \tableaustep=\tableauside\advance\tableaustep by-\tableaurule
 \kern\normallineskip\hbox
 {\kern\normallineskip\vbox
 {\gettableau#1 0 }%
 \kern\normallineskip\kern\tableaurule}%
 \kern\normallineskip\kern\tableaurule}}
\def\gettableau#1{\ifnum#1=0\let\next=\null\else
\squares{#1}\let\next=\gettableau\fi\next}

\counterwithin{equation}{section}

\newtheorem*{Theorem*}{Theorem}

 { \theoremstyle{definition}

 }

\begin{document}
\allowdisplaybreaks

\newcommand{\arXivNumber}{2201.11594}

\renewcommand{\PaperNumber}{064}

\FirstPageHeading

\ShortArticleName{Exponential Networks, WKB and Topological String}

\ArticleName{Exponential Networks, WKB and Topological String}

\Author{Alba GRASSI~$^{\rm ab}$, Qianyu HAO~$^{\rm c}$ and Andrew NEITZKE~$^{\rm d}$}

\AuthorNameForHeading{A.~Grassi, Q.~Hao and A.~Neitzke}

\Address{$^{\rm a)}$~Section de Math\'ematiques, Universit\'e de Gen\`eve, 1211 Gen\`eve 4, Switzerland}
\EmailD{\href{mailto:alba.grassi@cern.ch}{alba.grassi@cern.ch}}

\Address{$^{\rm b)}$~Theoretical Physics Department, CERN, 1211 Geneva 23, Switzerland}
\EmailD{\href{mailto:alba.grassi@cern.ch}{alba.grassi@cern.ch}}

\Address{$^{\rm c)}$~Department of Physics, University of Texas at Austin, \\
\hphantom{$^{\rm c)}$}~2515 Speedway, C1600, Austin, TX 78712-1992, USA}
\EmailD{\href{mailto:qhao@utexas.edu}{qhao@utexas.edu}}

\Address{$^{\rm d)}$~Department of Mathematics, Yale University,\\
\hphantom{$^{\rm d)}$}~PO Box 208283, New Haven, CT 06520-8283, USA}
\EmailD{\href{mailto:andrew.neitzke@yale.edu}{andrew.neitzke@yale.edu}}

\ArticleDates{Received March 07, 2023, in final form August 23, 2023; Published online September 13, 2023}

\Abstract{We propose a connection between 3d-5d exponential networks and exact WKB for difference equations associated to five dimensional Seiberg--Witten curves, or equivalently, to quantum mirror curves to toric Calabi--Yau threefolds $X$: the singularities in the Borel planes of local solutions to such difference equations correspond to central charges of 3d-5d BPS KK-modes. It follows that there should be distinguished local solutions of the difference equation in each domain of the complement of the exponential network, and these solutions jump at the walls of the network. We verify and explore this picture in two simple examples of 3d-5d systems, corresponding to taking the toric Calabi--Yau $X$ to be either $\mathbb{C}^3$ or the resolved conifold. We provide the full list of local solutions in each sector of the Borel plane and in each domain of the complement of the exponential network, and find that local solutions in disconnected domains correspond to non-perturbative open topological string amplitudes on $X$ with insertions of branes at different positions of the toric diagram. We also study the Borel summation of the closed refined topological string free energy on $X$ and the corresponding non-perturbative effects, finding that central charges of 5d BPS KK-modes are related to the singularities in the Borel plane.}


\Classification{39A70; 40G10; 81T30; 81T60}

\section{Introduction}\label{sec:intro}

This paper is motivated by the necessity to deepen our understanding of exact WKB methods for difference equations, non-perturbative effects in (refined) open and closed topological string amplitudes, and their relation to 5d/3d-5d BPS states and exponential networks.

We begin with the open string. In recent years, there has been considerable progress in the study of four-dimensional $\CN=2$ supersymmetric gauge theories and their connection to linear differential equations. One of the basic geometric objects in the story is
the spectral network, introduced in the context of 4d $\CN=2$ theories in \cite{gmn,Gaiotto:2011tf,Gaiotto:2010be,Gaiotto:2012rg,Gaiotto:2009hg}.
The spectral network captures the BPS spectrum of a surface defect in the 4d $\CN=2$ theory on the one hand, while on the other hand it is identified with the Stokes graph of the corresponding differential equation.

Various parts of this structure are modified when we lift from four-dimensional theories to five-dimensional ones (compactified on $S^1$).
In five-dimensional theories there is an analog of the spectral network, namely the
exponential network introduced in \cite{Banerjee:2018syt,eager}, which captures the BPS spectrum of the 5d theory coupled
to a 3d defect.
(See also \cite{Banerjee:2019apt,Banerjee:2020moh,Closset:2019juk,Longhi:2021qvz} for some other developments in this direction and \cite{Bershtein:2017swf,Bonelli:2020dcp,DelMonte:2021ytz} for connections between $q$-Painlev\'e equations, exponential networks and 5d BPS quivers.)
Moreover, it is expected that these 3d-5d systems should be related to $q$-difference equations, replacing the differential equations
which appeared in the 4d case. In particular, it
was pointed out in \cite{acdkv}, built upon \cite{mirmor,ns}, that the WKB expansion of a certain class of difference equations, known as higher genus quantum mirror curves, is closely related to the Nekrasov--Shatashvili (NS) limit of refined topological strings.\footnote{\label{fni}We should however stress that such WKB-like methods based on the refined topological strings involve a~second series expansion in $Q=\re^{-t}$ where, schematically, $t$ denotes the K\"ahler parameters of the underlying Calabi--Yau threefold. More precisely, at each order in $\hbar$, we have a series expansion in $Q$ which converges only in the large radius region of the moduli space. Hence this method does not allow us to have access to a~standard WKB expansion. This is very similar to what happens in four-dimensional theories: see discussion in~\cite[pp.~37--38]{Grassi:2021wpw} and references there.}\footnote{For difference equations associated with genus zero mirror curves, the WKB expansion of local solutions is usually expressed by using unrefined open topological strings \cite{adkmv,Kashani-Poor:2006puz}, also known as the Gopakumar--Vafa (GV) limit of refined topological strings. This might seem like a puzzle since above we stated that it is the NS limit which is related to difference equations. The resolution is that, in the genus zero case, the refined open topological string partition functions in the NS and GV limits are related in a very simple way; see for instance \cite{Cheng:2021nex,Dimofte:2010tz,ikv,Kozcaz:2018ndf}.
} Since topological strings can compute observables of 5d field theories (obtained via compactification of M-theory on a non-compact Calabi--Yau threefold)
this leads to the expectation that some observables of 5d field theories should obey $q$-difference equations.

The five-dimensional case has one important new feature: the WKB expansion for the difference equations has to be augmented by a new type of non-perturbative effects \cite{cgm,ghm, mz-wv}, which vanish when we implement the four-dimensional limit \cite{kkv,selfdual} leading to differential equations.
See also \cite{fhm,hm, Hatsuda:2018lnv, swh} for other applications to the quantization conditions of relativistic integrable systems, and \cite{Bonelli:2017gdk} for connections between 5d quantum mirror curves and tau functions of $q$-Painlev\'e equations \cite{bsu}.

In this paper, we clarify the relations between five-dimensional gauge theories, exponential networks, and difference equations:
\begin{itemize}\itemsep=0pt
\item
In Section~\ref{sec:stokes-bps},
we point out a direct connection between exponential networks and exact WKB-type solutions of difference equations.\footnote{Similar ideas have been explored by Fabrizio Del Monte and Pietro Longhi; we thank
them for a discussion about this.} We study the singularities in the Borel plane of local solutions to difference equations associated with Seiberg--Witten curves of 3d-5d systems, or
equivalently, quantum mirror curves to toric Calabi--Yau (CY) threefolds $X$. We focus on the first singularity in any given direction, and propose
that these singularities correspond to BPS particles living on the $S^1$
compactification of the 3d defect, with the positions of the singularities matching the central charges of the BPS particles.
It then follows from the definition of exponential network that there should be distinguished
local solutions of the difference equation in each domain of the complement of the
exponential network, and these solutions should jump at the walls of the exponential network.

All this is closely parallel to the story for more conventional 2d-4d systems \cite{Gaiotto:2011tf,Gaiotto:2012rg}, where it was conjectured in \cite{Grassi:2021wpw} that the positions of Borel plane singularities for local solutions of the differential equations are the central charges of BPS particles lying on the surface defect.
The 3d-5d setting however brings a few new features; in particular, in addition to the usual walls which carry labels $ij$,
there are new walls carrying labels $ii$, corresponding to BPS particles charged under the flavor symmetry.

\item In Sections~\ref{sec:c3} and \ref{sec:resconi}, we study this proposal in two specific examples of 3d-5d systems.
These examples correspond to taking $X$ to be either $\IC^3$ or the resolved conifold. In these examples the desired local solutions can be described
explicitly in terms of quantum dilogarithms, for which the Borel plane structure is
completely known thanks to the recent work \cite{Garoufalidis:2020pax}. Using the
techniques and results of \cite{Garoufalidis:2020pax} we show
that the proposed picture indeed holds. (We expect this structure holds also for other local CY manifolds with higher genus mirror curves. We comment more on this aspect in Section~\ref{hggen}.)

In these two examples, we give closed form expressions for the Borel transform and Borel summation of the local solutions in each domain of the complement of the exponential network. In the $\IC^3$ case the various solutions are described in Section~\ref{Locs} and summarized in Figure~\ref{fgsummaryls}, while the example of the resolved conifold is discussed in Section~\ref{coniopens}. In addition, we relate the resulting expressions to the open topological string partition function: local solutions in different domains of the exponential network correspond to open topological string partition functions with brane insertion at different positions (e.g., on the internal or on the external leg).

One interesting feature which appears in the $\IC^3$ case is that, for generic phase $\vartheta$, the complement of the spectral network is actually simply connected; thus
the jump of the local solution which occurs at a wall can also be obtained by analytic continuation of the solution along a path. We discuss this point in more detail in Section~\ref{jumpwall}.

In the conifold case, for each domain we give an analytic computation of the non-per\-tur\-ba\-tive effects and compare with some available results in the topological string literature~\cite{mz-wv}: see Section~\ref{coniopens}.
\end{itemize}

Now let us discuss the closed string. As in the open string case, our analysis is organized
around the theme of Borel plane singularities and their relation
to BPS particles --- now for the bulk theory rather than the theory with a defect.
Indeed, in 4d ${\mathcal N}=2$ theories the positions of Borel plane singularities for quantum periods are central charges of bulk BPS particles \cite{Grassi:2019coc, Grassi:2021wpw}.
The closed topological string amplitudes are analogues of quantum periods, now associated with difference equations rather than differential equations \cite{acdkv,mirmor};
with this in mind, we expect that the singularities in the Borel plane of the closed topological string amplitudes should be related to the central charges of 5d BPS KK modes. This is the prediction which we investigate.
%

For the unrefined limit of the resolved conifold, several studies in this direction have already been performed, for example in \cite{ho2,Krefl:2015vna,lv,ps09}.
Another interesting approach based on the Mellin--Barnes representation of the spectral zeta function can be found in \cite{hatsuda}. The closed topological string partition function on the resolved conifold in the $\epsilon_1+\epsilon_2=0$ phase was also discussed in \cite{Bridgeland:2017vbr}
as a solution to a certain Riemann--Hilbert problem, and in \cite{Bonelli:2017gdk} from the point of view of $q$-Painlev\'e equations.
See also \cite{Alexandrov:2021prq,Alim:2021gtw} for other interesting related work.

In this paper, we adopt an analytic approach and go beyond the unrefined case.
We can summarise our results for the closed sector as follows:
\begin{itemize}\itemsep=0pt
\item For the $\IC^3$ example our approach simply translates to the study of the resurgence properties of the McMahon function: see Section~\ref{McM}. It is nice to see that even in this toy model, as we go away from the imaginary axis, we have non-perturbative effects which are in fact encoded in the NS limit of the refined McMahon function, very much in line with what was found originally in the context of ABJM theory \cite{hmmo}.

\item For the resolved conifold we compute analytically the Borel transform and Borel summation, both in the unrefined and refined cases.
We give a detailed description of the non-perturbative effects in each sector, and compare with some previous results in the literature. In particular, for the unrefined case we can relate our picture to \cite{ho2}, where numerical studies have been performed, while for the refined case our results are new. See Section~\ref{closeconi}.

\item We find that there is a correspondence between the singularities in the Borel transform of the refined closed topological string free energy and the central charges of 5d BPS KK-modes. For the GV ($\espilon_1=-\espilon_2$) and the NS ($\espilon_2=0$) phases of the $\Omega$ background, the singularities lie precisely at the central charges of 5d BPS KK-modes. However,
for more generic phases of the $\Omega$ background ($\espilon_1=\alpha \espilon_2$)
the Borel plane has an additional series of poles: see \eqref{bpsa} and Section~\ref{closeconi}. When $\alpha\to 0$ these extra poles go to infinity, while when $\alpha\to -1$ they merge with the original series of poles.
This behavior suggests that in the refined topological string a BPS particle of
central charge $Z$ gives rise to two distinct nonperturbative effects,
of sizes $\E^{- |2\pi R Z / \epsilon_1|}$ and $\E^{-|2\pi R Z / \epsilon_2|}$.
\end{itemize}
We conclude this introduction by listing some open problems and future directions:
\begin{itemize}\itemsep=0pt
\item
In the case of 2d-4d systems, the theory of spectral networks and exact WKB are useful tools
for studying the hyperk\"ahler geometry of moduli spaces of solutions of Hitchin equations \cite{gmn,Gaiotto:2009hg}.
The basic reason why exact WKB has something to do with Hitchin equations is that
solutions of Hitchin equations can be identified with Higgs bundles and also with differential equations.

It is natural to imagine that this theory can be extended to 3d-5d systems. In
this extension, Higgs bundles would be replaced by group-valued Higgs bundles
\cite{Elliott:2018yqm},
differential equations would be replaced by difference equations ($q$-difference modules),
and solutions of Hitchin equations would be replaced by doubly periodic monopoles \cite{Cecotti:2013mba,Cherkis:2014vfa}. The correspondence between $q$-difference modules
and doubly periodic monopoles is carefully developed in \cite{mochizuki2019doubly}.

The results in this paper can be regarded as a step in this direction. It would
be very interesting to go further and give a twistorial construction of moduli spaces
of periodic monopoles in terms of central charge data and BPS degeneracies, in parallel
to \cite{gmn}.

\item
The local solutions which we consider are closely related to objects discussed in the
literature on boundary conditions and holomorphic blocks in 3d $\CN=2$ theories, e.g., \cite{Beem:2012mb,Cecotti:2013mba,Dimofte:2017tpi,Yoshida:2014ssa}. There is also closely related work
in 3d $\CN=4$ theories such as \cite{Bullimore:2020jdq,Bullimore:2021rnr}.
We do not develop this point of view much in the current paper; however, it seems
likely that in future developments this will be an important perspective.

\item
In a single 5d theory there are many different possible 3d defects which can be added.
For example, the defect we consider in the $\IC^3$ theory sits naturally in a family
parameterized by $f \in \IZ$, with corresponding Seiberg--Witten curves
\[
 XY^f-Y+1=0.\] The quantity $f$ is often called the ``framing''
following \cite{adkmv, akv}. The framing we are using in this paper is $f=0$;
for this defect there is only a single
vacuum, and thus a unique local solution to the difference equation
up to the flavor ambiguities, which substantially simplifies the analysis.
It would be interesting to extend our considerations to more general framings.
Some of the relevant exponential networks have already been described ---
e.g., see \cite{Banerjee:2018syt} for the case $f=-1$.

 \item Spectral/exponential networks drawn on a surface $C$
 can be used to study even BPS particles whose central charges do not vary along $C$.
 These particles are detected indirectly: the network $\CW^\vartheta$ depends on a phase $\vartheta = \arg \hbar$, and when there is a BPS particle whose central
 charge has $\arg \left(-Z\right) = \vartheta$, the network $\CW^\vartheta$ degenerates.
 From these degenerations one can try to read out
 the BPS spectrum, via wall-crossing methods described in \cite{Banerjee:2018syt, Gaiotto:2012rg}.
 One important instance of this is the use of spectral/exponential networks attached to
 coupled 2d-4d or 3d-5d systems,
 to study BPS particles in the 4d or 5d bulk which carry electromagnetic charge.

 In the cases we consider here, there are BPS particles in the bulk (corresponding in
 the Type IIA language to D0-branes or D0-D2 bound states), and there are corresponding
 degenerations of the exponential network. It would be very interesting to understand
 whether it is possible to
 compute the bulk BPS degeneracies directly by wall-crossing methods from these degenerations.
 For $X = \IC^3$ and the framing
 $f = -1$, such a computation was given in \cite{Banerjee:2018syt}.
\item As we mentioned above, it would be desirable to understand in detail the relation between exponential networks and exact WKB in more complicated examples,
involving higher-genus mirror curves. One of the main technical obstacles here is to develop an efficient way to compute the WKB expansion.
One can try to do this directly by writing a WKB ansatz like
one writes for differential equations, and then solving a Riccati-type equation order by
order in $\hbar$, as discussed, e.g., in \cite{dingle,kpamir,Zakany:2017txl}; applying this method to the
simple cases we consider in this paper indeed gives the correct series.
Alternatively, we could use the refined holomorphic anomaly equation in the NS limit \cite{cm-ha,hkpk,kw}, but then we still have to deal with the quantum mirror map, which at present we can compute only in a large radius expansion; see footnote~\ref{fni}. We comment more on this open direction in Section~\ref{hggen}.
\end{itemize}

\subsection*{Related work}
While this paper was in preparation, the independent work \cite{Alim:2021mhp} appeared;
among other things, this paper gives a clear and careful treatment of the Borel summation for the closed topological string in the conifold in the GV phase, using substantially the same techniques we used.
The results in Section~\ref{closedGV} match with \cite{Alim:2021mhp}. We understand that more recently Murad Alim, Lotte Hollands and Ivan Tulli have also studied the NS phase and independently obtained, among other things, results overlapping with Section~\ref{closedNS} \cite{toapp}.
We thank them for discussions about this.

\section{Stokes phenomena and BPS particles} \label{sec:stokes-bps}

\subsection{Stokes phenomena in 2d-4d systems}

Studying the large-order behavior of perturbation theory in quantum mechanics, quantum
field theory, or string theory has often given insight into the nature of discrete
objects in the theory such as instantons, particles, or branes. (For one remarkable
example, see \cite{Pol} where it was argued that D-branes are responsible for effects of order
$\E^{-1/g_s}$ in string perturbation theory. We refer to \cite{mmlargen} for a review and list of references.)

An interesting class of examples comes from supersymmetric coupled 2d-4d systems,
consisting of a 4d $\CN=2$ theory and a BPS surface defect
preserving 2d $\CN=(2,2)$ supersymmetry. In this case one has
observables $\psi_i$ in the defect theory, interpreted as boundary states
associated to vacua, with the index $i$ labeling
the choice of vacuum. The $\psi_i$ depend on a parameter~$\hbar \in \IC^\times$
which can be interpreted as an $\Omega$ background parameter in the NS phase.
They admit an asymptotic series expansion as $\hbar \to 0$, and
can be computed directly by Borel summation of that series.
This Borel summation suffers from Stokes phenomena, directly associated with BPS
particles in the defect theory. Indeed, one has a sharp relation \cite{Grassi:2021wpw}
\begin{equation} \label{eq:xi-Z-4d}
 \xi = -Z,
\end{equation}
where $\xi$ is the position of a singularity in the Borel plane for
$\psi_i$, and $Z$ is the central charge of a BPS particle which is in vacuum $i$ at $-\infty$. In general all of the quantities we
consider --- $\psi_i$, $\xi$ and $Z$ --- are holomorphic functions of
other parameters $x$, which represent couplings and moduli in the 2d-4d system.\footnote{More precisely,
in the Borel plane, there can be multiple singularities along any given ray emanating
from the origin, and there could be multiple BPS particles with central charges
along a given ray as well. In the
relation~\eqref{eq:xi-Z-4d}, and in similar relations we write below,
$\xi$ is to be interpreted
as the first pole along a given ray, and $Z$ similarly as the first central charge
along a given ray.}

Because of the Borel plane singularities, the observable
$\psi_i$ is only piecewise analytic as a~function of parameters $(\hbar,x)$;
it jumps when $\arg \hbar = \arg \xi(x)$, which using \eqref{eq:xi-Z-4d} means it
jumps at the loci
\begin{equation} \label{eq:jump-locus}
\arg \hbar = \arg \left(-Z(x)\right),
\end{equation}
where $Z$ ranges over the central charges of BPS particles in vacuum $i$ at
$-\infty$.

\subsection{Chiral couplings and spectral networks}

The best-explored examples of this story arise in UV complete 2d-4d systems,
such as Lagrangian theories and theories of class $S$ \cite{Gaiotto:2011tf}. In these examples,
one considers a surface defect with a~parameter space $C$ of chiral couplings,
where perturbation along $C$ is accomplished by
adding the descendant of a chiral operator on the defect.
Thus we now specialize to let $x$ denote a point of $C$, holding other
moduli fixed.

The observables $\psi_i = \psi_i(\hbar,x)$ are flat sections of a connection over $C$; more concretely, they are solutions of a linear ordinary differential
equation over $C$ (e.g., a meromorphic Schr\"odinger equation).
This equation can be viewed as a quantization of the Seiberg--Witten curve $\Sigma \subset T^*C$ determined
by the 2d-4d system; it has been discussed from many different points of view, e.g., \cite{acdkv,adkmv,Gaiotto:2014bza,Hollands:2019wbr,Hollands:2021itj,Jeong:2018qpc,Maruyoshi:2010iu,nrs,ns}.
The perturbation series in $\hbar$ is the usual
WKB series representing solutions of the differential equation,
and the Borel plane singularities
are responsible for Stokes phenomena, as familiar in the exact WKB theory.

Now suppose we consider the Borel summation for fixed $\hbar$, as a function of $x$.
Then \eqref{eq:jump-locus} says that Stokes phenomena occur at codimension-1
walls on $C$.
These walls make up the ``Stokes graph'' or ``spectral network''.
Each wall corresponds to a particular
BPS particle on the surface defect, and carries a label $ij$, where $i$ is the vacuum
at $-\infty$ and $j \neq i$ is the vacuum at $+\infty$.
At a wall with label $ij$, the solution $\psi_i$ jumps by adding
some multiple of the solution $\psi_j$.

\subsection{Flavor masses and exponential networks}

A variant of the above story arises for 2d-4d systems in which the
surface defect supports a~flavor symmetry. For simplicity let us discuss only
the case of a $U(1)$ symmetry. In this case the defect
theory can be deformed by a complex flavor mass $x$, parameterizing
the space $C = \IC$.
This situation is similar to the previous one, with a few new features:
\begin{itemize}\itemsep=0pt
\item
The Seiberg--Witten curve $\Sigma \subset C \times \IC^\times$ rather than
$T^* C$. Correspondingly,
the observables~$\psi_i$ are solutions of a difference equation in $x$
involving shifts $x \mapsto x - 2 \pi \I \hbar$, rather than a~differential equation.

\item Each $\psi_i$ depends on an additional $\IZ$-fold choice; changing this
choice multiplies $\psi_i$ by~$\E^{n x / \hbar}$ for some $n \in \IZ$.
This operation corresponds
to modifying the boundary condition by adding a supersymmetric flavor Wilson line with charge $n$.
\item As before, we expect that the local solutions $\psi_i$ experience
Stokes phenomena at walls in~$C$ determined by the
equation \eqref{eq:jump-locus}. These walls make
up a generalized kind of spectral network, which is a very
simple example of the notion of ``exponential network'' considered
in \cite{Banerjee:2018syt,Banerjee:2019apt,Banerjee:2020moh,eager,Longhi:2021qvz}.
In these
examples, any BPS particle which is charged under the defect flavor symmetry will
have mass depending on the parameter $x$. In particular, this can include
BPS particles which sit in a single
vacuum $i$ rather than interpolating from one vacuum to another.
Thus exponential networks generally include walls with labels $ii$, as well as
the more familiar ones with labels $ij$.
\end{itemize}

For a simple example, we could consider the case where the 4d system is actually trivial, and take the 2d-4d system to be a 2d $\mathcal{N}=(2,2)$ theory with one chiral multiplet. We turn on a~complex flavor mass $x$ for the $U(1)$ flavor symmetry. This theory has a Landau-Ginzburg model description, and one can get its Seiberg--Witten curve by minimizing the potential, found for example in \cite{Banerjee:2018syt,Hori:2000kt,Okuda:2015yra, Park:2015kra}.
The Seiberg--Witten curve is simply
\[
x=Y, \qquad Y\in\mathbb{C}^\times,\]
where $Y=\E^y$. The corresponding difference equation is
\[ x\psi(x)=\psi(x - 2 \pi \I \hbar), \]
which has a solution involving the gamma function (note that if $-2 \pi \I \hbar = 1$ this equation becomes
exactly the functional equation of the gamma function).

In this paper, we will not explore this kind of example in detail. Rather we
move on directly to the next case.

\subsection{Compactified 3d-5d systems and exponential networks}\label{intro3d5d}

Now we come to the type of examples we consider in this paper.
We start with a 3d-5d system,
consisting of a 5d $\CN=1$ theory coupled to a defect preserving 3d $\CN=2$,
with a $U(1)_f$ flavor symmetry on the defect.
After compactification on $S^1$ we obtain a 2d-4d system with~${U(1)_f \times U(1)_K}$
flavor symmetry, with $U(1)_K$ coming from shifts along the compactification
circle. Then, as above, we consider deforming by a flavor
mass for $U(1)_f$. In this case the imaginary part of the flavor
mass comes from the log-holonomy of a background $U(1)_f$ connection around the compactification circle, and invariance under
large gauge transformations of the background field implies
that the theory with mass $x/R$ is equivalent to the theory with mass
$(x + 2 \pi \I) / R$. Said otherwise, the parameter space of inequivalent theories
is actually $C = \IC^{\times}_X$, parameterized by~$X = \exp(x)$.\footnote{In passing
from pure 2d-4d systems to 3d-5d systems on a circle, we have renormalized our parameters:
the dictionary is $\hbar_{\text{3d-5d}} = 2 \pi R \hbar_{\text{2d-4d}}$ and $x_{\text{3d-5d}} = R x_{\text{2d-4d}}$.
This renormalization is not really necessary, but it matches the conventions in the literature,
which tends to use dimensionless parameters in the 3d-5d context. The difference
between the renormalizations of $x$ and $\hbar$ accounts for some shifts in factors
of $2 \pi$.}
Then, the picture we expect is
\begin{itemize}\itemsep=0pt
\item
The Seiberg--Witten curve $\Sigma \subset \IC^{\times}_X \times \IC^{\times}_Y$.
The observables $\psi_i$ are solutions of a
$q$-difference equation
in $X$, involving shifts $X \mapsto q X$ with $q = \E^{\I \hbar}$.
These difference equations have again been studied from various points of view,
e.g., \cite{adkmv,Beem:2012mb,dgg,mmrev}.
\item Each $\psi_i$ depends on an additional $\IZ^2$-fold choice; changing this
choice multiplies $\psi_i$ by~\smash{$\E^{(2 \pi n (x+\I \pi) + 4 \pi^2 \I m) / \hbar}$} for some
$(n,m) \in \IZ^2$.
This operation corresponds
to modifying the boundary condition by adding a supersymmetric $U(1)_f \times U(1)_K$ Wilson line with charges $(n,m)$.
\item\label{sec2:CS}
At phase $\vartheta=\pm \frac{\pi}{2}$ we have
some special features (arising ultimately from the fact that the supersymmetric
boundary conditions at this phase descend from Lorentz invariant boundary conditions
in 3d \cite{Cecotti:2013mba}.)
For example, under analytic continuation $x \to x + 2 \pi \I$, each $\psi_i$ is multiplied by
$\E^{2 \pi k_i (x+\pi \I) / \hbar}$, for some $k_i \in \IZ$. The constant $k_i$ is the
effective flavor Chern--Simons level of the 3d theory in vacuum $i$; this transformation law reflects the
fact that a supersymmetric domain wall in which $x$ shifts by~$2\pi\I$ is
equivalent to a~supersymmetric $U(1)_f$ Wilson line with charge $k_i$.\footnote{We can roughly understand this as follows. We consider the 3-dimensional theory in $D \times S^1$, where $D$ denotes the disc, with a domain-wall background, where $\oint_{S^1} A$ shifts
by $2 \pi$ in a small neighborhood of a loop $\ell \subset D$.
This shift requires that the background field has nontrivial curvature, and
in the presence of this curvature the Chern--Simons term $k \int_{S^1 \times D} A \wedge F$ becomes $k \oint_{\ell} A$, a flavor Wilson line with
charge $k$. The statement we are after is a supersymmetrization of this one.}
The $k_i$ can depend on ${\mathrm {Re}}\,x$, since the effective Chern--Simons level
can jump as we vary the flavor mass parameter of the 3d theory.
\item
The $\psi_i$ suffer Stokes phenomena
associated to BPS particles living on the compactified defect.
Precisely, we expect the positions $\xi$ of the Borel plane poles to be given by
\[
 \xi = -2 \pi R Z.
\]
Since $R$ is real, this would lead to the jump locus for the $\psi_i$ being given by
\eqref{eq:jump-locus} just as in the 2d-4d case. Thus again
we expect that the $\psi_i$ jump at the walls of the exponential network
determined by the 3d-5d BPS spectrum.
Said otherwise, we expect that the exponential network plays the role of a Stokes
graph for the difference equation obeyed by the $\psi_i$.

\end{itemize}

This is the picture we will check below, in two simple examples, where the
defect theory has only a single vacuum. These examples
isolate one of the key new phenomena in the cases with flavor mass,
namely the walls of type $ii$ ---
indeed they have only the walls of type $ii$!

In the examples we consider, the local solutions
$\psi_i$ are combinations of variants
of the quantum dilogarithm function. The key technical
advance which makes our study possible is the work~\cite{Garoufalidis:2020pax},
where the Borel poles for the $\hbar$-expansion of this function are
determined.

The difference equations we consider also arise in a different context,
that of $A$ model topological strings on a Calabi--Yau threefold $X$ with a
D-brane placed on a Lagrangian submanifold~$L \subset X$, e.g., \cite{acdkv,adkmv,Dijkgraaf:2007sw}.
In this language $\Sigma$ is the mirror curve of $X$.
One can try to connect this directly to a 3d-5d system by considering M-theory on $X \times \IR^5$ with an M5-brane
on $L \times \IR^3$,
with some appropriate regularization to take care of the non-compactness of $X$ and~$L$;
this setup has been used often in the literature beginning with
\cite{Ooguri:1999bv}.
We will not try to make the
connection between the two pictures directly here, but freely use both languages.

\section[A simple model: C\^{}3]{A simple model: $\boldsymbol{\mathbb{C}^3}$}\label{sec:c3}

A simple example of the difference equations that we study in this paper arises as
the quantized mirror curve of $\mathbb{C}^3$.
We first remind the readers of the basic setup. The mirror curve is
\[
\Sigma=\{X-Y+1=0\}\subset\mathbb{C}^\times_X\times\mathbb{C}^\times_Y. \]
$\Sigma$ is a thrice-punctured sphere, with the punctures at
\[ \{(X,Y)\}_{\rm sing}=\left\{(0,1),(-1,0),(\infty,\infty)\right\}.\]
We will also use the logarithmic variables
\[
x=\log(X),\qquad y=\log(Y).\]
Below we will sometimes need to pick a specific branch, e.g., in writing explicit formulas
for local solutions; we will always take the principal branch, i.e.,
\be\label{imx}-\pi<\operatorname{Im}(x)\leq \pi.\ee
In these variables the mirror curve becomes
\[ \E^x-\E^y+1=0.\]
We choose $ \mathbb{C}^{\times}_X$ to be the base of $\Sigma$, and will later introduce the exponential network on this base.

Quantization of this curve gives rise to the quantum mirror curve.
Our convention for the quantum mirror,
\be\label{c3diff} \bigl(\E^{\hat{p}}-1-\E^{\hat{x}}q^{-\frac{1}{2}}\bigr)\psi(x, \hbar)=0,\qquad \left[\hat x, \hat p\right]=\ri \hbar, \ee
is the same as the one used in \cite{Garoufalidis:2020pax}.\footnote{If we had chosen the alternative convention
$\bigl(\E^{\hat{p}}-1-\E^{\hat{x}}\bigr)\psi(x, \hbar)=0$, we would get slightly different series, but all the structure we
consider in this paper, particularly the positions of the Borel plane singularities, would be the same.}

\subsection{All-orders WKB expansion of local solutions}\label{openc3}

For convenience, rather than studying local solutions $\psi$ of \eqref{c3diff} directly, we study $\phi(x,\hbar) = \log \psi(x,\hbar)$ which satisfies the difference equation
\[\phi\left(x+\ri\frac{\hbar}{2},\hbar\right)-\phi\left(x-\ri\frac{\hbar}{2},\hbar\right)=-\log\left(1+\E^x\right).\]
This equation is solved by a formal series (see, for example, \cite[Section~2.1]{Garoufalidis:2020pax}):\footnote{We did not obtain the series solution \eqref{formallocal} by a direct WKB-type expansion of a solution of the difference equation; rather we just lifted it from
\cite{Garoufalidis:2020pax}. However, using the difference equation and the boundary condition obeyed by \eqref{formallocal}, one can show that it does match with the result of a WKB-type expansion.}
\be\label{formallocal}\phi(x,\hbar) \sim \frac{1}{\ri\hbar}\operatorname{Li}_2(-\E^x)+\sum_{k\geq 1}\frac{B_{2 k}\left(\frac{1}{2}\right) (\ri \hbar )^{2 k-1}}{(2 k)!}\operatorname{Li}_{2-2 k}\left(-\E^{x}\right).\ee
There is a $\IZ \times \IZ$ ambiguity here associated with the choice of branch for $\Li_2$.
The resulting ambiguity of the local solutions
is the one discussed in Section~\ref{intro3d5d}.
When we write explicit formulas, we will always resolve this
ambiguity by choosing the principal branch of $\Li_2$.

The formal series is not the actual analytic solution that we seek, but the Borel summation of it is.
The Borel transform of \eqref{formallocal}, as defined in \eqref{bdefo}, can be rewritten as \cite{Garoufalidis:2020pax}
\be\label{hadamardc3}{\bf \mathcal{B}}\phi(x,\xi)=-\frac{\ri}{4\pi^2}\sum_{n=1}^\infty\frac{(-1)^n}{n^2}\biggl(\frac{1}{1+\E^{\frac{\xi}{2\pi n}-x}}+\frac{1}{1+\E^{-\frac{\xi}{2\pi n}-x}}\biggr).\ee
We can see that ${\bf \mathcal{B}}\phi(x,\xi)$ has singularities in the Borel plane located at
\be\label{c3pole}
 \xi^*(x,m,n)\equiv 2 \pi n \left(x +\pi \ri (2m+1) \right),\qquad m\in\mathbb{Z},\qquad n \in \IZ/\{0\},\ee
with residues \be\label{C3residue}-\frac{(-1)^n}{2\pi\ri n}.\ee

Hence when
 \[
 \arg(\hbar)=\arg(\xi^*(x,m,n)),\]
 the Borel summation \eqref{boreldef} is not defined.
Nevertheless, it is analytic for sufficiently small variations of $\hbar$, those for
which the contour of integration in the Borel summation
does not go through any of the poles in the Borel plane.

For example, it has been proved in \cite{Garoufalidis:2020pax} that
\begin{align}\label{matching}
s(\phi(x, \hbar))&=\log\left(\Phi(x,\hbar)\right)
, \qquad \arg(\hbar)=0,\qquad -\pi<\operatorname{Im}(x)\leq \pi,\end{align}
 where we define
\be\label{definePhix}\Phi(x,\hbar)\equiv\Phi_{\bf b}\left(\frac{x}{2\pi {\bf b}}\right),\ee
with $\Phi_{\bf b}(x)$ the quantum dilogarithm function of Faddeev \cite{Faddeev:1995nb}, and
\[\hbar=2 \pi {\bf b}^2.\]
More about the quantum dilogarithm can be found in Appendix~\ref{ap:fad}. For $\operatorname{Im}(\hbar)\neq 0$, we can express the Faddeev quantum dilogarithm in terms of $q$-Pochhammer symbols; for example we use \eqref{phiq} for $\operatorname{Im}(\hbar)> 0$. For $\operatorname{Im}(\hbar)= 0$, we need to use the integral expression \eqref{definePhixint}.

 Likewise, it was shown in \cite{Garoufalidis:2020pax} that
\begin{align}\label{phpi}
s(\phi(x,\hbar))&=\log\bigl(\bigl(-q^{\frac{1}{2}}\E^x;q\bigr)_\infty\bigr),
 \end{align}
 for $\arg(\hbar)=\frac{\pi}{2}$, $-\pi<\operatorname{Im}(x)\leq \pi$, $\operatorname{Re}(x)<0$.

We will discuss the Borel summation and its analytic structure in the full Borel plane in Section~\ref{Locs}.

\subsection{Field theory and BPS states}\label{3d5d}

Now we recall that
$\mathbb{C}^\times_X$ is not only the base of the 5d Seiberg--Witten curve: it also plays the role of a parameter space of flavor mass couplings in the
$S^1$ compactification of a 3d-5d system.
In the example we are discussing now, the 5d system is actually trivial, so rather than a 3d defect we are just
considering a 3d $\CN=2$ field theory with a $U(1)$ flavor symmetry. The 3d
theory is the ``tetrahedron'' theory of \cite{dgg},
which can be described as the Lagrangian field theory of~a~single 3d chiral multiplet with charge $1$ under
the $U(1)$ flavor symmetry, plus a background
Chern--Simons coupling at level $-\frac{1}{2}$.

We compactify the theory on $S^1$ and consider the spectrum of BPS particles
in the compactified theory.
Since the theory is free, this spectrum can be described simply: the single chiral multiplet of the 3d theory
gives rise to an infinite Kaluza-Klein tower of chiral multiplets in 2d, and each one of these leads
to a single BPS particle and its corresponding antiparticle. The central charges are
\begin{equation} \label{C33dZ}
 Z = \pm R^{-1} (x + \pi \I(2m+1)),
\end{equation}
where the integer $m$ keeps track of the KK momentum.

\subsection{Exponential network}

The exponential network $\mathcal{W}^\vartheta$ on $\mathbb{C}^\times_X$ is defined as the set of points $X\in \mathbb{C}^\times_X$, such that in the theory with parameter $X$ there exists a BPS particle satisfying
\be\label{2dcc} \arg(-Z(X))=\vartheta.\ee
Combining \eqref{C33dZ} and \eqref{2dcc}, we see that $X\in\mathcal{W}^\vartheta$ if and only if
\be\label{c3trajectory} X=-\E^{\pm s \E^{\ri\vartheta}} \ee
for some $s \in {\mathbb R}_{\geq 0}$.
In Figure~\ref{ENtop}, we show the networks $\CW^\vartheta$ for
several choices of $\vartheta$.\footnote{See \cite{Banerjee:2018syt,eager} for previous studies in another framing.} Three distinct shapes occur:

\begin{figure}[t]
 \centering
 \begin{subfigure}[b]{0.24\textwidth}
 \centering
 \includegraphics[width=\textwidth]{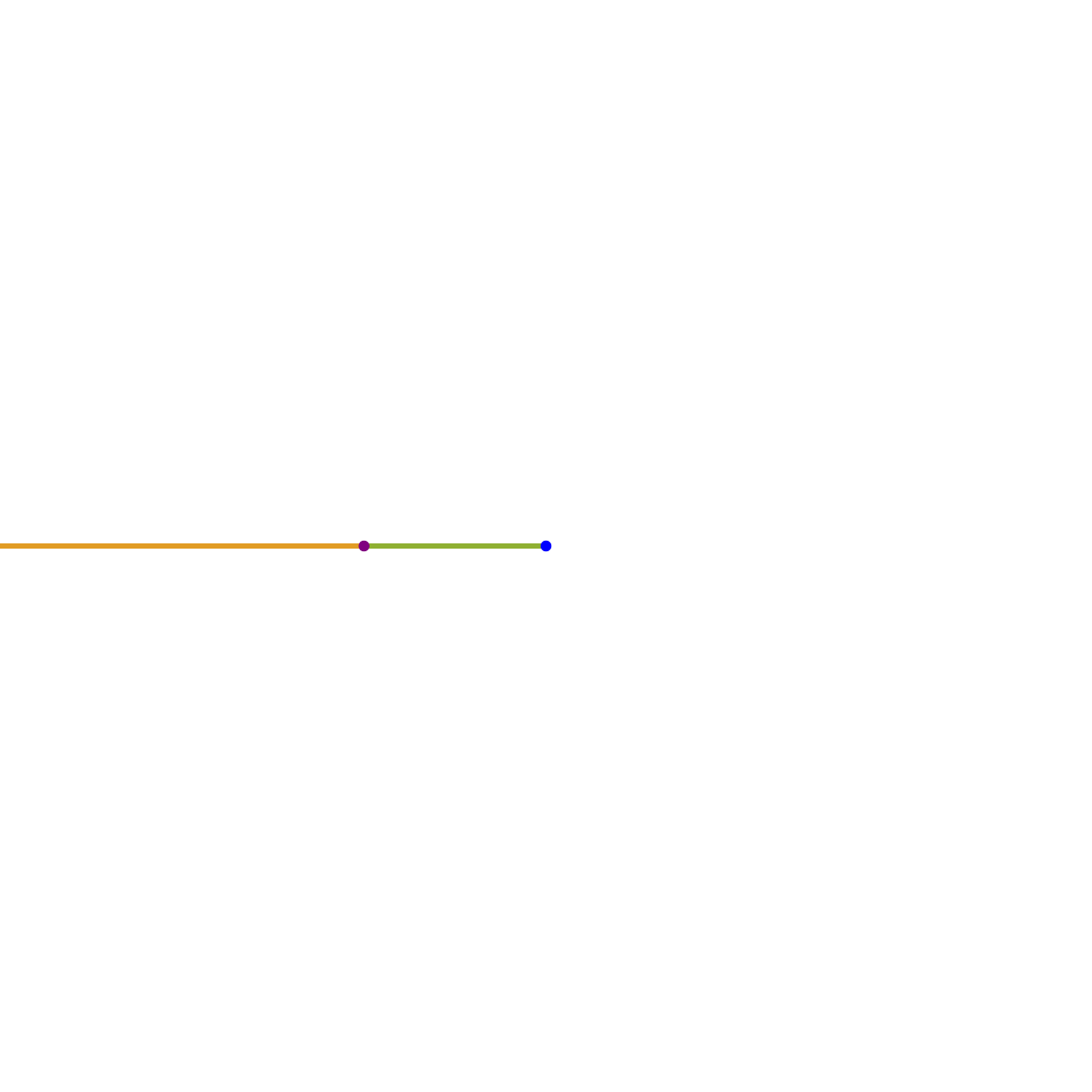}
 \caption{$\vartheta=0$}
 \label{theta0}
 \end{subfigure}
 \begin{subfigure}[b]{0.24\textwidth}
 \centering
 \includegraphics[width=\textwidth]{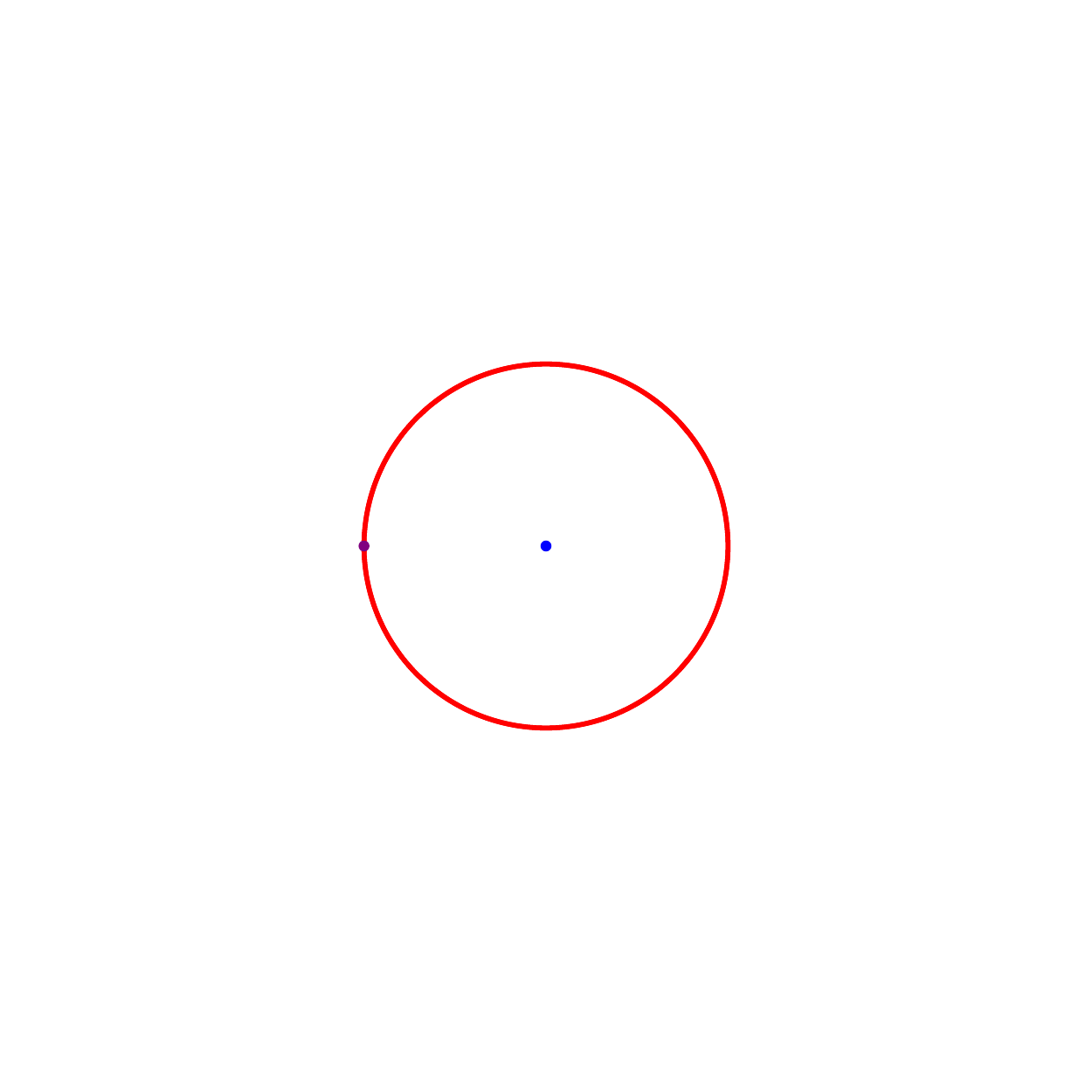}
 \caption{$\vartheta=\pi/2$}
 \label{thetapi2}
 \end{subfigure}
 \begin{subfigure}[b]{0.24\textwidth}
 \centering
 \includegraphics[width=\textwidth]{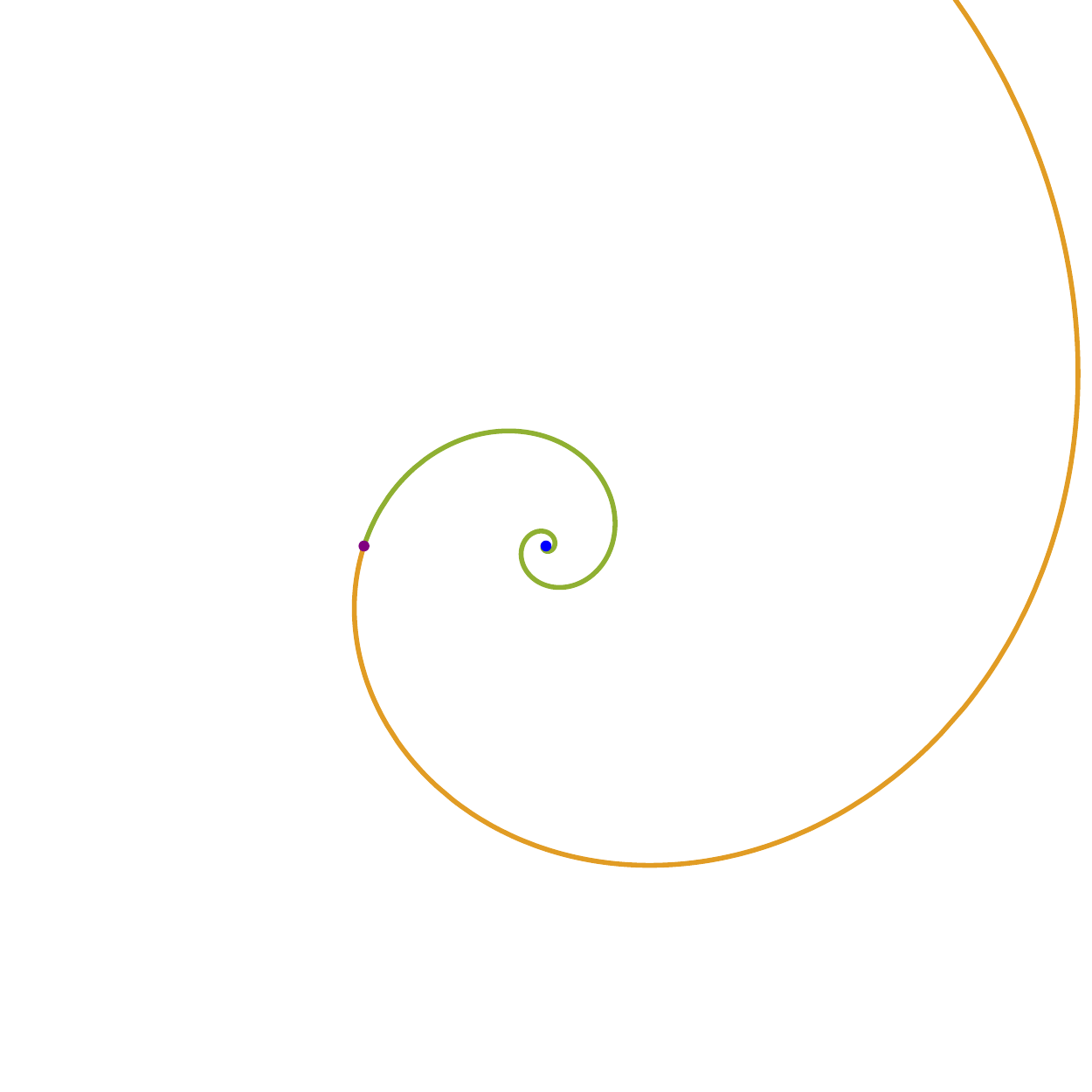}
 \caption{$\vartheta=\frac{2\pi}{5}$}
 \label{theta25}
 \end{subfigure}
 \begin{subfigure}[b]{0.24\textwidth}
 \centering
 \includegraphics[width=\textwidth]{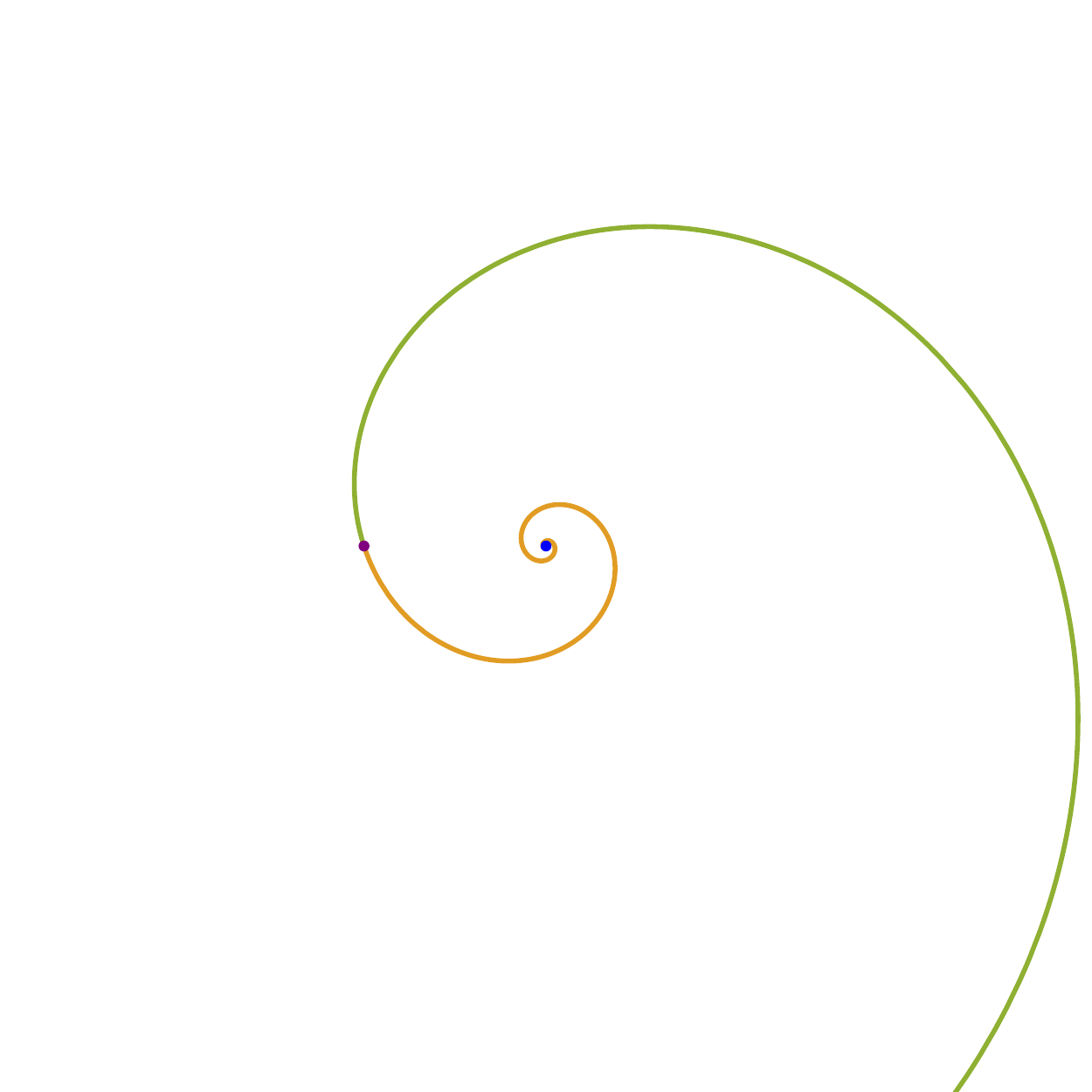}
 \caption{$\vartheta=\frac{3\pi}{5}$}
 \label{theta35}
 \end{subfigure}
           \caption{The exponential networks $\mathcal{W}^{\vartheta}$ on $\mathbb{C}^\times_X$ at various phases $\vartheta$. The blue dot at $X=0$ and purple dot at $X=-1$ represent punctures of $\Sigma$. The orange wall is the locus $X=-\E^{ s \E^{\ri\vartheta}}$ and green wall is the locus $X=-\E^{ -s \E^{\ri\vartheta}}$. The degenerate wall at phase $\vartheta=\pi/2$ is painted in~red. }
 \label{ENtop}
 \end{figure}
\begin{itemize}\itemsep=0pt
\item[(1)] $\vartheta=0$ or $\vartheta = \pi$:
$\mathcal{W}^{\vartheta}$ consists of 2 straight walls ending at $X=0$ and $X=\infty$ respectively. See Figure~\ref{theta0}.
\item[(2)] $\vartheta=\frac{\pi}{2}$ or $\vartheta = - \frac{\pi}{2}$:
$\mathcal{W}^{\vartheta}$ consists of a degenerate wall which is a circle of radius $1$ around~${X=0}$. See Figure~\ref{thetapi2}.
\item[(3)] $\vartheta\neq n\frac{\pi}{2}$, $n\in\mathbb{Z}$:
$\mathcal{W}^{\vartheta}$ consists of two spirals. One spiral ends at $X=0$ and is contained in the region~$0<|X|<1$. The other spiral ends at $X=\infty$ and is contained in the region~$|X|>1$. The correspondence between walls in \eqref{c3trajectory} and spirals, as well as the orientation of the spirals, depend on the phase $\vartheta$. Examples are shown in Figures~\ref{theta25} and~\ref{theta35}.
\end{itemize}

\subsection{The exponential network and exact WKB}

Now recall the basic picture we proposed in Section~\ref{sec:stokes-bps}:
there are canonical formal WKB solutions, and the
poles $\xi$ in the Borel plane for these formal solutions are
related to the central charges~$Z$
of BPS particles in the 3d-5d system on $S^1$, via the formula
\begin{equation} \label{eq:R-xi-relation}
 \xi =- 2 \pi R Z.
\end{equation}
More precisely,
in the Borel plane, there can be multiple poles in each direction, and
in the relation \eqref{eq:R-xi-relation}, $\xi$ is to be interpreted
as the first pole in any given direction.

Let us see whether this relation holds in the $\IC^3$ example.
In \eqref{c3pole}, we see infinite sequences of poles,
distinguished by the multiplicity $n\in\mathbb{Z}\backslash\{0\}$;
we consider the first pole in each sequence,~i.e.,
\[
\xi = \xi^*(x, m, n= \pm 1) = \pm 2\pi (x + \pi \I(2m+1)).
\]
On the other hand, \eqref{C33dZ} says that
\[
 Z = \pm R^{-1} (x + \pi \I(2m+1)).
\]
Thus we see directly
 that the relation \eqref{eq:R-xi-relation} indeed holds in this
 example.

\begin{figure}[t]
 \centering
 \includegraphics[width=0.7\textwidth]{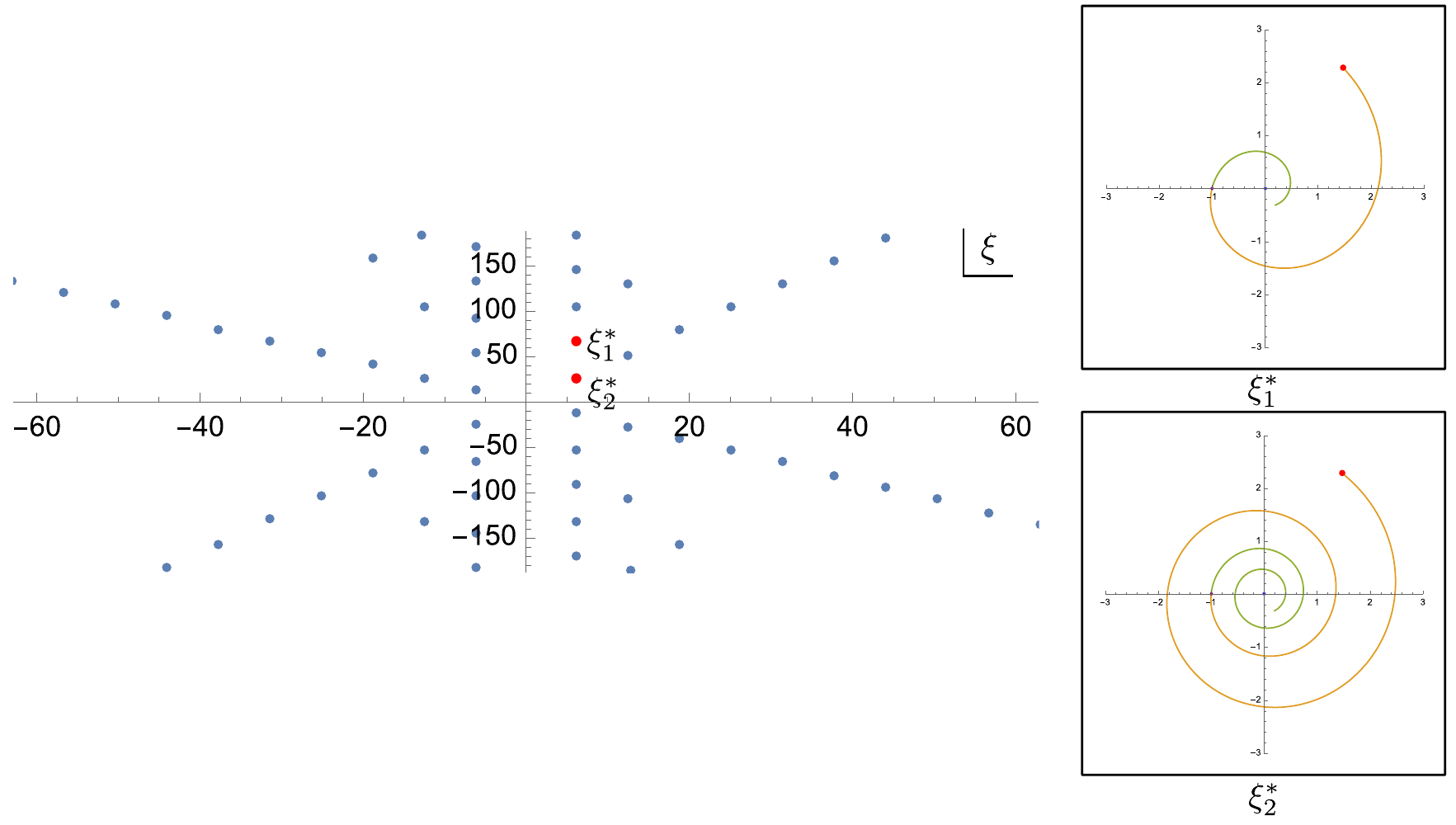}

 \caption{Left: Singularities of the Borel transform in the Borel plane when $x = 1 + \I$. The two points in red are $\xi^*_1=\xi^*(x,0,1)$ and $\xi^*_2=\xi^*(x,1,1)$. Right: Truncated exponential network~$\mathcal{W}^{\arg(\xi^*_i)}(|\xi^*_i|)$, $i=1,2,$ on $\mathbb{C}^{\times}_X$. The red dots are at the point $X(x)$.}
 \label{pole3d5d}
 \end{figure}

A graphical interpretation of this statement is that, for any $m \in \IZ$,
the truncated exponential network $\mathcal{W}^{\arg(\xi^*(x,m,\pm 1))}(|\xi^*(x,m,\pm 1)|)$ ends at the point $X=\E^x$.
We illustrate this in two examples in Figure~\ref{pole3d5d} by plotting the
truncated networks directly.

\subsection{Local solutions in each sector} \label{Locs}

In this section we review the Borel summation \eqref{boreldef} of the local solution \eqref{formallocal} and discuss the corresponding jumps as we move in the Borel plane. Ultimately the reason for these jumps is due to the fact that the integration contour in equation \eqref{boreldef} depends on $\hbar$.
 Consequently, when we change the contour, we cannot do so continuously because of the existence of poles \eqref{c3pole} in the Borel transform \eqref{hadamardc3}.
In addition, since these poles $\xi^*(x,m,n)$ depend on $x$, $m$, $n$, we need to separate the discussion by quadrants in the Borel plane and the sign of $\operatorname{Re}(x)$.

Borel resummation of local solutions for the case $\operatorname{Re}(x)<0$ and ${\rm arg}(\hbar)\in\big[0,\frac{\pi}{2}\big]$ was discussed in
 \cite{Garoufalidis:2020pax}. Here we complete the analysis by exploring other regions as well.
 This extended analysis will also be useful in the study of the resolved conifold: see Section~\ref{sec:resconi}.

We will elaborate only on the case of the first quadrant; all other cases can be found in Appendix~\ref{summaryls} and are summarised in Figure~\ref{fgsummaryls}.

\begin{figure}[t]
\begin{center}
\includegraphics[width=0.8\textwidth]{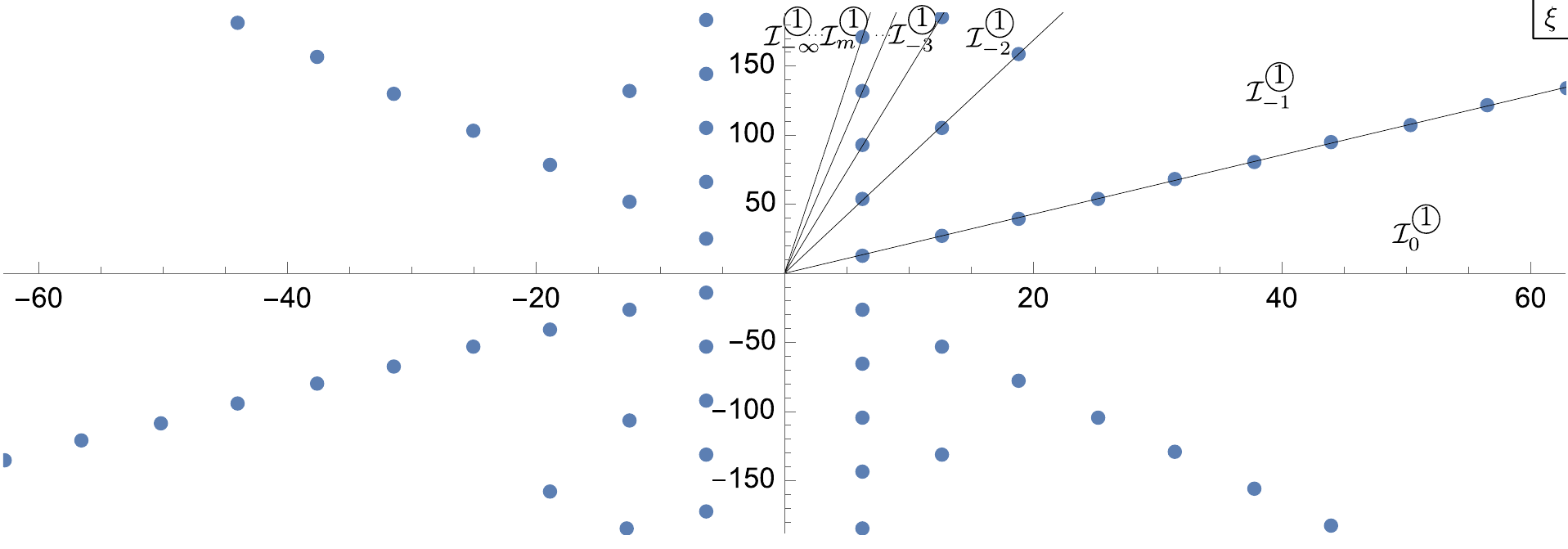}
\caption{ The Borel plane of the local solution \eqref{formallocal} for $ x=-1+\ri.$
 The rays of poles in the first quadrant separate the Borel plane into sectors ${\mathcal I}^{\tcircled{1}}_m$, $m\leq 0$.}
\label{c3borel}
\end{center}
\end{figure}

A sample Borel plane is shown in Figure~\ref{c3borel}.
We can see that there is an infinite number of rays containing poles, with phases
\[
 \vartheta_{x,m}^\pm=\arg\left(\pm\left(x+\pi\ri\left(2m+1\right)\right)\right), \qquad m \in\mathbb{Z}.\]
These rays divide the Borel plane into sectors.
We define the sector containing the positive real axis as
\[
\mathcal{I}_{0}^{\tcircled{i}},\]
 where $\tcircled{i}$ denotes the quadrant.\footnote{$\mathcal{I}_{0}^{\tcircled{1}}$ and $\mathcal{I}_{0}^{\tcircled{4}}$ refer to the same sector. Same for $\mathcal{I}_{0}^{\tcircled{2}}$ and $\mathcal{I}_{0}^{\tcircled{3}}$.}

\subsubsection[First quadrant of the Borel plane and Re(x)<0]{First quadrant of the Borel plane and $\boldsymbol{\operatorname{Re}(x)<0}$}
\label{sec:1strexl0}

In this case the relevant sectors of the Borel plane are
\begin{align*}
 {\mathcal I}^{\tcircled{1}}_{x,m}= \bigl(\vartheta_{x,m}^{ -}; \vartheta_{x,m-1}^{-}\bigr), \qquad m\leq -1.\end{align*}
For convenience, we neglect the subscript $x$ and simply use
\[ {\mathcal I}^{\tcircled{i}}_{m}\equiv{\mathcal I}^{\tcircled{i}}_{x,m}.\]

To calculate the jump of the solution in the first quadrant, we assume that $\vartheta$ is independent of~$\hbar$ and study the two contour integrals along rays whose phases are given by
 \begin{align*}
 \vartheta^-&\in{\mathcal I}_{m+1}^{\tcircled{1}},\qquad \vartheta^+\in {\mathcal I}_{m}^{\tcircled{1}},
 \end{align*}
 respectively.

 The jump of the solution when crossing the ray of phase $\vartheta_{x,m}^-$ is
 \begin{align*}
&-2\pi\ri \sum_{n=1}^\infty\res\bigl({\bf \mathcal{B}}\phi(x,\xi)\E^{-\frac{\xi}{\hbar}},-2\pi n(x+(2m+1)\pi\ri)\bigr)\\
&\qquad=-2\pi\ri \sum_{n=1}^\infty\frac{(-1)^n}{2\pi\ri n}\E^{\frac{2\pi n(x+(2m+1)\pi\ri)}{\hbar}}=\log\bigl(1+\E^{\frac{2\pi (x+(2m+1)\pi\ri)}{\hbar}}\bigr),\qquad m\leq -1,
 \end{align*}
where we have used \eqref{C3residue}.

Using the explicit expressions of the jumps and the solution along the positive real axis \eqref{matching}, the Borel summation of \eqref{formallocal} for $\arg (\hbar)\in\mathcal{I}_m^{\tcircled{1}}$ is given by
\begin{align*} s(\phi)(x,\hbar)=&\log\Bigg(\frac{\bigl(-q^{\frac{1}{2}}\E^x;q\bigr)_\infty}{\bigl(-\tilde{q}^{\frac{1}{2}}\E^{\frac{2\pi x}{\hbar}};\tilde{q}\bigr)_{\infty}}\prod_{i=0}^{-(m+1)}\bigl(1+\E^{\frac{2\pi x}{\hbar}}\tilde{q}^{\frac{1}{2}}\tilde{q}^{i}\bigr)\Bigg)\\
=&\log\Bigg(\frac{\bigl(-q^{\frac{1}{2}}\mathrm{e}^x;q\bigr)_\infty}{\bigl(-\tilde{q}^{\frac{1}{2}}\mathrm{e}^{\frac{2\pi x}{\hbar}};\tilde{q}\bigr)_{\infty}}\bigl(-\mathrm{e}^{\frac{2\pi x}{\hbar}}\tilde{q}^{\frac{1}{2}},\tilde{q}\bigr)_{-m}\Bigg),
\end{align*}
where the $q$-Pochhammer symbol is defined as \be\label{defineqpoch}(a;q)_n=\prod_{i=0}^{n-1}\bigl(1-aq^i\bigr).\ee
Using \eqref{definePhix} and \eqref{sh2}, this can also be written as
\be\label{1stlocal}\boxed{s(\phi)(x,\hbar)=\log\Phi(x+2\pi\mathrm{i} m,\hbar),\qquad \arg(\hbar)\in \mathcal{I}_m^{\tcircled{1}}, \qquad m\leq -1.}\ee
We will discuss the appearance of $\Phi(x+2\pi\ri m)$ from the point of view of analytic continuation in Section~\ref{jumpwall}.
Note that
\be\label{pisol}
\lim_{m\rightarrow-\infty}\frac{\bigl(-q^{\frac{1}{2}}\E^x;q\bigr)_\infty}{\bigl(-\tilde{q}^{\frac{1}{2}}\E^{\frac{2\pi x}{\hbar}};\tilde{q}\bigr)_{\infty}}\bigl(-\E^{\frac{2\pi x}{\hbar}}\tilde{q}^{\frac{1}{2}},\tilde{q}\bigr)_{-m}=\bigl(-q^{\frac{1}{2}}\mathrm{e}^x;q\bigr)_\infty.\ee
Hence we get \eqref{phpi}
\be \label{sne}\boxed{s(\phi)(x,\hbar)=\log \bigl(-q^{\frac{1}{2}}\mathrm{e}^x;q\bigr)_\infty, \qquad \hbar \in \ri\IR_+, \qquad \operatorname{Re}(x)<0.}\ee
So far this was as in \cite{Garoufalidis:2020pax}.
Let us now look instead at $\operatorname{Re}(x)>0$.

\subsubsection[First quadrant of the Borel plane and Re(x)>0]{First quadrant of the Borel plane and $\boldsymbol{\operatorname{Re}(x)>0}$}
 \label{sec:1strexg0}

In this case the relevant
sectors of the Borel plane are
\begin{align*}
{\mathcal I}_{m}^{\tcircled{1}}= \left(\vartheta_{m-1}^+; \vartheta_{m}^+\right),\qquad m\geq 1,\end{align*}
with the understanding that $ {\mathcal I}_0^{\tcircled{1}}$ is the sector containing the real axis.

We use the sum of residues to get the jump
\begin{align*}
(-2\pi\ri)\sum_{n=1}^\infty\sum_{k=0}^{m-1}-\frac{(-1)^n}{2\pi\ri n}\E^{-\frac{n(2\pi(x+(1+2k)\pi\ri))}{\hbar}}=\log\left(\frac{1}{\bigl(-\tilde{q}^{\frac{1}{2}}\mathrm{e}^{-\frac{2\pi x}{\hbar}};\tilde{q}\bigr)_m}\right),\qquad m\geq 1,
\end{align*}
for $\mathcal{I}_m^{\tcircled{1}}$ from the real axis. The local solution in $\mathcal{I}_m^{\tcircled{1}} $ is thus
\be\label{intx}\frac{\bigl(-q^{\frac{1}{2}}\mathrm{e}^x;q\bigr)_\infty}{\bigl(-\tilde{q}^{\frac{1}{2}}\mathrm{e}^{\frac{2\pi x}{\hbar}};\tilde{q}\bigr)_{\infty}}\frac{1}{\bigl(-\tilde{q}^{\frac{1}{2}}\mathrm{e}^{-\frac{2\pi x}{\hbar}};\tilde{q}\bigr)_m},\qquad m\geq 1.\ee
Alternatively, we can express \eqref{intx} as
\be \label{2ndlocal} \boxed{s(\phi)(x,\hbar)= \log \Phi(x+2\pi\ri m,\hbar)+\frac{2\pi m(x+m\pi\ri)}{\hbar}, \qquad \arg(\hbar) \in \mathcal{I}_m^{\tcircled{1}}, \qquad m\geq 1.} \ee
We will discuss more about this expression from the point of view of analytic continuation in Section~\ref{jumpwall}.

The solution along positive imaginary axis for $\operatorname{Re}(x)>0$ is obtained from
\begin{align*}\lim_{m\rightarrow \infty}\frac{\bigl(-q^{\frac{1}{2}}\mathrm{e}^x;q\bigr)_\infty}{\bigl(-\tilde{q}^{\frac{1}{2}}\mathrm{e}^{\frac{2\pi x}{\hbar}};\tilde{q}\bigr)_{\infty}}\frac{1}{\bigl(-\tilde{q}^{\frac{1}{2}}\mathrm{e}^{-\frac{2\pi x}{\hbar}};\tilde{q}\bigr)_m}&=\frac{\bigl(-q^{\frac{1}{2}}\mathrm{e}^x;q\bigr)_\infty}{\bigl(-\tilde{q}^{\frac{1}{2}}\mathrm{e}^{\frac{2\pi x}{\hbar}};\tilde{q}\bigr)_{\infty}}\frac{1}{\bigl(-\tilde{q}^{\frac{1}{2}}\mathrm{e}^{-\frac{2\pi x}{\hbar}};\tilde{q}\bigr)_\infty}\\
&=\frac{\E^{\frac{x^2\ri}{2\hbar}}q^{\frac{1}{24}}}{\tilde{q}^{\frac{1}{24}}}\frac{1}{\bigl(-q^{\frac{1}{2}}\E^{-x};q\bigr)_{\infty}}.
\end{align*}
Hence
\be\label{spo} \boxed{s(\phi)(x,\hbar)=\log\left( \frac{\E^{\frac{x^2\ri}{2\hbar}}q^{\frac{1}{24}}}{\tilde{q}^{\frac{1}{24}}}\frac{1}{(-q^{\frac{1}{2}}\E^{-x};q)_{\infty}}\right), \qquad \hbar \in \ri \IR_+, \qquad \operatorname{Re}(x)>0}. \ee

 \subsubsection[First quadrant of the Borel plane and Re(x)=0]{First quadrant of the Borel plane and $\boldsymbol{\operatorname{Re}(x)=0}$}
 \label{sec:rex0}

 When $\operatorname{Re}(x)=0$ the situation is special since all the poles in the Borel plane lie on the imaginary axis.
 We found that in this case the median summation is
 \begin{gather} \label{re0} s(\phi)(x,\hbar)=\frac{1}{2}\! \left(\frac{\ri \bigl(12 x^2\!+\hbar ^2\!+4 \pi ^2\bigr)}{24 \hbar }-\log\frac{\bigl(-\E^{-x}q^{\frac{1}{2}};q \bigr){}_{\infty }}{\bigl(-\E^{x}q^{\frac{1}{2}};q\bigr){}_{\infty }}\right), \qquad\!\! \operatorname{Re}(x)=0, \quad\! \hbar\in \ri \IR_+.\!\!\!\!\!
 \end{gather}
Note also that $\eqref{re0}=\frac{1}{2}\left(\eqref{sne} + \eqref{spo}\right)$.

 For all the other values of $\hbar$, the Borel summation matches with $\Phi(x)$:
\begin{align*}
s(\phi)(x,\hbar)= \log\Phi(x,\hbar), \qquad \operatorname{Re}(x)=0, \qquad
\operatorname{Re} (\hbar)> 0,\qquad{\operatorname{Im} } (\hbar)\geq 0.
\end{align*}

\subsubsection{Summary and comments}

\begin{figure}[t]\centering
\includegraphics[width=.9\linewidth]{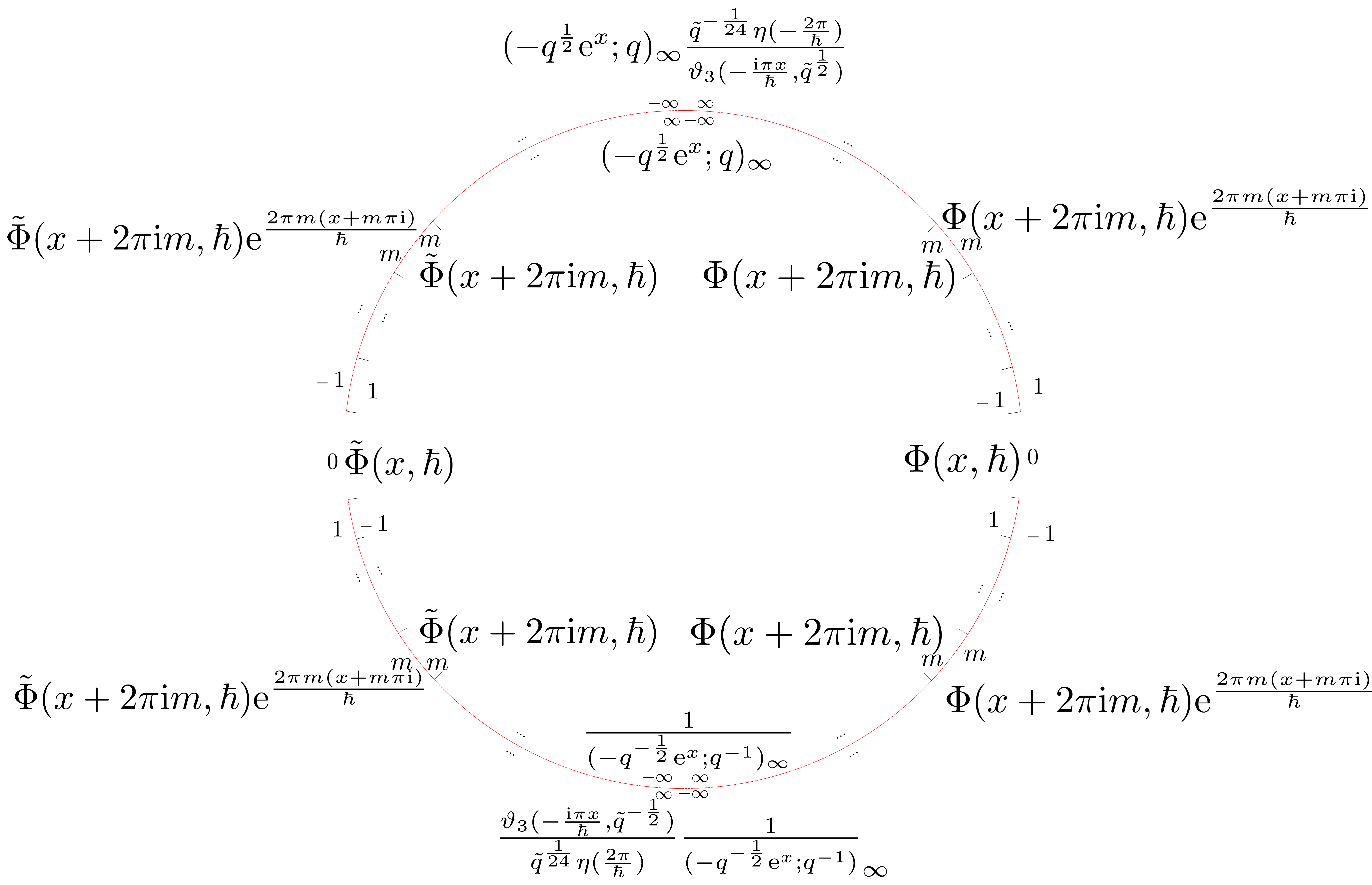}
\caption{A summary of local solutions for all the sectors in the Borel plane. The angle represents~$\arg(\hbar)$. The circle is cut into sectors $\mathcal{I}^{\tcircled{i}}_m$ where $m$ is listed for each sector. If the function is written in the interior or exterior of the red circle, it represents the solution for $\operatorname{Re}(x)<0$ or~$\operatorname{Re}(x)>0$ respectively. In $\mathcal{I}^{\tcircled{i}}_0$, there is a unique solution for both $\operatorname{Re}(x)<0$ and $\operatorname{Re}(x)>0$. For $\mathcal{I}^{\tcircled{i}}_0$, $i=1,4$, this unique solution is $\Phi(x,\hbar)$ and for $\mathcal{I}^{\tcircled{i}}_0$, $i=2,3$, this solution is $\tilde{\Phi}(x,\hbar)$. The functions $\Phi(x,\hbar)$ and $\tilde \Phi(x,\hbar)$ are defined in \eqref{pdef} and \eqref{tpdef}, respectively.
For practical reasons the solution at $\operatorname{Re}(x)=0$ is not shown on the figure but can be found in the main text, see, e.g., \eqref{re0} and item~\ref{appen:rex0}.}
\label{fgsummaryls}
\end{figure}

Calculations for all the other quadrants can be found in Appendix~\ref{summaryls}; we summarize all the local solutions in the whole Borel plane in Figure~\ref{fgsummaryls}.
It is interesting to note that for generic~$\hbar$ there are two kinds of solutions depending on whether
 $\operatorname{Re}(x) < 0$ or $\operatorname{Re}(x) > 0$, which nevertheless coincide when $\hbar\in \IR$. Let us look at the imaginary axis $\hbar\in \ri \IR_+$: the two solutions are~\eqref{sne} and~\eqref{spo}.
Physically $q$-Pochhammer in~\eqref{sne} gives the open topological string partition function on~$\IC^3$ corresponding to an anti-brane where $\ri \hbar=g_s$, see for instance \cite[p.~24]{Kashani-Poor:2006puz}. The other solution~\eqref{spo} can be schematically obtained from~\eqref{sne} using an S transformation, up to an overall \smash{${\bigl(q\tilde{q}^{-1}\bigr)^{\frac{1}{24}}}$} and a shift in the argument.

\subsection{Jumps of local solutions via analytic continuation}
\label{jumpwall}
In this subsection, we use the exponential network to explain why the jumps of local solutions have the form of an analytic continuation, for example as in \eqref{1stlocal}. The discussion depends on which quadrant of the Borel plane we are in. However, they are all similar, so we will only consider the first quadrant, i.e., $\vartheta\in\bigl(0,\frac{\pi}{2}\bigr)$.

We start by noticing that varying $\vartheta$ rotates the integral contour in the Laplace transform, while varying $x$ shifts the poles. However, as long as the poles passing through the integral contour are the same, the jumps obtained by varying $\vartheta$ or $x$ are equivalent. Therefore, we can equally well study the behavior of the solutions on $\mathbb{C}^{\times}_X\backslash\mathcal{W}^\vartheta$, instead of in the Borel plane.

\begin{figure}[t]\centering
\includegraphics[width=0.3\linewidth]{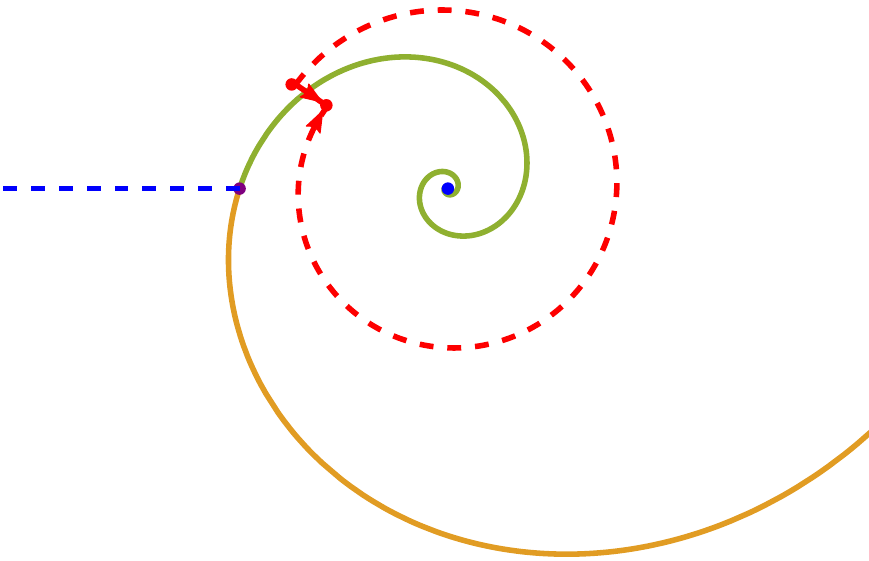}
\caption{Two paths tracking the change of solutions on $\mathbb{C}^{\times}_X$. The blue dashed line is the branch cut for $\operatorname{Li}_2(-X)$. The short red path going through the green wall corresponds to a jump of the solution. The dashed red path corresponds to analytic continuation. Since the two paths have the same starting and ending points, the transformations of the solution obtained from the two paths must be the same.}
\label{annew}
\end{figure}

The exponential network on $\mathbb{C}^{\times}_X$ encodes information on the solution $\Phi(X,\hbar)$\footnote{We are considering the solution on $\IC^\times_X$ directly, rather than writing it in terms of the logarithmic variable $x$.} to the $q$-difference equation. For $\vartheta\in\bigl(0,\frac{\pi}{2}\bigr)$, the exponential networks have two constituent walls in the region $|X|<1$ and $|X|>1$, respectively; see green and orange walls in Figure~\ref{theta25} or in Figure~\ref{annew}. There are some subtleties in the discussion of $|X|>1$; thus we first discuss the case $|X|<1$.

For $\vartheta\in\bigl(0,\frac{\pi}{2}\bigr)$, $\mathbb{C}^{\times}_X\backslash\mathcal{W}^\vartheta$ is simply connected. By analytic continuation, we assign a single solution $\Phi(X,\hbar)$ to the whole complement of the exponential network for $|X|<1$. Now, when we follow the short solid red path in Figure~\ref{annew}, the solution $\Phi(X,\hbar)$ jumps;
the jump is captured by changing $m \to m - 1$ in Figure~\ref{annew}. On the other hand, this jump must be equivalent
to performing analytic continuation along the dashed red path in Figure~\ref{annew}. In terms of the variable~$x$, this continuation is $x \to x - 2 \pi \I$; indeed, in \eqref{1stlocal} we see that changing $m \to m-1$ is
equivalent to shifting $x \to x - 2 \pi \I$.

For $|X|>1$, the situation is slightly more complicated: we need to consider the branch cut of the dilogarithm $\operatorname{Li}_2(-X)$ from $X=-1$ to $X=\infty$, plotted as a dashed blue line in Figure~\ref{annew}.
Every time a path on $\mathbb{C}^{\times}_X$ crosses this branch cut, in order to match with the Borel resummation which chooses the principal branch of quantum dilogarithm, the solution acquires a factor\footnote{The jump factor can be checked by
\begin{gather*}
\lim_{x\rightarrow \log(X)+\pi\ri}\operatorname{Li}_2(-X)=\lim_{x\rightarrow \operatorname{Re}(\log(X))-\pi\ri}\operatorname{Li}_2(-X)-2\pi\ri \operatorname{Re}(\log(X)).
 \end{gather*} }
\begin{gather*}
 \E^{\frac{2\pi(\log(X)+\ri\pi)}{\hbar}},
 \end{gather*}
where $\log(X)$ also has a branch cut along the negative real axis, fixed as in \eqref{imx}. This factor is also subject to analytic continuation.
Taking these factors into account we see that the shift~$m \to m+1$ in \eqref{2ndlocal} is indeed
equivalent to continuing $x \to x + 2 \pi \I$.

When $\vartheta= \frac{\pi}{2}$, the exponential network consists of a degenerate wall lying on the unit circle. It separates $\mathbb{C}^{\times}_X$ into two domains: $|X|<1$ and $|X|>1$. So there are two solutions for the two domains; they can be thought of as $m\rightarrow\mp\infty$ limits of the exponentials of \eqref{1stlocal} and \eqref{2ndlocal}. The explicit form of the solutions is
 \begin{align*}
& \bigl(-q^{\frac{1}{2}}\mathrm{e}^x;q\bigr)_\infty, \qquad \hbar \in \ri\IR_+, \qquad \operatorname{Re}(x)<0,\\
& \frac{\E^{\frac{x^2\ri}{2\hbar}}q^{\frac{1}{24}}}{\tilde{q}^{\frac{1}{24}}}\frac{1}{\bigl(-q^{\frac{1}{2}}\E^{-x};q\bigr)_{\infty}}, \qquad \hbar \in \ri \IR_+, \qquad \operatorname{Re}(x)>0.
 \end{align*}
In the low energy effective theory of the defect described in Section~\ref{3d5d}, integrating out the chiral shifts the flavor Chern--Simons coupling level by $\pm \frac{1}{2}$ for $\mp x>0$. Thus the effective defect theory for $x<0$ does not have the effective flavor Chern--Simons term and the one with~$x>0$ has an effective flavor Chern--Simons term with level $-1$.\footnote{These background Chern--Simons levels can be conveniently understood in terms of a Type IIB $(p,q)$-fivebrane construction, as discussed in \cite{Cecotti:2013mba}. The defect comes from a D3-brane with one end on a spectator brane and the other end on the $(p,q)$-fivebrane web. The two domains $\operatorname{Re}(x) < 0$ and $\operatorname{Re}(x) > 0$ correspond to D3-branes ending on different legs of the $(p,q)$-fivebrane web. The Chern--Simons levels obtained from the $(p,q)$ charges of the legs and the spectator brane are as above.} Thus that under analytic continuation~${x\rightarrow x+2\pi\ri}$, \eqref{1stlocal} stays fixed and \eqref{2ndlocal} is multiplied by \smash{$\E^{-\frac{2\pi(x+\pi\ri)}{\hbar}}$} is exactly what we expected according to Section~\ref{sec2:CS}. The solutions in different domains and their connections to the Chern--Simons terms have also been discussed in \cite{Beem:2012mb}.

\subsection{The closed sector and the McMahon function}\label{McM}
Let us now briefly discuss the resurgence structure of the closed string free energies associated to $\IC^3$. This is parallel to the study of exact WKB of quantum periods in 4d.
We have
 \be \label{c3wkb} F^{\rm \IC^3}(\hbar)=\sum_{g\geq 0} F_g \hbar^{2g-2},\ee
where
\begin{align*}
& F_0 =-\zeta(3),\qquad
F_1= \frac{1}{12}\log (-\ri \hbar )+\zeta'(-1),\qquad F_g= -\frac{ (-1)^g B_{2 g} B_{2 g-2} }{2 g (2 g-2) (2 g-2)!},\qquad g\geq 2.
 \end{align*}
Note that
\begin{align*}
 F_g\sim (2g-3)!,\qquad g\gg1.
\end{align*}
Hence \eqref{c3wkb} is obviously divergent.
We compute the Borel transform using the definition \eqref{bdefo} and get
\begin{align*}
\mathcal{B}F^{\rm \IC^3}(\xi)=-\sum_{n\geq1} \frac{\xi ^2 \operatorname{csch}^2\big(\frac{\xi }{4 \pi n}\big)+8 \pi n \big(2 \pi n-\xi \coth \big(\frac{\xi }{4 \pi n}\big)\big)}{32 \pi ^4 \xi n^4},\end{align*}
 which has poles on the imaginary axis at
 \begin{align*}
 \xi= 4\pi^2 \ri m n, \qquad m\in \IZ/\{0\}.
 \end{align*}
Physically these poles correspond to BPS states arising from D0-branes in type $\rm IIA$ description (their position can be identified with the central charge associated to such an object), see Section~\ref{sec:intro}.

We now consider the Borel summation as defined in \eqref{boreldef}. The jump obtained from the sum over all the residues at the poles on positive imaginary axis with $m\geq 1$ and $n\geq 1$ are
\be \label{dr}
\Delta (\hbar)=2\pi\ri\sum_{m> 0}\left(\frac{\operatorname{Li}_2\big(\E^{-\frac{4 \ri m \pi ^2}{\hbar}}\big)}{2 \pi ^2}- \frac{2 \ri m \log \big(1-\re^{-\frac{4 \ri \pi ^2 m}{\hbar}}\big)}{\hbar}\right).\ee
We can express \eqref{dr} using the NS limit of the refined McMahon function, namely \cite{ikv}
 \begin{align*}
 {\rm McM}^{\rm NS}(\hbar)=\sum_{n\geq 1}\frac{\ri \re^{-\frac{1}{2} \ri \hbar n} \csc \left(\frac{\hbar n}{2}\right)}{2 n^2}.
 \end{align*}
In particular, it is easy to check that
 \begin{align*}
 \Delta (\hbar)=\frac{1}{\ri \pi}{\frac{\partial }{\partial \hbar}\left(\hbar{\rm McM}^{\rm NS}\left(\frac{4 \pi ^2}{\hbar}\right)\right)}.
 \end{align*}
 This is in line with the expectation from non-perturbative strings of \cite{ghm, hmmo}. We have following results:
 \begin{itemize}\itemsep=0pt
 \item[(1)]
 If $\hbar\in \ri \IR_+$, the Borel summation of \eqref{c3wkb} reproduces the logarithm of the McMahon function,
 namely
 \begin{align*}
\log {\rm Mc
M}(\hbar)&=s\big(F^{\rm \IC^3}\big)(\hbar),\qquad \hbar\in \ri \IR_+,
 \end{align*}
where
 \begin{align*}
 {\rm McM}(\hbar)=\Bigg(\prod_{ k \geq 1}\bigl(1-\re^{ \ri \hbar k}\bigr)^{-k}\Bigg). \end{align*}
 \item[(2)]
 If $\hbar$ is not imaginary, then we have to take into account the contribution of the poles along the imaginary axes, namely \eqref{dr}.
 We then find
 \begin{align*} \log {\rm Mc
M}(\hbar)+ \frac{\Delta(\hbar)}{2}=s\big(F^{\rm \IC^3}\big)(\hbar),\qquad \operatorname{Re} (\hbar)> 0. \end{align*}
 We can also check that the r.h.s.\ in the above equation matches
 \begin{align*} A_c(\hbar)= \frac{8}{\hbar^2} \int_0^{\infty}{\rm d}x \frac{x }{\re^{\frac{4 \pi x}{\hbar}}-1} \log \bigl(1-\re^{-2 x}\bigr)\rd x-\frac{\zeta (3)}{\hbar^2}+\frac{\hbar \zeta (3)}{4 \pi ^3}-\frac{\log (\ri)}{12},
 \end{align*}
 as in \cite{ho, ho2}.\footnote{In these references $A_c(\hbar)$ was used to resumm the constant map contribution in a particular slice of local~${\IP^1\times \IP^1}$.}
 \item[(3)] By using
 \begin{align*} s\big(F^{\rm \IC^3}\big)(\hbar)=s\big(F^{\rm \IC^3}\big)(-\hbar)+\frac{\pi}{12},
 \end{align*}
 we can reach the rest of the $\hbar$ plane which is not discussed in items~(1) and (2) above.
\end{itemize}

\section{The resolved conifold}\label{sec:resconi}
We now move to our second example which is the resolved conifold.
The Seiberg--Witten curve of the resolved conifold is
\[
\Sigma=\{1-Y+X-QXY=0\}\subset \mathbb{C}_X^\times\times \mathbb{C}_Y^\times,\]
where $Q=\re^{-t}$ and $t$ is the K\"{a}hler parameter of the resolved conifold.
$\Sigma$ is a four-punctured sphere with punctures at
\[ \{(X,Y)\}_{\rm sing}=\left\{(0,1),(-1,0),\left(-\frac{1}{Q},\infty\right),\left(\infty,\frac{1}{Q}\right)\right\}.\]
In this paper, we choose the following quantum mirror curve for the resolved conifold
\be\label{coni} \bigl(1-\re^{\hat p}+q^{-1/2}\re^{\hat x}-q^{-1/2}Q\re^{\hat x}\re^{\hat p}\bigr)\Psi(x,\hbar,t)=0,\ee
where again, $q=\re^{\ri \hbar}$.
Our convention here is such that the resolved conifold behaves as two copies of $\mathbb{C}^3$ in the convention we used in Section~\ref{sec:c3}, with the variable $x$ shifted by $t$ in one copy.

 We recall that formal solutions to \eqref{coni} and their connection with open topological strings were discussed previously in the literature, for example in \cite{acdkv,Hyun:2006dr,Kashani-Poor:2006puz}. Here we are interested in Borel summation of formal solutions and the corresponding non-perturbative effects in the open string amplitudes. The relation with exponential networks will also play an important role in our analysis.

\subsection{All-orders WKB expansion of local solutions}

We can work out the formal series expansion for log of the solution using the same technique as in \cite{Garoufalidis:2020pax}. We find
\begin{gather}\label{conio}\varphi(x,t,\hbar)=\log \Psi(x,t,\hbar)=-\ri \hbar \sum_{k\geq 0}\frac{B_{2 k}\left(\frac{1}{2}\right) (\ri \hbar )^{2 k-1}}{(2 k)!}\left(\operatorname{Li}_{2-2 k}\left(-\E^{x}\right)-\operatorname{Li}_{2-2 k}\left(-\E^{x-t}\right) \right).\!\!\!\end{gather}
The series \eqref{conio} is the difference of two pieces: one is the series \eqref{formallocal} for the $\IC^3$ introduced in Section~\ref{sec:c3} and the other is \eqref{formallocal} for the $\IC^3$ with the shift $x\rightarrow x-t$. Hence the Borel summation of the local solution also decouples into two pieces.
Parallel to \eqref{imx}, we assume
\begin{align*}& -\pi< \operatorname{Im}(x)\leq\pi,\qquad  -\pi  < \operatorname{Im}(x-t)\leq \pi.\end{align*}
Therefore the Borel transform of \eqref{conio} is simply
\be \label{Bconi}
 {\bf \mathcal{B}}\varphi(x,t,\xi)={\bf\mathcal{B}}\phi(x, \xi)-{\bf \mathcal{B}}\phi(x-t, \xi),
\ee
where ${\bf \mathcal{B}}\phi(x, \xi)$ is defined in \eqref{hadamardc3}. Hence
\eqref{Bconi} has two sets of singularities coming from ${\bf \mathcal{B}}\phi(x, \xi)$ and ${\bf \mathcal{B}}\phi(x-t, \xi)$, respectively,
\begin{gather}
 \xi^*(x,m,n)= n ( 2 \pi (x + (2m+1)\pi \ri) ), \qquad m\in\IZ,\qquad n \in \IZ/\{0\}, \nonumber\\
  \xi_t^*(x,t,m,n)= n ( 2 \pi ((x-t) + (2m+1)\pi\ri ) ), \qquad m\in\IZ, \qquad n \in \IZ/\{0\}.\label{poleconi}
\end{gather}
These singularities correspond to the central charges of the 3d-5d BPS KK-modes as we discuss below.

\subsection{BPS states in 3d-5d system}

We have not studied the BPS spectrum in this case directly from
a 3d-5d field theory description; instead we use the M-theory point of view,
along the lines of \cite{gv,Ooguri:1999bv}.
This leads to the prediction that there are two 3d
particles, corresponding to
two M2-brane discs ending on the M5-brane (Ooguri--Vafa invariants).
The areas of these two discs sum to the area of the compact $\IC\IP^1$ in $X$,
and there should be one 5d particle corresponding to an M2-brane wrapping
this cycle (Gopakumar--Vafa invariant).

Thus we expect that there should be an effective
description of the 3d-5d system in which the
field content on the defect is two chiral multiplets
with charges $(+1,0)$ and $(-1,+1)$ under a~$U(1) \times U(1)$ flavor symmetry.
When the system is reduced on $S^1$, the theory has an extra~$U(1)$ flavor symmetry coming from the rotation of the circle. These two fields
give rise to two infinite towers of KK modes corresponding to the third $U(1)$, with central charges
\begin{gather} \pm R^{-1} (x + (2m+1)\pi \ri) , \qquad m\in\IZ, \nonumber\\
  \pm R^{-1} ( (x-t) + (2m+1)\pi\ri ),\qquad m\in\IZ.\label{coni3d5dc}
\end{gather}
Here $(x,t)$ are the two complex flavor masses, complexifying the two flavor masses of the 3d theory. In addition $t$ is identified with the complexified vev of the scalar in the 5d vector multiplet.
When the system is reduced on $S^1$ the 5d particle gives rise to two towers of KK modes with central charges
\begin{equation}
\label{5dcconi} \pm R^{-1} ( t + 2m \pi\ri ), \qquad m\in\IZ.
\end{equation}

\subsection{The exponential network and exact WKB}\label{sec:enconi}

Once again, we note that the central charges of the BPS KK-modes
\eqref{coni3d5dc} are related to positions~\eqref{poleconi} of the first poles
along each ray, by the relation
\begin{equation*}
 \xi =- 2 \pi R Z.
\end{equation*}

\begin{figure}[t]
 \centering
 \begin{subfigure}[b]{0.24\textwidth}
 \centering
 \includegraphics[width=\textwidth]{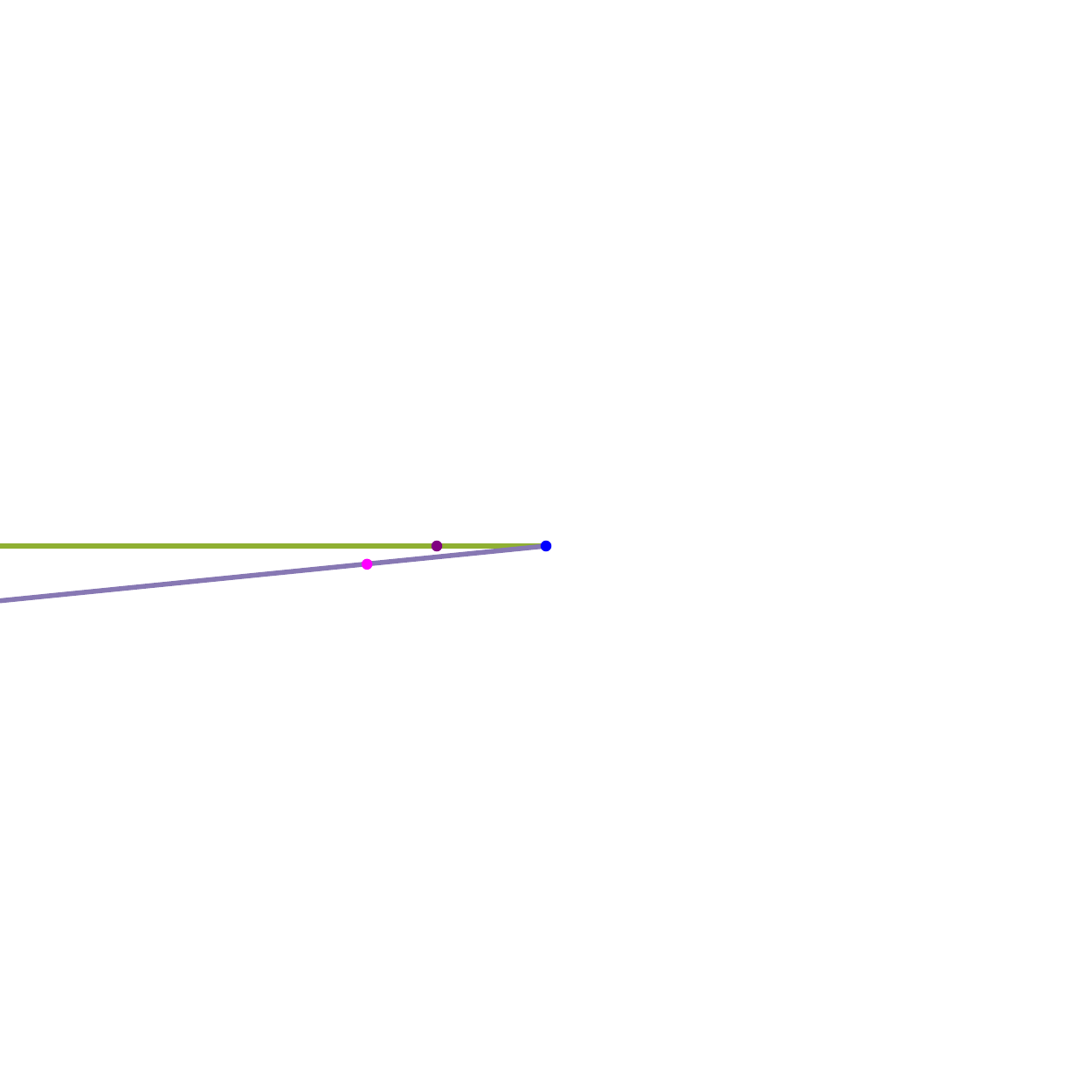}
 \caption{$\vartheta=0$}
 \label{coni0}
 \end{subfigure}
 \begin{subfigure}[b]{0.24\textwidth}
 \centering
 \includegraphics[width=\textwidth]{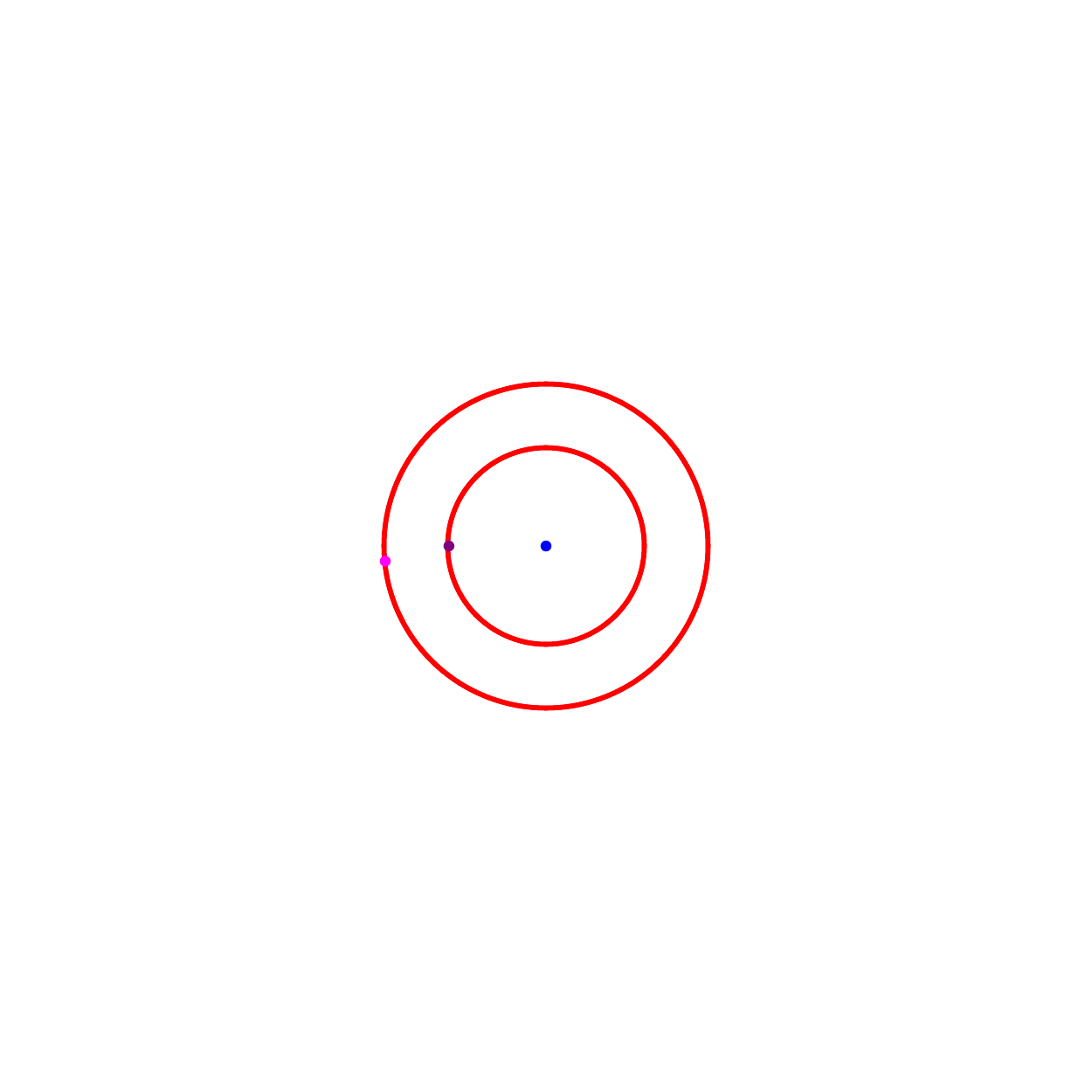}
 \caption{$\vartheta=\pi/2$}
 \label{conipi2}
 \end{subfigure}
 \begin{subfigure}[b]{0.24\textwidth}
 \centering
 \includegraphics[width=\textwidth]{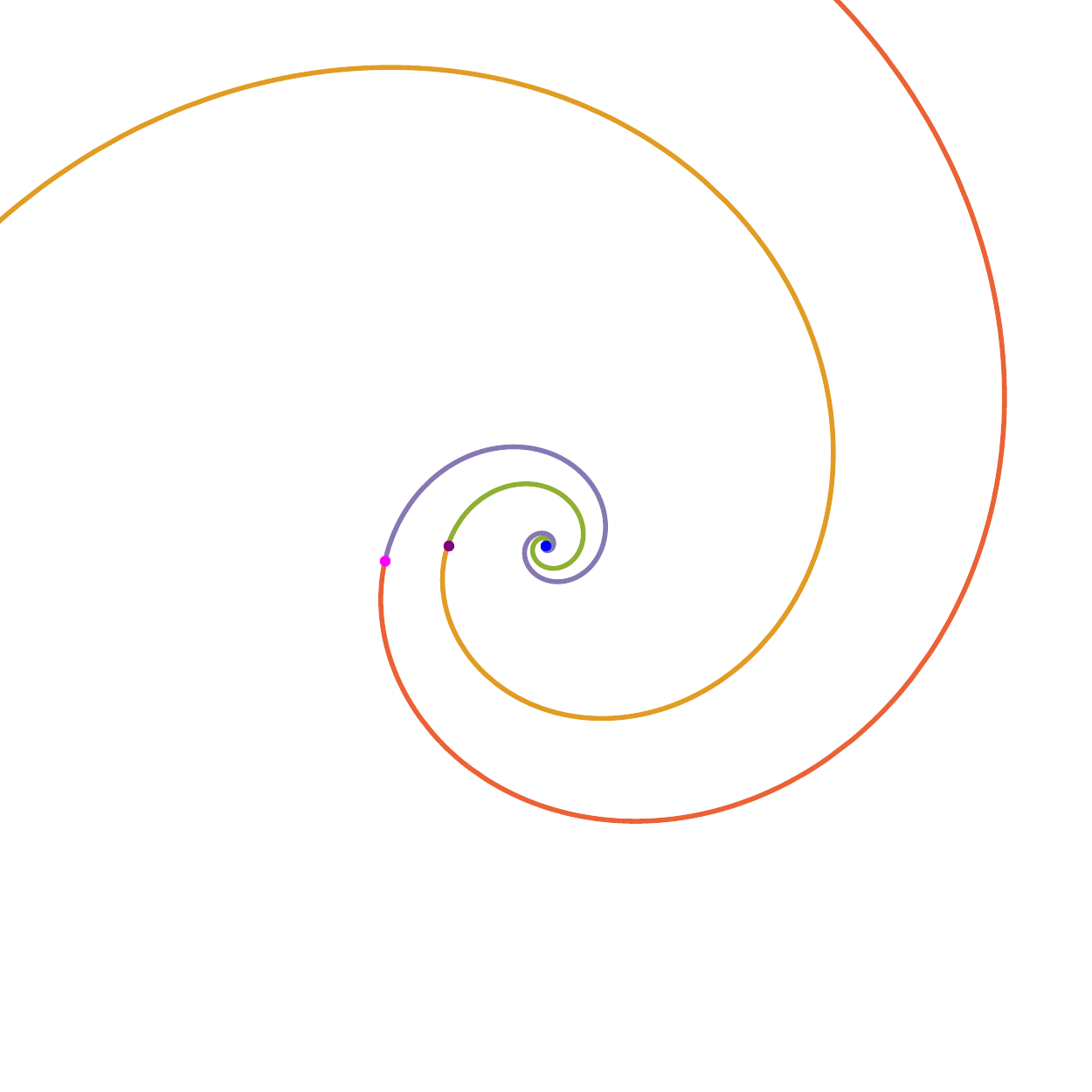}
 \caption{$\vartheta=\frac{2\pi}{5}$}
 \label{coni2pi5}
 \end{subfigure}
 \begin{subfigure}[b]{0.24\textwidth}
 \centering
 \includegraphics[width=\textwidth]{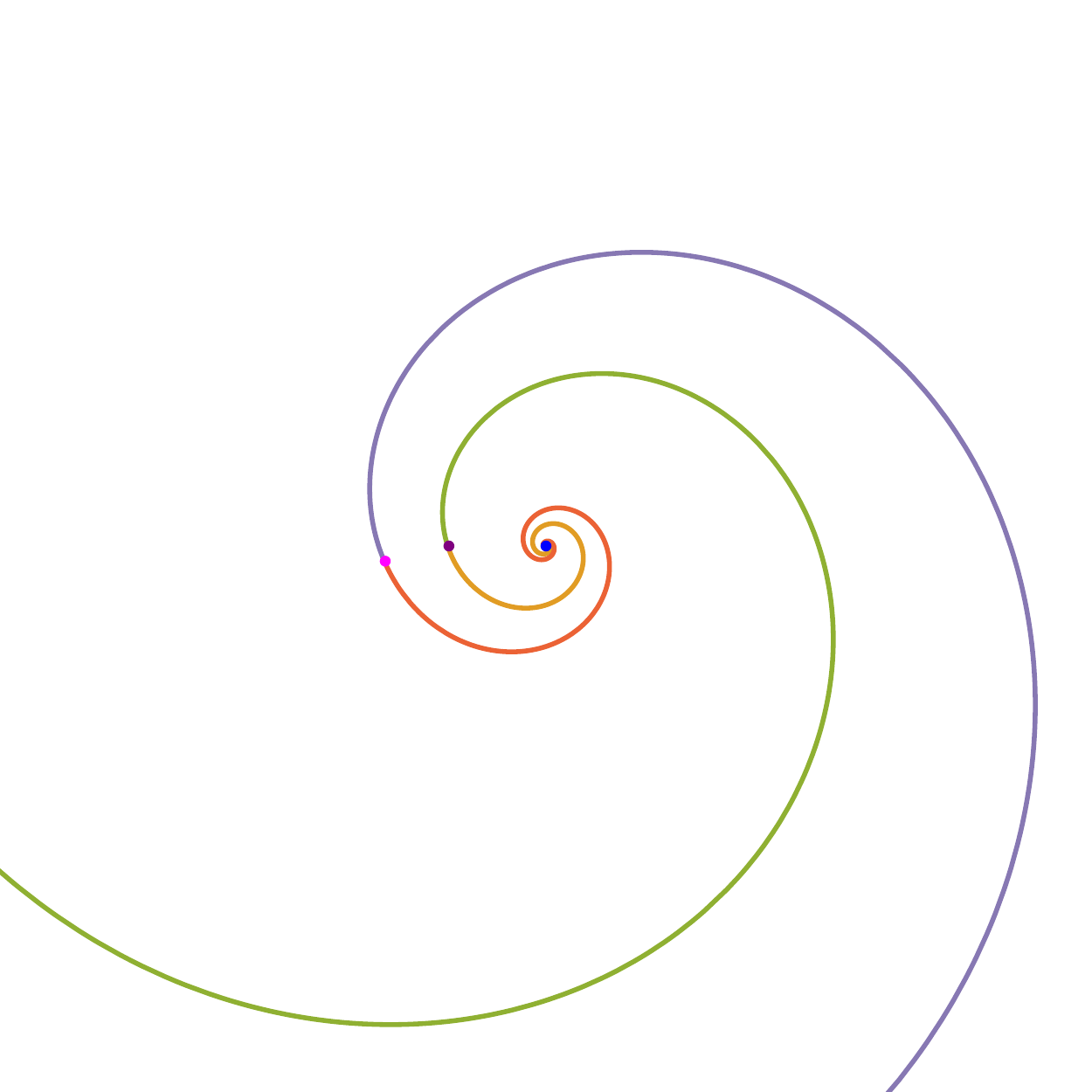}
 \caption{$\vartheta=\frac{3\pi}{5}$}
 \label{coni3pi5}
 \end{subfigure}
           \caption{The exponential networks $\mathcal{W}^{\vartheta}$ on $\mathbb{C}^\times_X$ at different $\vartheta$'s for $t=\frac{1}{2}+\frac{\ri}{10}$. The blue, purple and magenta dots represent $X=0$, $X=-1$ and $X=-\frac{1}{Q}$ respectively. The orange, green, red and purple walls are given by \smash{$X=-\E^{ s \E^{\ri\vartheta}}$}, \smash{$X=-\E^{ -s\E^{\ri\vartheta}}$}, \smash{$ X=-\frac{1}{Q}\E^{ s\E^{\ri\vartheta}}$} and \smash{$X=-\frac{1}{Q}\E^{- s\E^{\ri\vartheta}}$} respectively. The degenerate walls at phase $\vartheta=\pi/2$ are painted in red.}
 \label{conitop}
 \end{figure}

\begin{figure}[t]
 \centering
 \begin{subfigure}[b]{0.3\textwidth}
 \centering
 \includegraphics[width=\textwidth]{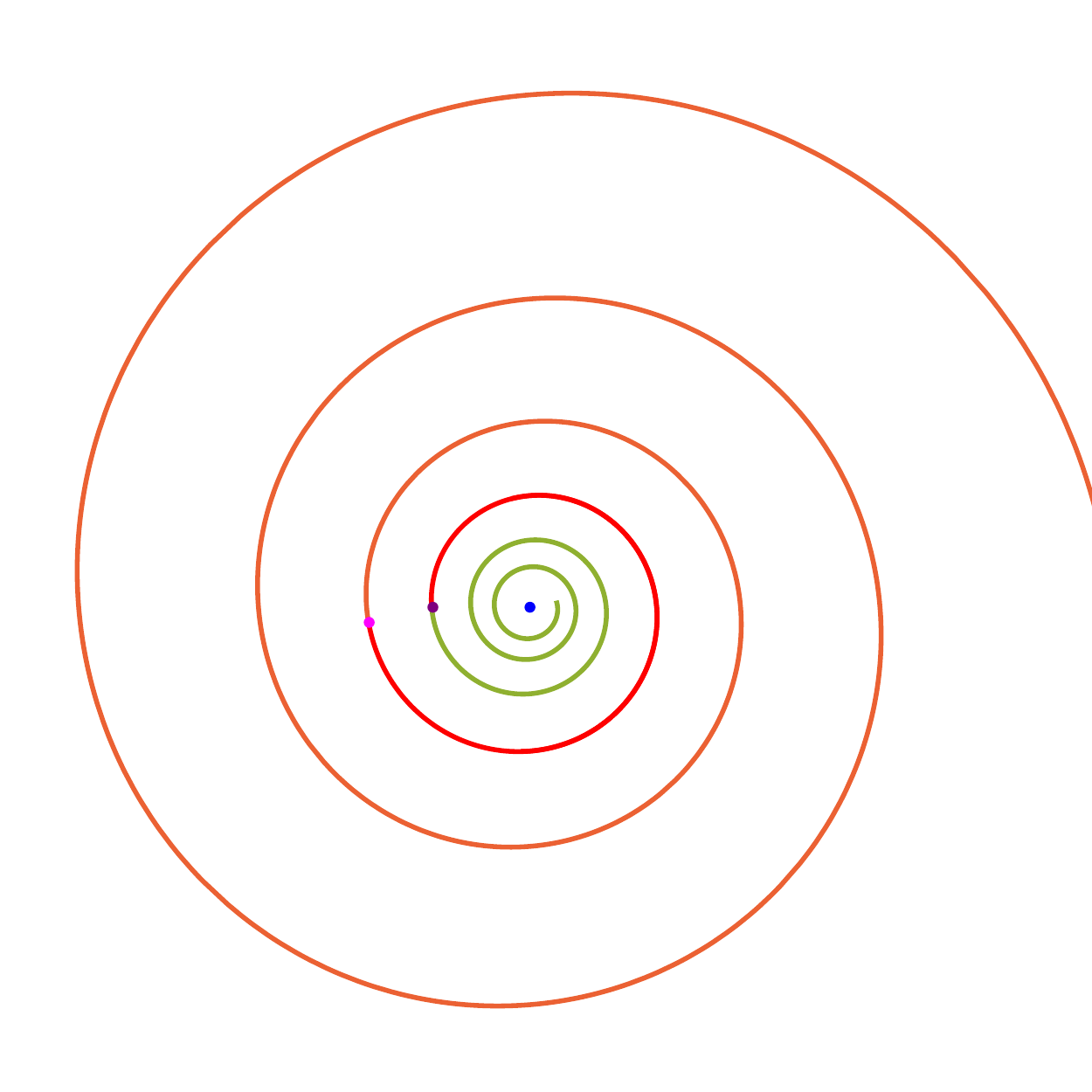}
 \caption{$\vartheta=\operatorname{ArcTan}(2(\frac{1}{10}-2\pi))$}
 \label{conidm1}
 \end{subfigure}
 \begin{subfigure}[b]{0.3\textwidth}
 \centering
 \includegraphics[width=\textwidth]{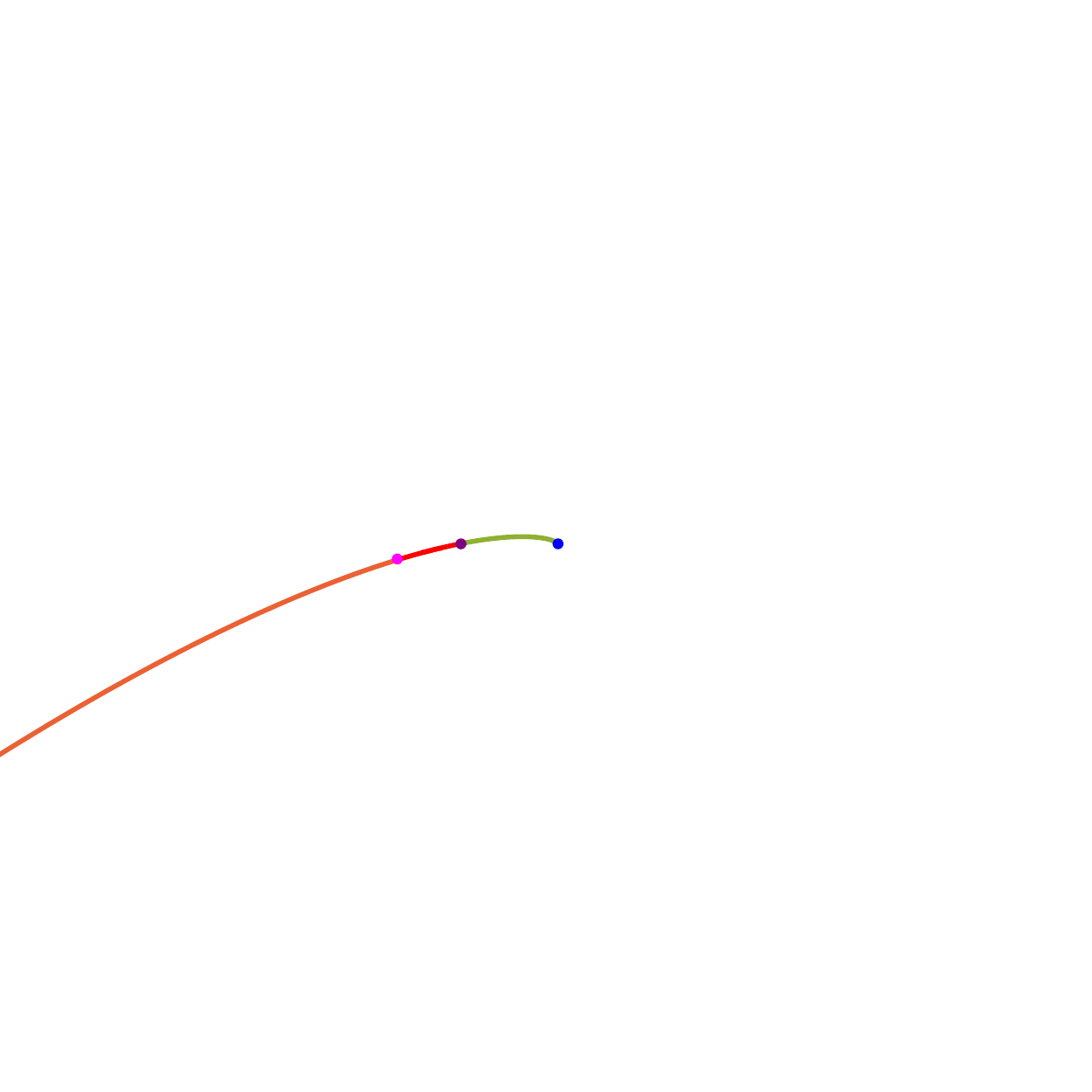}
 \caption{$\vartheta=\operatorname{ArcTan}(\frac{1}{5})$}
 \label{conid0}
 \end{subfigure}
 \begin{subfigure}[b]{0.3\textwidth}
 \centering
 \includegraphics[width=\textwidth]{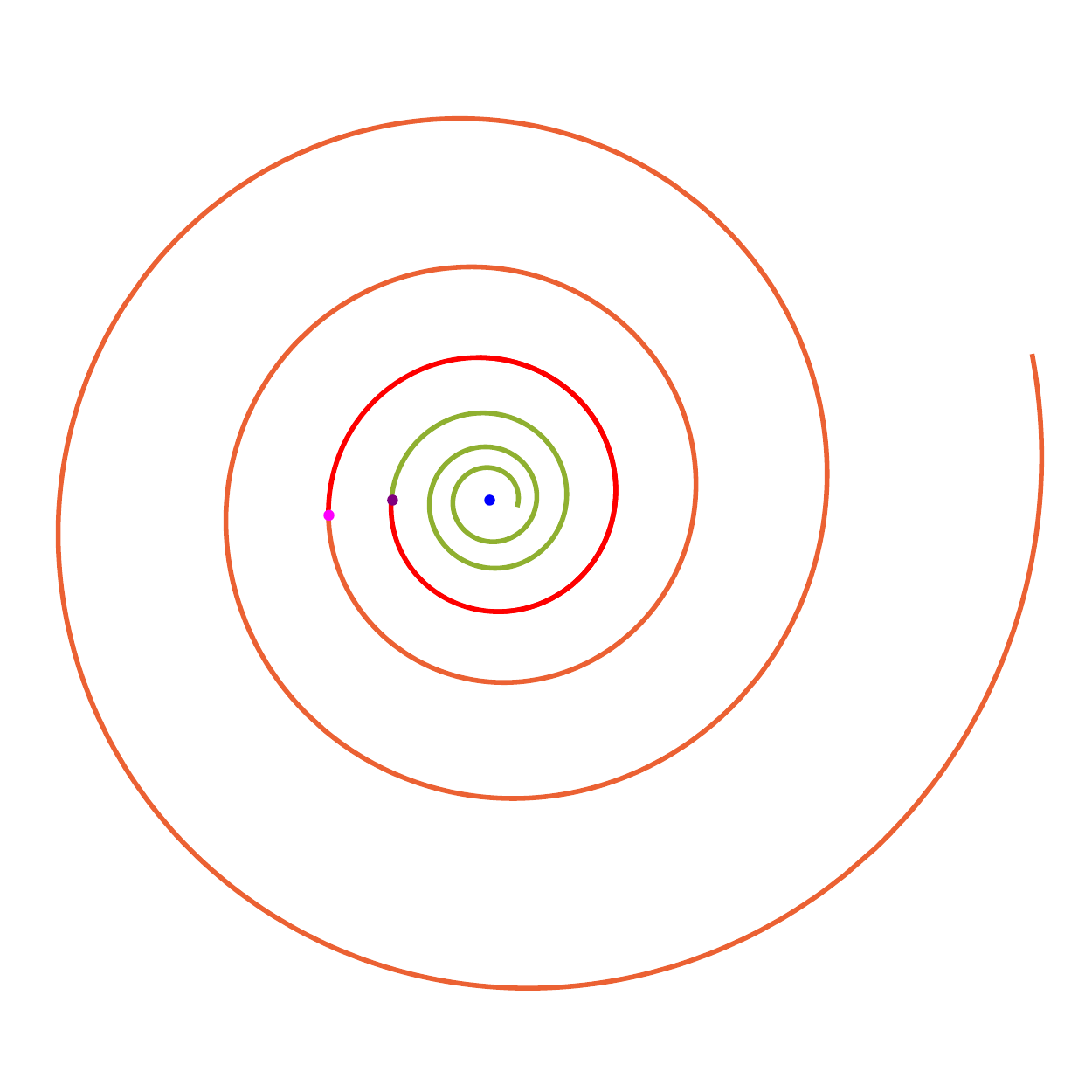}
 \caption{$\vartheta=\operatorname{ArcTan}(2(\frac{1}{10}+2\pi))$}
 \label{conid1}
 \end{subfigure}
                  \caption{Examples of exponential networks $\mathcal{W}^{\vartheta}$'s on $\mathbb{C}^\times_X$ with a degenerate wall (painted in red) at different $\vartheta$'s for $t=\frac{1}{2}+\frac{\ri}{10}$. The blue, purple and magenta dots are punctures of $\Sigma$ at $X=0$, $X=-1$, $X=-\frac{1}{Q}$, respectively. In each of the figures, there is a degenerate wall connecting~$X=-1$ and $X=-\frac{1}{Q}$.}
 \label{conitopd}
 \end{figure}

Recall that the exponential network $\mathcal{W}^\vartheta$ consists of those $X$ such that there are KK-modes obeying
\[ \arg(-Z(X))=\vartheta.\]
In this case $X\in\CW^\vartheta$ if and only if
\[
 X=-\E^{\pm s \E^{\ri\vartheta}}\qquad \text{or} \qquad 
 X=-\frac{1}{Q}\E^{\pm s \E^{\ri\vartheta}}\]
for some $s \in \IR$.
Thus, for a generic phase $\vartheta$, the exponential network for the conifold consists of two copies of the exponential network for $\mathbb{C}^3$, as shown in Figure~\ref{conitop}. One copy emanates from~${X=-1}$, and the other copy emanates from \smash{$X=-\frac{1}{Q}$}. A new phenomenon for the resolved conifold is that there can exist degenerate walls with two ends at $X=-1$ and \smash{$X=-\frac{1}{Q}$}, which occur
when $\vartheta$ is the phase of one of the 5d BPS KK-modes \eqref{5dcconi}. Examples of such degenerate walls can be found in Figure~\ref{conitopd}.

\begin{figure}[t]
 \centering
 \includegraphics[width=0.75\textwidth]{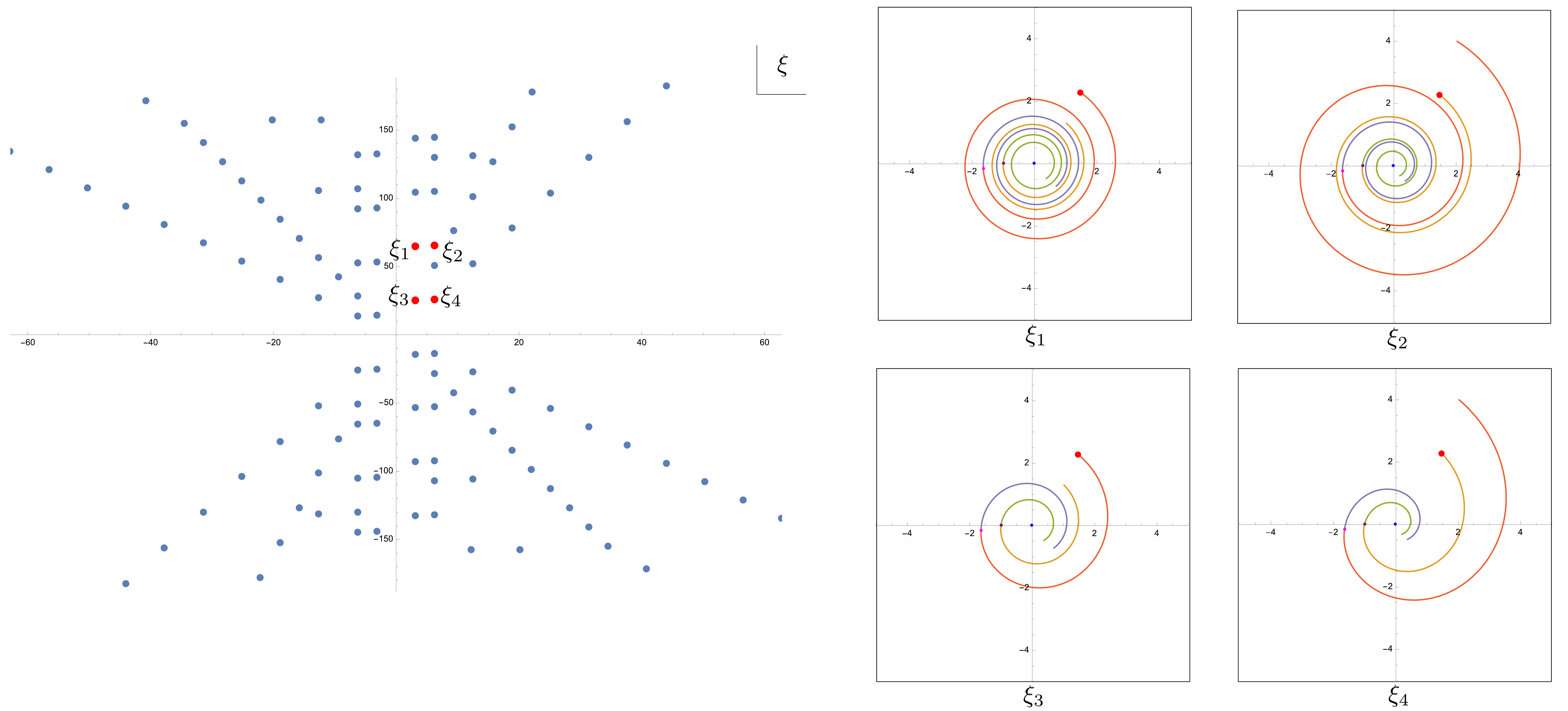}

           \caption{Left: Poles \eqref{poleconi} of the Borel transform \eqref{Bconi}. The four points in red correspond to~$\xi_1=\xi_t^\ast\bigl(1+\ri,\frac{1}{2}+\frac{\ri}{10},1,1\bigr)$, $\xi_2=\xi^\ast(1+\ri,1,1)$, $\xi_3=\xi_t^\ast\bigl(1+\ri,\frac{1}{2}+\frac{\ri}{10},0,1\bigr)$, $\xi_4=\xi^\ast(1+\ri,0,1).$ Right: Truncated exponential networks $\mathcal{W}^{\arg(\xi_j)}(|\xi_j|),~j=1,\dots,4,$ corresponding to the poles~$\xi_j$ in the Borel plane. The red dots are the point $X(x)=\re^{1+\ri}$ on $\mathbb{C}^{\times}_X$.}
 \label{conitrunc}
 \end{figure}

The graphical interpretation of that
 the truncated exponential network \[
 \mathcal{W}^{\arg(\xi_{t}^*(x,t,m,\pm 1))}(|\xi_{t}^*(x,t,m,\pm 1)|)\]
 $\forall m\in\mathbb{Z}$ should end at the point $X=\E^x$ for some examples is shown in Figure~\ref{conitrunc}.

\begin{figure}[htbp]
\begin{center}
\includegraphics[width=0.7\linewidth]{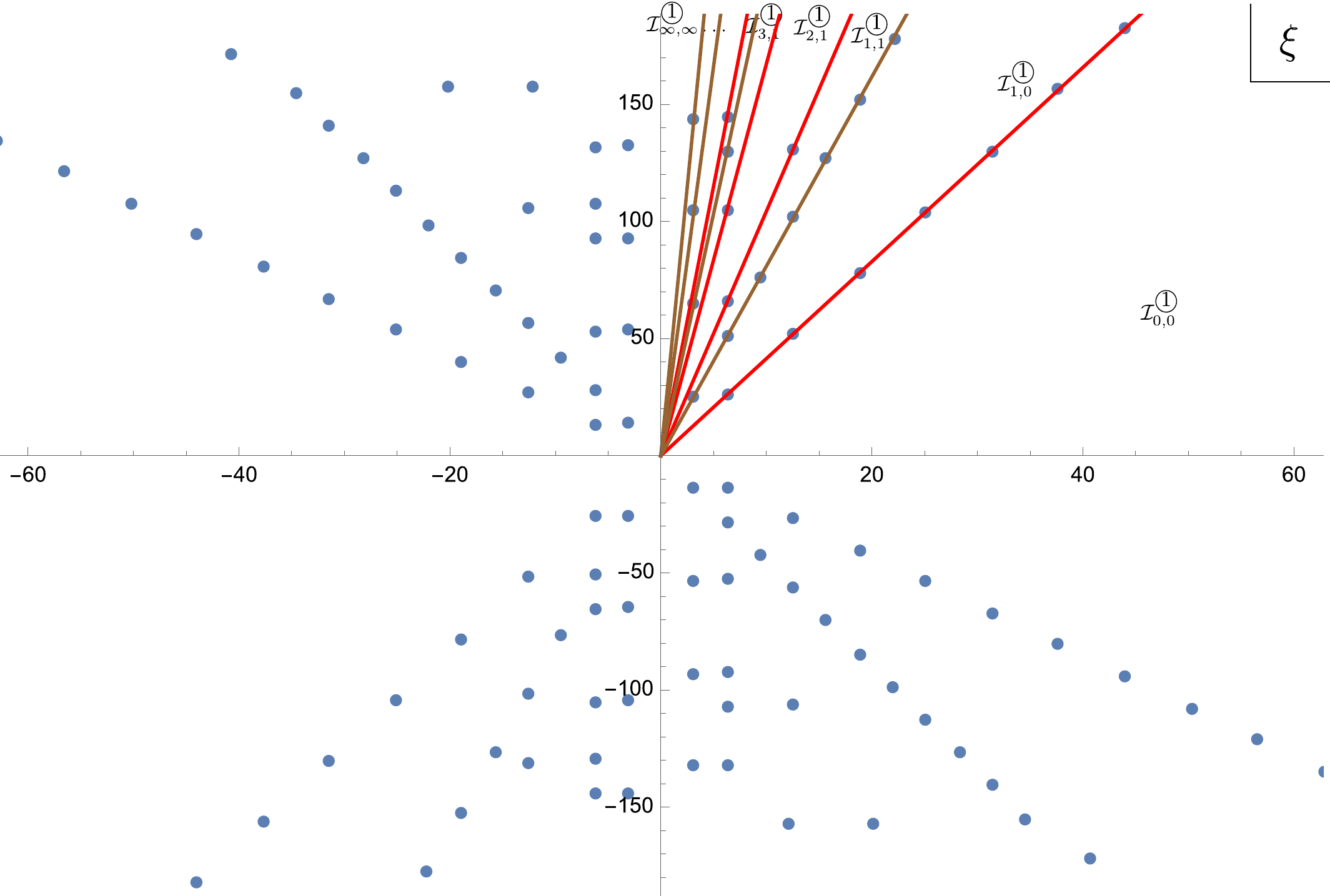}
\caption{We show as an example the case $0<\operatorname{Re}(t)<\operatorname{Re}(x)$ and $\arg{\hbar}\in\big[0,\frac{\pi}{2}\big]$. The red and brown rays of poles $\xi^*(1+\ri,m_1,n_1)$ and $\xi_t^*\bigl(1+\ri,\frac{1}{2}+\frac{\ri}{10},m_2,n_2\bigr)$, where $m_1,m_2\geq 0$, $n_1,n_2>0$ separate the Borel plane into sectors ${\mathcal I}^{\tcircled{1}}_{m_1,m_2}$, which is defined in \eqref{sectorconi}.}
\label{conidomain}
\end{center}
\end{figure}

\subsection{Local solutions in each sector }\label{coniopens}
The rays of poles as shown in Figure~\ref{conidomain} divide the Borel plane of the resolved conifold into sectors. We define the sector
\be\label{sectorconi}\mathcal{I}^{\tcircled{i}}_{x,t,m_1,m_2}=\CI^{\tcircled{i}}_{x,m_1}\cap\CI^{\tcircled{i}}_{x-t,m_2},\ee
where $\CI_{x,m_1}^{\tcircled{i}}$ are as in Section~\ref{Locs}. For convenience, we neglect the subscripts $x$ and $t$ and label the sectors by the 2 integers $m_1$ and $m_2$.
Nevertheless the dependence on $x$ and $t$ is important to keep in mind, especially in relation to the various domains of the exponential network.

For instance, let us consider $\operatorname{Re}(t)=\frac{1}{2}$. The exponential network at $\vartheta=\pi/2$ is shown on Figure~\ref{conipi2}, where we can clearly see two degenerate walls. In particular, we have 3 domains corresponding to $\operatorname{Re}(x)>\operatorname{Re}{(t)}=\frac 1 2$ (outer domain), $0<\operatorname{Re}(x)<\operatorname{Re}{(t)}=\frac 1 2$ (domain between the two circles) and $\operatorname{Re}(x)<0$ (inner domain). This means that we have 3 corresponding solutions which, on the physics side, correspond to insertion of branes at different locations (external brane, internal brane, external anti-brane). The discussion of Borel summation for each local solution follows directly from the discussion of the $\IC^3$ example in Section~\ref{Locs} and Appendix~\ref{summaryls}, as we discuss below. We focus on the first quadrant of the Borel plane for~${\arg(\hbar)\in \big[0,\frac\pi 2\big]}$.

\subsubsection[Re(x) and Re(x-t) same sign: brane on external leg]{$\boldsymbol{\operatorname{Re}(x)}$ and $\boldsymbol{\operatorname{Re}(x-t)}$ same sign: brane on external leg}
Let us assume
\[ \operatorname{Re}(x)<0, \qquad \operatorname{Re}(x-t)<0. \]
By using Section~\ref{sec:1strexl0},
we find that in the $\mathcal{I}^{\tcircled{1}}_{m_1,m_2}$ sector the Borel summation of \eqref{conio} agrees with
\begin{gather}
s(\varphi)(x,t,\hbar)=\log \Phi(x+2\pi\ri m_1,\hbar)-\log \Phi(x-t+2\pi\ri m_2,\hbar),\nonumber\\
\arg\hbar \in \mathcal{I}_{m_1,m_2}, \qquad m_1,m_2 \leq 0.\label{conifed}
 \end{gather}
 In the limit $m_1,~m_2\to- \infty$, we get
\be \label{consol1} s(\varphi)(x,t,\hbar) =\log\Bigg( {\prod_{r\geq 0}\bigl(1+q^{r+1/2}\re^{x}\bigr) }\Bigg)- \log \Bigg({\prod_{r\geq 0}\bigl(1+q^{r+1/2}Q\re^{x}\bigr)}\Bigg),\qquad \hbar \in \ri \mathbb{R}_+.\ee
This can be viewed as the log of the resolved conifold open string partition function with a brane insertion on the external leg. Indeed, this partition function reads
\be \label{topopenex}Z_{\rm ext}^{\rm open}(z, t, \hbar)
=\frac{\bigl(-q^{1/2}\re^{z},q \bigr)_{\infty}}{\bigl(-q^{1/2} Q\re^{z},q \bigr)_{\infty}}, \qquad q=\re^{\ri\hbar},\qquad Q=\E^{-t},\ee
where ``ext'' is to stress that here we are considering a brane on the external leg of the toric diagram, see for example \cite{Kashani-Poor:2006puz}.
Hence
\be \label{coni1}\boxed{ s(\varphi)(x,t,\hbar) =\log Z_{\rm ext}^{\rm open}(x, t, \hbar), \qquad \hbar \in \ri \mathbb{R}_+, \qquad \operatorname{Re}(x)<0,\qquad\operatorname{Re}(x-t)<0. }\ee
 Notice that the difference between the two solutions along the real and imaginary axis is
 \be\label{differenceA} \eqref{consol1}-\eqref{conifed}\big|_{m_1=m_2=0}=\log\left({\bigl(-\re^{\frac{2\pi x}{\hbar}}\tilde q^{\frac{1}{2}};\tilde q\bigr)_\infty\over\bigl(-\re^{ \frac{2\pi(x-t)}{\hbar}}\tilde q^{\frac{1}{2}};\tilde q\bigr)_\infty}\right).\ee
Keeping in mind the analogy with the open TS/ST framework of \cite{mz-wv, mz-wv2}, we may wonder how to relate \eqref{differenceA} to the NS limit of the open string partition function for the resolved conifold. From~\cite{Cheng:2021nex,ikv}, it is easy to see that the refined open partition function of the resolved conifold in the NS and GV limits are related by some simple shifts in the arguments. Hence we express~\eqref{differenceA} simply by using \eqref{topopenex}. We have
\be \label{ratio}{\re^{\eqref{consol1}}\over\re^{ \eqref{conifed}}\big|_{m_1=m_2=0} }= {Z_{\rm ext}^{\rm open}\left({2\pi \over \hbar} x,\frac{2\pi }{\hbar}t, -{4 \pi^2\over \hbar}\right)},\ee
which is consistent with the open TS/ST framework of \cite{mz-wv, mz-wv2}. However here we do not have an honest spectral theory side so this analogy is only partial.
In particular, in this simple example it all reduces to a simple transformation
\be \label{pertnonpert} \left(x, t, \hbar \right) \text{in \eqref{coni1}} \qquad \circled{\text{vs}} \qquad \left({2\pi\over \hbar} x, {2\pi\over \hbar} t, -{4 \pi^2\over \hbar }\right)\qquad \text{in \eqref{ratio}}. \ee
Similar computations can be done for $\operatorname{Re}(x)>0$, $\operatorname{Re}(x-t)>0 $. In this case the local solution on the imaginary axis corresponds to a $\overline {q}$-brane (or anti-brane) inserted on the external leg. This means that we have a replacement $Z_{\rm ext}^{\rm open}\to \bigl(Z_{\rm ext}^{\rm open}\bigr)^{-1}$ and some other modifications coming from the extra polynomial factors besides the $q$-Pochhammer symbol in \eqref{spo}.
 The difference between the solution along real and imaginary axis is then given by \be\label{oppos}
 \log\left({\bigl(-\re^{-\frac{2\pi(x-t)}{\hbar}}\tilde q^{\frac{1}{2}};\tilde q\bigr)_\infty\over\bigl(-\re^{-\frac{2\pi x}{\hbar}}\tilde q^{\frac{1}{2}};\tilde q\bigr)_\infty}\right),
 \ee
 which we can express using \eqref{topopenex} as
 \[ \eqref{oppos}=\log\left(\frac{1}{ {Z_{\rm ext}^{\rm open}\bigl(-{2\pi \over \hbar} x,-\frac{2\pi }{\hbar}t, -{4 \pi^2\over \hbar}\bigr)}}\right).\]

\subsubsection[Re(x) and Re(x-t) with opposite signs: internal leg]{ $\boldsymbol{\operatorname{Re}(x)}$ and $\boldsymbol{\operatorname{Re}(x-t)}$ with opposite signs: internal leg}

Let us now take
\[
0<\operatorname{Re}(x)<\operatorname{Re}(t).
\]
By using Section~\ref{sec:1strexl0},
we find that in the $\mathcal{I}^{\tcircled{1}}_{m_1,m_2}$ sector the Borel summation of \eqref{conio} agrees with
\[ 
 \boxed{s(\varphi)(x,t,\hbar)=\log \Phi(x+2\pi\ri m_1,\hbar)+{2 \pi m_1\over \hbar}(x+m_1 \pi \ri)-\log \Phi(x-t+2\pi\ri m_2,\hbar), }
\]
where
\[
 \arg\hbar \in \mathcal{I}^{\tcircled{1}}_{m_1,m_2}, \qquad m_1\geq 0,\qquad m_2 \leq 0.
 \]
 In the limit $m_1,-m_2\to \infty$ ($\hbar$ is on positive imaginary axis), we get
\begin{align} s(\varphi)(x,t,\hbar) ={}&-\log\Bigg( {\prod_{r\geq 0}\bigl(1+q^{r+1/2}\re^{-x}\bigr) }\Bigg)- \log \Bigg({\prod_{r\geq 0}\bigl(1+q^{r+1/2}Q\re^{x}\bigr)}\Bigg)\nonumber\\
&{}+{1\over 24}\log \bigl(q\tilde q^{-1}\bigr)+{\ri\over 2 \hbar}x^2,\qquad \hbar \in \ri \mathbb{R}_+. \label{consol2}
\end{align}
Also in this case we can express \eqref{consol2} using the open topological string free energy, but in this case the brane has to be on the internal leg. The open string partition function of the resolved conifold with internal leg insertion can be found for example in \cite{Kozcaz:2018ndf} where the refined open amplitudes are also discussed. We have
\be \label{topopenint}
 Z^{\rm open}_{\rm int}(t_L, t_R, y, \hbar)= \bigl( - q^{1/2} \re^{-t_L} \re^y; q\bigr)_{\infty} \bigl(- q^{1/2} \re^{-t_R} \re^{-y};q\bigr)_{\infty}, \qquad q=\re^{\ri \hbar}.\ee
Hence we have ($\operatorname{Re}(x)>0$, $\operatorname{Re}(x-t)<0$)\footnote{We chose $t_L=t$, $t_R=0$, $y=x$ but this is obviously not the only choice. }
\[
\boxed{\ba s(\varphi)(x,t,\hbar) =-\log Z^{\rm open}_{\rm int}(t, 0, x, \hbar)&+{1\over 24}\log \bigl(q\tilde q^{-1}\bigr)+{\ri\over 2 \hbar}x^2,\qquad\hbar \in \ri \mathbb{R}_+.\ea}
\]
In this case the difference between the solution on the imaginary and real axis is
 \be\label{difr}\frac{1}{\bigl(-\tilde{q}^{\frac{1}{2}}\mathrm{e}^{\frac{2\pi (x-t)}{\hbar}};\tilde{q}\bigr)_\infty \bigl(-\tilde{q}^{\frac{1}{2}}\mathrm{e}^{-\frac{2\pi x}{\hbar}};\tilde{q}\bigr)_\infty},\ee
 which can again be expressed using \eqref{topopenint}, we have
 \[ \eqref{difr}={1\over Z^{\rm open}_{\rm int}\bigl({{2 \pi t \over \hbar}}, 0, {2\pi\over \hbar}x,- {4 \pi^2 \over \hbar}\bigr) }.\]
Likewise for $\operatorname{Re}(x)<0$, $\operatorname{Re}(x-t)>0 $ the situation is very similar to the one we just discussed upon replacement to $Z_{\rm \dots}^{\text{open}}\to \frac{1}{Z_{\rm \dots}^{\text{open}}}$ (brane $\to$ anti-brane).

\subsubsection[Either Re(x) or Re(x-t) is zero]{ Either $\boldsymbol{\operatorname{Re}(x)}$ or $\boldsymbol{\operatorname{Re}(x-t)}$ is zero}
Without loss of generality, we choose the following example to elucidate this case:
\[ \operatorname{Re}(x)=0,\qquad \operatorname{Re}(t)>0. \]
For this choice of parameters, the solution for $\hbar \in \ri \IR_+$ is
\[
 s(\varphi)(x,t,\hbar) =\frac{1}{2} \Bigg(\log\Bigg(\frac{\E^{\frac{x^2\ri}{2\hbar}}q^{\frac{1}{24}}}{\tilde{q}^{\frac{1}{24}}}\frac{1}{\bigl(-q^{\frac{1}{2}}\E^{-x};q\bigr)_{\infty}}\Bigg)+\log \bigl(-q^{\frac{1}{2}}\mathrm{e}^x;q\bigr)_\infty\Bigg)- \log \bigl(-q^{1/2}Q\re^x;q\bigr)_{\infty}.
 \]
Using the topological partition functions \eqref{topopenex} and \eqref{topopenint}, this can be expressed as
\begin{align*}
& s(\varphi)(x,t,\hbar) =\frac{1}{2}\left(-\log Z^{\rm open}_{\rm int}(t, 0, x, \hbar)+{1\over 24}\log \bigl(q\tilde q^{-1}\bigr)+{\ri\over 2 \hbar}x^2\right)+\frac{1}{2}\log\bigl(Z_{\rm ext}^{\rm open}(x, t, \hbar)\bigr).
\end{align*}
This is the average between the open string amplitude with a brane inserted in an external leg~\eqref{topopenex}, and the open string amplitude with a brane inserted in an internal leg \eqref{topopenint}.

The jump from the imaginary to the real axis is
 \[
 \frac{1}{2}\log\left(\frac{\bigl(-\E^{\frac{2\pi x}{\hbar}}\tilde{q}^{\frac{1}{2}},\tilde{q}\bigr)_{\infty}}{\bigl(-\tilde{q}^{\frac{1}{2}}\mathrm{e}^{-\frac{2\pi x}{\hbar}};\tilde{q}\bigr)_{\infty}}\right)-\log\bigl(\bigl(-\E^{\frac{2\pi (x-t)}{\hbar}}\tilde{q}^{\frac{1}{2}},\tilde{q}\bigr)_{\infty}\bigr),\]
which can again be expressed using both $Z^{\text{open}}_{\text{int}}$ and $Z_{\text{ext}}^{\text{open}}$ with a transformation of the argument of the form \eqref{pertnonpert}.

\subsubsection[Re(x)=0 and Re(t)=0]{$\boldsymbol{\operatorname{Re}(x)=0}$ and $\boldsymbol{\operatorname{Re}(t)=0}$}
 In this case all the poles are on the imaginary axis. Median summation along the imaginary axis gives
 \begin{align*} s(\varphi)(x,\hbar)=&-\frac{\ri t (t-2 x)}{4 \hbar }+\frac{1}{2} \log\left(\frac{\bigl(-\E^{x}q^{1/2};q\bigr)_{\infty }}{\bigl(-\E^{-x}q^{1/2};q\bigr)_{\infty }}\frac{\bigl(-\E^{-(x-t)}q^{1/2};q\bigr)_{\infty }}{\bigl(-\E^{(x-t)}q^{1/2};q\bigr)_{\infty }}\right), \qquad \hbar\in \ri \IR_+.\end{align*}
 In term of brane insertions, the expression inside the logarithm can be written either as a product of two external brane or as a product of two internal brane. This is in line with the diagrammatic picture of the toric diagram which, in this situation, is very degenerate.

 The jump in the first quadrant to go from the imaginary to the real axis, is given by
 \[
 \frac{1}{2}\log\left(\frac{\bigl(-\E^{\frac{2\pi x}{\hbar}}\tilde{q}^{\frac{1}{2}},\tilde{q}\bigr)_{\infty}}{\bigl(-\mathrm{e}^{-\frac{2\pi x}{\hbar}}\tilde{q}^{\frac{1}{2}};\tilde{q}\bigr)_{\infty}}\frac{\bigl(-\mathrm{e}^{-\frac{2\pi (x-t)}{\hbar}}\tilde{q}^{\frac{1}{2}};\tilde{q}\bigr)_{\infty}}{\bigl(-\E^{\frac{2\pi (x-t)}{\hbar}}\tilde{q}^{\frac{1}{2}},\tilde{q}\bigr)_{\infty}}\right).
 \]
This is in line with our expectations.

\subsection{The closed sector}\label{closeconi}
We now move to the study of the (refined) closed topological string free energy. This part is analogous to the exact WKB of the 4d quantum periods. We are going to show that it is closely related to the 5d BPS states in comparison to the relation between the open topological string free energy and the 3d-5d BPS states in the open sector.
The (refined) closed topological string free energy quantity depends on two set of parameters: the $\Omega$ background parameters ($\espilon_1$ and~$\espilon_2$) and the K\"ahler parameter $t$. We will further restrict to
\[
\espilon_2=\alpha \espilon_1, \qquad\epsilon_1=\hbar. \]
As discussed around \eqref{symhmh} one has a simple symmetry $\hbar\to -\hbar$.
Hence one can study without loss of generality the case $\operatorname{Re}(\hbar)\geq0$.
Likewise, since
\[
{\rm Li}_{-n}(z)=(-1)^{n-1}{\rm Li}_{-n}\left({1\over z}\right),\qquad n\in \IN_+\]
the Borel summation has the symmetry
\[ t\to -t.\]
Hence we can restrict without loss of generality to the case
\[ \operatorname{Re}(t)\geq 0.\]

In the rest of the section, we compute analytically the Borel transform and Borel summantion of the (refined) closed topological string free energy. We also find that the singularity structure of the Borel plane for $\alpha=0$ and $\alpha=-1$ is identical and, as we predicted, the Borel singularities correspond to the central charges of 5d BPS KK-modes for the resolved conifold \eqref{5dcconi}. This is the 5d generalization of \cite{Grassi:2019coc, Grassi:2021wpw}. From the stringy perspective such 5d BPS KK-modes come from $\text{D2}\pm m \text{D0}$ branes in the Type IIA theory compactified on $X$.\footnote{We do not see the purely $\text{D0}$ brane contribution in the Borel plane because we are not including the constant map contribution in the (refined) free energy.}

When $\alpha\neq 0,-1$
we have also an additional series of poles whose positions are at \be \label{bpsa}-2\pi n \frac{\text{(the central charges of 5d BPS KK-modes)}}{\alpha}.\ee
When $\alpha\to 0$ these poles go to infinity while when $\alpha\to -1$ they merge with the other series of poles.

Note also that one should be able to obtain information on the closed sector starting from the open sector. For example in the $\epsilon_1=-\espilon_2$ phase of the $\Omega$ background this can be done using the topological recursion framework, see for instance \cite[equation~(3.9)]{bkmp}. For the NS phase see for instance \cite{acdkv} and references there. Nevertheless our analysis of the closed sector will be carried out independently of the discussion of the open sector.

\subsubsection[The NS sector alpha=0]{The NS sector $\boldsymbol{\alpha=0}$} \label{closedNS}

The first case that we study is the one where $\espilon_1=\hbar$ and $\espilon_2=0$. This is the so-called NS phase of the $\Omega$ background. The corresponding perturbative free energy is\footnote{Usually there is also an overall piece which is the analogous of the "constant map" contribution in the standard topological string. In turn this is given by the closed $\IC^3$ free energy. Here we will omit this contribution and, as a consequence, in the Borel plane we won't see the contribution from purely D0-branes.}
\be \label{eq:FNSWKB}F^{\text{NS}}_{\text{WKB}}=\hbar \sum_{g\geq 0}{\hbar}^{2g-2}{ F}_g^{\text{NS}}(t),\ee
where
\be \label{FNSWKBg}{ F}^{\rm NS}_g(t)=\frac{(\ri)^{2g-1}B_{2g}\big(\frac{1}{2}\big)}{(2g)!}\operatorname{Li}_{3-2g}\bigl(\re^{-t}\bigr).\ee
The NS free energy is \cite{ikv,ns}
\be \label{FNSe}F^{\text{NS}}({\hbar},t)=\frac{1}{2\ri}\sum_{k\geq 1}\frac{\re^{-k t}}{k^2}\frac{1}{\sin(k{\hbar/2})} .\ee
Further expansion of \eqref{eq:FNSWKB} with respect to $Q=\re^{-t}$ agrees with the expansion of \eqref{FNSe} with respect to ${\hbar}$.
The Borel transform of the series \eqref{FNSWKBg} is
\[ \mathcal{B}F^{\rm NS}_{\rm WKB}(t,\xi)=\sum_{g\geq 2} F_{g}^{\rm NS}(t) {\xi^{2g-3}\over (2g-3)!}=f_1(\xi)\star f_2(\xi,t),\]
where $\star$ stands for the Hadamard product, and we use
\begin{align*}
 &f_1(\xi)=\sum_{g=2}^{\infty}\frac{B_{2g}(1/2)}{(2g)!}(\xi)^{2g-3}=\frac{\xi ^2+12 \xi \operatorname{csch}\bigl(\frac{\xi }{2}\bigr)-24}{24 \xi ^3},\\
&f_2(\xi,t)=\sum_{g\geq 2}{\rm Li}_{3-2g}\bigl(\re^{-t}\bigr) (\ri)^{2g-1} {\xi^{2g-3}\over (2g-3)!}=\ri \frac{\sin ( \xi )}{2 \cos ( \xi )-2 \cosh (t)}.
\end{align*}
It follows from the definition of the Hadamard product that
\[ \mathcal{B}F^{\rm NS}_{\rm WKB}(t,\xi)=\frac{1}{2\pi\ri}\oint_\gamma f_1(s)f_2\left(\frac{\xi}{s},t\right)\frac{\rd s}{s}, \]
where $\gamma$ is a contour around $0$ including only poles of $f_2\bigl(\frac{\xi}{s}\bigr)$,\footnote{We assume $t$ is valued in the domain such that the absolute values of poles of $f_2\bigl(\frac{\xi}{s}\bigr)$ are always smaller than the absolute values of poles of $f_1(s)$. The result for generic values of $t$ is defined by analytic continuation.
} located at
\[ s=\pm \frac{\ri \xi }{t+2 \ri \pi m},\qquad m\in \IZ. \]
 Computing the integral by residues, the Borel transform can be expressed as an exact function in $\xi$,
\[\label{gns}
 \mathcal{B}F^{\rm NS}_{\rm WKB}(t,\xi)= \sum_{m\in \IZ}\frac{\ri \xi ^2+12 \xi (t+2 \pi \ri m) \operatorname{csch}\bigl(\frac{\ri\xi }{2(t+2\pi\ri m)}\bigr)+24 i (t+2\pi\ri m)^2}{24 \xi ^3}.\]
This expression shows explicitly that the poles of the Borel transform are located at
\be\label{nspoles}\xi=2\pi n\left(t+2\pi\ri m\right), \qquad m\in\mathbb{Z},\qquad n\in \IZ/\{0\}.\ee
As we discussed at the beginning of Section~\ref{closeconi},
these indeed correspond to the central charges of 5d BPS KK-modes for the resolved conifold: see \eqref{5dcconi}.

Let us now look at the Borel summation as defined in \eqref{boreldef2}. We start from the positive imaginary axis. Along this axis we find that
\be \label{Nsim}s\bigl(F^{\rm NS}_{\rm WKB}\bigr)(t,\hbar)=F^{\text{NS}}({\hbar},t), \qquad \hbar\in \ri \IR_+.\ee
Note that if ${\rm {Re}}(t)=0$ all the poles in the Borel plane are along the imaginary axis. Hence we should understand the l.h.s.\ of \eqref{Nsim} as median summation.
To obtain the exact expression of Borel summation in
other sectors, we simply sum over the contributions coming from the poles lying along the rays that we cross when moving from one sector to another.
For example, if we want to obtain the expression for Borel summation on the real axis, we sum over the contributions from all the poles in the first quadrant of the Borel plane. Let assume for example that $\operatorname{Re}(t)>0$, $\operatorname{Im}(t)\in(0, 2\pi$) and $\hbar\in \IR_+$. Since $m\in \IZ$ the relevant poles in the first quadrant are at \be \label{polesns}\ba
\xi=&2 \pi n (t+2 \ri \pi m), \qquad m\geq0, \qquad n> 0.\\
\ea\ee
The corresponding contribution is
\begin{align*}
&2\pi\ri\sum_{m\geq0,n\geq 1}\operatorname{Res}\bigl( \mathcal{B}F^{\rm NS}_{\rm WKB}(t,\xi),2 \pi n (t+2 \ri \pi m)\bigr)\\
&\qquad=\frac{\hbar}{4\pi\ri}\sum_{n\geq 1}\frac{(-1)^n}{n^2 \sin\bigl(\frac{2\pi^2 }{\hbar}n\bigr)}\E^{-\frac{2\pi n}{\hbar}(t-\pi\ri)}={ \hbar \over 2\pi} F^{\text{NS}}\left({4 \pi^2\over \hbar},\E^{{2\pi \over \hbar}(t-\ri \pi)+\ri \pi)}\right).\end{align*}
Hence the Borel summation for $\hbar\in\IR_+$ is\footnote{If $\operatorname{Re}(t)=0$ and $\operatorname{Im}(t)\in (0,2\pi)$ all the poles of the Borel transform are on the imaginary axis. Hence, away from this axis, Borel summation agrees with $F_{\rm NS}(\hbar,t)+{ \hbar \over 2\pi} F^{\text{NS}}\bigl({4 \pi^2\over \hbar},\re^{{2\pi \over \hbar}(t-\ri \pi)+\ri \pi)}\bigr)$. There is only one subtlety which is that $ F^{\text{NS}}$ is not well defined if $\operatorname{Re}(t)=0$ \textit{and} $\operatorname{Im}(\hbar)=0$.}
\begin{gather} \label{f2ns}s\bigl(F^{\rm NS}_{\rm WKB}\bigr)(t,\hbar)=\begin{cases}
F^{\rm NS}(\hbar,t)+{ \hbar \over 2\pi} F^{\text{NS}}\bigl({4 \pi^2\over \hbar},\re^{{2\pi \over \hbar}(t-\ri \pi)+\ri \pi}\bigr), \quad \operatorname{Im}(t)\in (0,2\pi),\quad 0< \operatorname{Re}(t),\vspace{2mm}\\
F^{\rm NS}(\hbar,t)+{ \hbar \over 2\pi} F^{\text{NS}}\bigl({4 \pi^2\over \hbar},\re^{{2\pi \over \hbar}(t-\ri \pi)+\ri \pi}\bigr)-\frac{\hbar \operatorname{Li}_2\bigl(-\E^{-\frac{2\pi t} \hbar}\bigr)}{4\pi}, \quad t \in \IR_+.
\end{cases}\hspace{-20mm}
\end{gather}
This is very much expected from the point of view of the spectral theory for relativistic integrable systems \cite{ggu, ghm, swh, wzh}.
\subsubsection[The GV sector alpha=-1]{The GV sector $\boldsymbol{\alpha=-1}$}\label{closedGV}

We now study the case where $\espilon_1=-\epsilon_2=\hbar$. This is the so-called GV phase of the $\Omega$ background (also known as self-dual or standard topological string phase).
The perturbative expansion of the free energy is \be \label{fwkb} F^{\rm GV}_{\rm WKB}(t)=\sum_{g\geq 0}\hbar^{2g-2}F_g(t),\ee
 where
 \begin{align*} &F_0(t)=-\operatorname{Li}_3(\exp (-t)),\qquad
 F_1(t)= - \frac{1}{12} \operatorname{Li}_1(\exp (-t)),\\
& F_g(t)=- \frac{(-1)^{g-1} B_{2 g} \operatorname{Li}_{3-2 g}(\exp (-t))}{2 g (2 g-2)!}, \qquad g\geq 2,
 \end{align*}
and $B_k$ is the standard Bernoulli number. The Gopakumar--Vafa free energy is
 \be\label{f1coni} F^{\rm GV}(t,\hbar)=-\sum_{m\geq 1}{\re^{-m t}\over m} \left(2\sin\left({m \hbar\over 2}\right)\right)^{-2}.\ee
 If we expand \eqref{fwkb} with respect to $Q=\re^{-t}$ and \eqref{f1coni} with respect to ${\hbar}$, we find agreement between the two series.

The Borel transform of \eqref{fwkb} is
\begin{align*} &\mathcal{B}F_{\rm WKB}^{\rm GV}(t,\xi)= \!\sum_{g\geq 1} \xi^{2g-1}{F_{g+1}(t)\over (2g-1)!}=-\!\sum_{g\geq 1}\xi^{2g-1}\frac{(-1)^g B_{2 (g+1)} }{2 (g+1) (2 g)!} {f^{2g-1}(t)\over (2g-1)!}
=-f_1(\xi)\star f_2(\xi, t),
\end{align*}
where $\star$ is the Hadamard product and we choose
\begin{align*} &f_1(\xi)=\frac{-\xi^2+3 \xi^2 \csc ^2\bigl(\frac{\xi}{2}\bigr)-12}{12 \xi^3},\qquad f_2(\xi,t)={1\over 2}\left(f(t+\xi)-f(t-\xi)\right),
\end{align*}
where $ f(t)=\frac{1}{1-\re^t}.$
By using the integral representation of the Hadamard product we get
\[ \mathcal{B}F_{\rm WKB}^{\rm GV}(t,\xi)=-{1\over 2\pi \ri}\oint_\gamma f_1(s) f_2\left({\xi\over s},t\right){\rd s \over s}.\]
As before the integral contour $\gamma$ is chosen such that it only includes the contribution from
the poles of $f_2(\xi/s)$ at
\[ s=\pm \frac{\xi }{ t+2 \ri \pi m},\qquad m\in \IZ.\]
Hence we get
\be \label{Gres} \ba
 \mathcal{B}F_{\rm WKB}^{\rm GV}(t,\xi)=
 2\sum_{m\in \IZ} \frac{\xi ^2+3 \xi ^2 \operatorname{csch}^2\bigl(\frac{\xi }{4 \pi m-2 \ri t}\bigr)-12 (2 \pi m-\ri t)^2}{24 \xi ^3}.
\ea\ee
The singularities of the Borel transform \eqref{Gres} are at
\be \label{polesb} \xi=2\pi n \left( t+2\pi\ri m\right), \qquad m\in \IZ, \quad n \in {\IZ}/\{0\}. \ee
This is exactly as in \eqref{polesns}, and in agreement with the 5d BPS KK-modes central charges \eqref{5dcconi}.
Let us look at the Borel summation.
When $\hbar$ is purely imaginary we find
\[ s\bigl(F^{\rm GV}_{\rm WKB}\bigr)(t,\hbar)= F^{\rm GV}(\hbar,t),\qquad \hbar\in \ri \IR_+.\]
As before, to obtain the exact expression in other sectors we need to take into account the residue contribution from the poles in the Laplace transform \eqref{boreldef}. For example, for $\hbar\in \IR_+$ with a bit of algebra, we obtain
\[ 2\pi\ri\sum_{m\geq0,n\geq 1}\operatorname{Res}\left(\mathcal{B}F_{\rm WKB}^{\rm GV}(t,\xi),2 \pi n (t+2 \ri \pi m)\right)=-{1\over 2 \pi \ri}{\partial\over\partial \hbar}\left( \hbar F^{{\rm NS}}\left({2\pi^2\over \hbar},{2\pi (t-\ri \pi)\over \hbar}\right) \right).\]
We thus get\footnote{If $\operatorname{Re}(t)=0$ and $\operatorname{Im}(t)\in (0,2\pi)$ all the poles of the Borel transform are on the imaginary axis. Hence, away from this axis, Borel summation agrees with $F^{\rm GV}(\hbar,t)-{1\over 2 \pi \ri}{\partial\over\partial \hbar}\big( \hbar F^{{\rm NS}}\big({2\pi^2\over \hbar},{2\pi (t-\ri \pi)\over \hbar}\big) \big)$. There is only one subtlety which is that $ F^{\text{GV}}$ is not well defined if $\operatorname{Re}(t)=0$ and $\operatorname{Im}(\hbar)=0$.}
\begin{align}\label{fingv}
s\bigl(F^{\rm GV}_{\rm WKB}\bigr)(t,\hbar)=\begin{cases}
  \displaystyle F^{\rm GV}(\hbar,t)-{1\over 2 \pi \ri}{\partial\over\partial \hbar}\left( \hbar F^{{\rm NS}}\left({2\pi^2\over \hbar},{2\pi (t-\ri \pi)\over \hbar}\right) \right), \\
\qquad \qquad\qquad \operatorname{Im} (t)\in (0,2\pi),  \qquad 0< \operatorname{Re}(t),\\
  \displaystyle F^{\rm GV}(\hbar,t)-{1\over 2 \pi \ri}{\partial\over\partial \hbar}\left( \hbar F^{{\rm NS}}\left({2\pi^2\over \hbar},{2\pi (t-\ri \pi)\over \hbar}\right) \right)\\
\displaystyle \phantom{F^{\rm GV}(\hbar,t)}{} -\frac{\ri\bigl(-2\pi t\log\bigl(1-\E^{-\frac{2\pi t}{\hbar}}\bigr)+\hbar\operatorname{Li}_2\bigl(\E^{-\frac{2\pi t}{\hbar}}\bigr)\bigr)}{4 \hbar\pi}, \qquad t \in \IR_+,
\end{cases}\hspace{-10mm} \end{align}
which is in agreement with \cite{ho2}.

\subsubsection[The refined sector epsilon\_2=alpha epsilon\_1, alpha not in Q]{The refined sector $\boldsymbol{\epsilon_2=\alpha \epsilon_1}$, $\boldsymbol{\alpha \not\in \mathbb{Q}}$}

The refined free energy for the resolved conifold is (see, for instance, \cite[equation~(67)]{ikv})
\be\label{ref} F^{\rm ref}(q_1,q_2,t)=-\sum_{n\geq 1} {\re^{-n t} \over n \bigl(q_1^{n/2}-q_1^{-n/2}\bigr)\bigl(q_2^{n/2}-q_2^{-n/2}\bigr)}.\ee
If $q_1=q_2^{-1}=\re^{\ri \hbar}$, we recover \eqref{f1coni}.
If $q_1=\re^{\ri \hbar}$, $q_2=\re^{\ri\epsilon_2}$, then we get
\[
-\lim_{\epsilon_2\to 0} \ri\epsilon_2 F^{\rm ref}\bigl(\re^{\ri \hbar},\re^{\ri\epsilon_2},t\bigr)=F^{\rm NS}(\hbar,t),\]
where $F^{\rm NS}(\hbar,t)$ is as in \eqref{FNSe}.
In this section, we study the case
\[ q_1=\re^{\ri \hbar}, \qquad q_2=\re^{\ri \alpha \hbar},\]
and we denote the corresponding free energy by
\[
F(\alpha, \hbar,t)\equiv F^{\rm ref}\bigl(\re^{\ri \hbar},\re^{\ri \alpha \hbar},t\bigr).\]
The perturbative refined free energy is \be\label{frefp} F^{\alpha}_{\rm WKB }(\alpha,\hbar,t)=\frac{\operatorname{Li}_3\bigl(\re^{-t}\bigr)}{\alpha \hbar^2}-\frac{\bigl(\alpha ^2+1\bigr) \log \bigl(1-\re^{-t}\bigr)}{24 \alpha }+\sum_{g\geq 2}\hbar^{2g-2}F_g^{\alpha}(t), \ee
where
\begin{gather*} F_g^{\alpha}(t)=c_g(\alpha)\operatorname{Li}_{3-2 g}\bigl(\re^{-t}\bigr),\qquad
c_g(\alpha)={(-1)^g \sum _{k=0}^g \hat B_{2 k} \hat B_{2 g-2 k} (\alpha)^{2g-2 k-1}},\\
\hat B_{ m}=\left(\frac{1}{2^{m-1}}-1\right) \frac{B_m}{m!}.
\end{gather*}
The Borel transform of \eqref{frefp} is
\begin{align}
&\mathcal{B}F^\alpha_{\rm WKB}(t,\xi)=\sum_{g\geq 2} \xi^{2g-3}{F_{g}^\alpha(t)\over (2g-3)!} =\sum_{k\geq 0}\hat B_{2k}\alpha^{2k-1}\sum_{g\geq k} \xi^{2g-3}
(-1)^{g}\hat B_{2g-2k}{f^{2g-3}(t)\over (2g-3)!},\label{Galpha}
\end{align}
where
\[ f(t)=\frac{1}{1-\re^t}.\]
By the calculation in Appendix~\ref{Ga}, we find
\begin{gather}\label{gafin} \mathcal{B}F^\alpha_{\rm WKB} (t,\xi)= -\!\sum_{n\in \IZ}\!\frac{\bigl(\alpha ^2+1\bigr) \xi ^2+6 \alpha \xi ^2 \operatorname{csch}\bigl(\frac{\xi }{4 \pi n-2 \ri t}\bigr) \operatorname{csch}\bigl(\frac{\alpha \xi }{4 \pi n-2 \ri t}\bigr)-24 (2 \pi n-\ri t)^2}{24 \alpha \xi ^3},\!\!\!
\end{gather}
which has poles at
\begin{align}
\xi&=2 \pi m (t+2 \ri \pi n), \qquad n\in \IZ,\quad m\in \IZ/\{0\},\nonumber \\
\xi&=\frac{2 \pi}{ \alpha} m (t+2 \ri \pi n), \qquad n\in \IZ,\quad m\in \IZ/\{0\},\label{polesa}
\end{align}
in agreement with our discussion about 5d BPS KK-modes central charges, see \eqref{bpsa} and \eqref{5dcconi}.
We start from the Borel summation along the imaginary axis, we find
\[ s\bigl( F^{\alpha}_{\rm WKB }\bigr)(t,\hbar) = F(\alpha,\hbar, t), \qquad \hbar=\ri \IR_+. \]
We now wish to go to the Borel summation along the real axis. For this we have to properly take into account the residues of the Borel transform in the first quadrant of the Borel plane. We find that the residue at the pole listed in \eqref{polesa} with fixed $m$, $n$ is
\begin{align*}
 & {\rm{Res}}_{m,n}\bigl({\mathcal{B}F^\alpha_{\rm WKB} \bigl(t,\xi \re^{\ri {\rm arg}(\hbar)}\bigr)\re^{-\xi/\hbar}}\bigr)\\
 &\qquad {}\equiv{\rm{Res}}\left({\mathcal{B}F^\alpha_{\rm WKB} \bigl(t,\xi \re^{\ri {\rm arg}(\hbar)}\bigr)\re^{-\xi/\hbar}},2 \pi m (t+2 \ri \pi n)\right) \\
 &\qquad\quad{}
 +{\rm{Res}}\Bigr({\mathcal{B}F^\alpha_{\rm WKB} \bigl(t,\xi \re^{\ri {\rm arg}(\hbar)}\bigr)\re^{-\xi/\hbar}},\frac{2 \pi}{ \alpha} m (t+2 \ri \pi n)\Bigl)\\
&\qquad{} ={(-1)^m\over 4 \pi m}\Bigr(\csc \left(\frac{\pi m}{\alpha }\right) \E^{-\frac{2 \pi m (t+2 \ri \pi n)}{\alpha \hbar}}
 + \csc (\pi \alpha m) \E^{-\frac{2 \pi m (t+2 \ri \pi n)}{\hbar}}\Bigl).
 \end{align*}
Note that we are considering the case $\alpha \not\in {\mathbb{Q}}$.
Let us take $ \operatorname{Re}(\alpha) >0$ as an example.
Since we care about the poles in the first quadrant we take $m>0$ and $n\geq 0$. Hence we have to consider
\begin{align*}
 & \sum_{m>0}\sum_{n\geq 0}{\rm{Res}}_{m,n}\bigl({\mathcal{B}F^\alpha_{\rm WKB} \bigl(t,\xi \re^{\ri {\rm arg}(\hbar)}\bigr)\re^{-\xi/\hbar}}\bigr)\\
& \qquad{} =-\sum_{m> 0} \frac{\ri (-1)^m \csc \bigl(\frac{\pi m}{\alpha }\bigr) \csc \bigl(\frac{2 \pi ^2 m}{\alpha \hbar}\bigr) \E^{\frac{2 \ri \pi m (\pi +\ri t)}{\alpha \hbar}}}{8 \pi m}\\
&\qquad\quad {}-\sum_{m> 0} \frac{\ri (-1)^m \csc (\pi \alpha m) \csc \bigl(\frac{2 \pi ^2 m}{\hbar}\bigr)\E^{\frac{2 \ri \pi m (\pi +\ri t)}{\hbar}} }{8 \pi m}\\
&\qquad{}= {1\over 2 \pi \ri} \bigg(F\bigg(\frac{2 \pi }{\hbar},\frac{2 \pi }{\alpha },\frac{ 2\pi }{\alpha \hbar}( t-\ri \pi )\bigg)
+F\bigg(\frac{2 \pi }{\alpha \hbar}, 2\pi\alpha,\frac{2 \pi }{\hbar} (t-\ri\pi )\bigg) \bigg).
\end{align*}
Therefore for $\hbar\in \IR_+$, we have
\begin{align}\label{finA}
 s\bigl( F^{\alpha}_{\rm WKB }\bigr)(t,\hbar) =\begin{cases}
 \displaystyle F(\alpha,\hbar, t)+F\left(\frac{2 \pi }{\hbar},\frac{2 \pi }{\alpha },\frac{ 2\pi }{\alpha \hbar}( t-\ri \pi )\right)\\
\displaystyle \phantom{F(\alpha,\hbar, t)}{} + F\left(\frac{2 \pi }{\alpha \hbar}, 2\pi\alpha,\frac{2 \pi }{\hbar} (t-\ri\pi )\right) \qquad\operatorname{Im}(t)\in (0,2\pi), \quad 0< \operatorname{Re}(t),
\\
\displaystyle F(\alpha,\hbar, t)+F\left(\frac{2 \pi }{\hbar},\frac{2 \pi }{\alpha },\frac{ 2\pi }{\alpha \hbar}( t-\ri \pi )\right) + F\left(\frac{2 \pi }{\alpha \hbar}, 2\pi\alpha,\frac{2 \pi }{\hbar} (t-\ri\pi )\right) \\
\displaystyle \phantom{F(\alpha,\hbar, t)}{}-\pi\ri \sum_{m=1}^{\infty}\frac{(-1)^m\bigl(\operatorname{csc}\bigl(\frac{\pi m}{\alpha}\bigr)\E^{-\frac{2\pi m t}{\alpha\hbar}}+\operatorname{csc}(\pi \alpha m)\E^{-\frac{2\pi m t}{\hbar}}\bigr)}{4\pi m}\\
 \qquad\qquad\qquad 0<t \in \IR.
 \end{cases}\hspace{-20mm}\end{align}
We also cross-checked this result numerically.
Some observations:
\begin{itemize}\itemsep=0pt
\item Each term on the r.h.s.\ of equation \eqref{finA} has a dense set of poles on the real $\hbar$ axis. However in the full expression these poles cancel, this is a generalization of the HMO cancellation mechanism \cite{hmo2} to the refined topological string setup. After the cancellation, the remaining regular part matches with the Borel summation.
\item Even though the general structure of the r.h.s.\ of \eqref{finA} resembles \cite{hmmo, lv}, the details of the expression are different (e.g., different shift in the K\"ahler parameter and in the $\epsilon$'s).
\item It would be interesting to study more in details the relation between the l.h.s.\ of \eqref{finA} and the refined CS matrix model \cite{Aganagic:2011sg} similar to what was done in \cite{ho2} for the unrefined case.
\end{itemize}

\section{Comment on higher genus geometries}\label{hggen}
In this paper, we tested our proposal in two concrete examples in which the underlying mirror curves have genus zero.
Therefore, it is natural to ask to what extent our proposal can be generalized
 to difference equations corresponding to higher genus geometries such as, say, local~$\IP^2$ or local $\IP^1\times \IP^1$.
We expect the relation between exponential network and exact WKB, particularly the connection between singularities in the Borel plane and BPS central charges, still to hold also in this more general framework.\footnote{We also recall that,
when $\epsilon_1=-\espilon_2$, the singularities in the Borel plane of the closed string free energy have been studied for instance in \cite{cesv2,cms,dmp-np, gmz}. More precisely, it was observed that these singularities are related to combinations of periods of the underlying CY manifold. This is of course consistent with our results since the the central charges of 5d BPS KK-modes correspond to a particular combinations of periods.} Likewise, we also expect different domains of the exponential network to be related to open strings with brane insertions at different places. There are nevertheless some differences which we discuss below.
\begin{itemize}\itemsep=0pt
\item One important difference is the fact that in genus zero geometries we can express the local solutions to the difference equations either by using the NS or by using the GV open topological string free partition function. Switching between these two phases is very straightforward.

This is no longer the case for difference equations arising in quantization of mirror curves to higher genus geometries. In this case the WKB solution to the quantum curve is encoded in the Nekrasov--Shatashvili phase $\espilon_2=0$, $\espilon_1={ \hbar}$ \cite{acdkv} while the non-perturbative corrections are encoded in the $\espilon_2=-\epsilon_1={1\over \hbar}$ phase \cite{cgm, ghm, mz-wv, mz-wv2}. One may argue in favour of a connection between the NS and the GV phase using blowup equation as in~\cite{Bershtein:2021uts,ggu,Jeong:2020uxz,Lencses:2017dgf,Nekrasov:2020qcq}. However this is much more subtle than for the case of the resolved conifold.

\item In the WKB solution for quantum curves of higher genus an important role is also played by the quantum mirror map. This is a new ingredient which is absent in the resolved conifold example (the mirror map does not get quantum corrections in this case). In particular, even though we can compute the genus $g$ free energy in the NS phase efficiently via the holomorphic anomaly equation \cite{coms}, we do not know an efficient way to compute the quantum mirror map away from the large radius region of the moduli space. This is one of the main technical obstacles we encounter when trying to construct an efficient algorithm computing WKB for difference equations.
\item Some further comments on the closed sector:
\begin{itemize}\itemsep=0pt
\item In the case of the resolved conifold, the structure of the Borel singularities for the NS and GV phase of the $\Omega$ background is in fact identical (see poles structure in~\eqref{nspoles} and~\eqref{polesb}). In higher genus geometries it could be that this relation is more complicated. Nevertheless we know that these two $\Omega$ background phases are related by blowup equations \cite{ggu}, see also \cite{Bershtein:2021uts,Nekrasov:2020qcq} and references there. Hence it should be possible to find a relation between the two Borel planes. It would be interesting to investigate this further.

\item The topological vertex expression for the free energy of the resolved conifold is well defined also for complex values of $\epsilon$ parameters. In particular, \eqref{FNSe}, \eqref{f1coni} and \eqref{ref}, as series expansion in $Q=\re^{-t}$, are convergent even when the $\epsilon$'s are complex.

This is not the case for CY of higher genus. For example if we consider local $\IP^2$ and we take $\epsilon_1=-\epsilon_2$ to be complex, then the topological vertex expression (as series expansion in $Q=\re^{-t}$) is divergent. See for example \cite{hmmo} for some numerical studies.\footnote{For some geometries, like local $\IP^1\times \IP^1$, one can nevertheless perform a partial resummation of the topological vertex expression with respect to one of the K\"ahler parameters. This gives the Nekrasov type of expression. For complex values of the coupling the latter is better behaving, see \cite{bsu, gm17} for related discussions. Such Nekrasov expression is however not always available. For example we currently do not have it for local $\IP^2$. }

\item For the resolved conifold we saw that, on the axis where the $\epsilon$'s are real, the Borel summation of the $F_g$'s matches a suitable combination of free energies in different phases of the $\Omega$ background, see \eqref{f2ns}, \eqref{fingv} and \eqref{finA}.

In the case of higher genus geometries this is no longer the case. Explicit tests have been performed in \cite{gmz}\footnote{To be precise in \cite{gmz} the authors also consider the quantum mirror map. Here we do not consider the quantum mirror map, but nevertheless we have checked that Borel summation does not agree with expressions like \eqref{fingv}.} and further investigations were done in \cite{cms}. So in higher genus examples the Borel summation does not match the non-perturbative completion of topological string coming from the spectral theory of quantum mirror curves: there are additional non-perturbative effects which are not captured by Borel summation\footnote{This happens also in simpler quantum mechanical examples like the pure quartic oscillator \cite{gmz}.} (at least not in the chamber connected to the real $\epsilon$ axis). It would be interesting to understand this using the framework of exponential networks.
 \end{itemize}
 \end{itemize}

\appendix
\section{Conventions}
In this work we study asymptotic series of the form
\be \label{fgen}\phi(x,\hbar) \sim \hbar^{a+1} \sum_{g\geq 0} c_g(x)\hbar^{2g-b}, \qquad c_g\approx (2g-b)!, \qquad g\gg1,\ee
where $a$, $b$ are some fixed constants.
For example, when considering the asymptotic series of local solutions we have
\[ (a, b)= (0,2), \]
while for the closed topological string free energy we have $b=3$.
Our convention for the Borel transform of \eqref{fgen} is to take
\be \label{bdefo}{\bf \mathcal{B}}\phi(x,\xi)=\sum_{g\geq \ceil{b/2}}c_g(x) {\xi^{2g-b} \over (2g-b)!}. \ee
Let
\[ \hbar=\re^{\ri \vartheta} |\hbar|,\]
and let $\rho_\vartheta$ be the ray along $\vartheta$.
We define the Borel summation of \eqref{fgen} as
\be \label{boreldef}
 \mathcal{L}\mathcal{B}\phi(x,\hbar)=\int_{\rho_\vartheta}{\bf \mathcal{B}}\phi(x,\xi)\E^{-\frac{\xi}{\hbar}}\rd\xi = \E^{\ri\vartheta}\int_{0}^\infty{\bf \mathcal{B}}\phi\bigl(x,\xi\E^{\ri\vartheta}\bigr)\E^{-\frac{\xi}{|\hbar|}}\rd\xi,\ee
with the understanding that, if ${\bf \mathcal{B}}\phi(x,\hbar)$ has poles along the integration contour, we take median summation.
We will also use
\be \label{boreldef2}s(\phi)(x,\hbar)=\hbar^ a \left(\sum_{g=0}^{ \floor{b/2}} c_g(x)\hbar^{2g-b+1}+ \mathcal{L}\mathcal{B}\phi(x,\hbar) \right).\ee

\section[Quantum dilogarithm and q-Pochhammer functions]{Quantum dilogarithm and $\boldsymbol{q}$-Pochhammer functions}
\label{ap:fad}

The Faddeev's quantum dilogarithm \cite{Faddeev:1995nb, fk} admits an integral representation
 \be\label{definePhixint}\Phi_{\bf b}(x)=\exp\left(\int_{\mathbb{R}+\ri\epsilon}\frac{\E^{-2\ri xz}}{4\sinh(z{\bf b})\sinh\bigl(z {\bf b}^{-1}\bigr)}\frac{\rd z}{z}\right), \qquad {{\bf b}^2}\notin \IR_-, \qquad |\operatorname{Im}(x)|<|\operatorname{Im}(c_{\bf b})|,\ee
 where
 \[c_{\bf b} ={\ri\over 2}\bigl({\bf b}+{\bf b}^{-1}\bigr).\]
When $ \operatorname{Im}\bigl({\bf b^2}\bigr)>0,~-\pi<\operatorname{Im}(x)\leq\pi$, the
Faddeev's quantum dilogarithm admit an alternative representation as
\be\label{phiq}\Phi_{\bf b}(x)=\frac{\bigl(\E^{2\pi {\bf b}(x+\frac{\ri}{2}({\bf b}+{\bf b}^{-1}))};q\bigr)_{\infty}}{\bigl(\E^{2\pi {\bf b}^{-1}(x-\frac{\ri}{2}({\bf b}+{\bf b}^{-1}))};\tilde{q}\bigr)_{\infty}},\qquad \operatorname{Im}\bigl({{\bf b}^2}\bigr)>0,\ee
where
\begin{align*}
q=\E^{\ri\hbar},\qquad \tilde{q}=\E^{-\frac{4\pi^2\ri}{\hbar}},\qquad \hbar=2\pi {\bf b}^2
\end{align*}
and we used \eqref{defineqpoch}.
When $\operatorname{Im}\bigl({\bf b}^2\bigr)<0$, \eqref{definePhixint} still makes sense, but \eqref{phiq} is not well defined. In order to get an expression in terms of the $q$-Pochhammer symbol, we can use the symmetry property
\[ 
\Phi_{\bf b}(z)=\Phi_{{\bf b}^{-1}}(z).\]
Since
\begin{align*}
&\Phi_{\bf b}(z)=\frac{\bigl(\E^{2\pi {\bf b} z+\pi \ri {\bf b}^2+\pi \ri};\E^{2\pi\ri {\bf b}^2}\bigr)_\infty}{\bigl(\E^{2\pi {\bf b}^{-1}z-\pi\ri-\pi\ri {\bf b}^{-2}};\E^{-2\pi\ri {\bf b}^{-2}}\bigr)_\infty},\qquad
&\Phi_{{\bf b}^{-1}}(z)=\frac{\bigl(\E^{2\pi {\bf b}^{-1} z+\pi \ri {\bf b}^{-2}+\pi \ri};\E^{2\pi\ri {\bf b}^{-2}}\bigr)_\infty}{\bigl(\E^{2\pi {\bf b}z-\pi\ri-\pi\ri {\bf b}^{2}};\E^{-2\pi\ri {\bf b}^{2}}\bigr)_\infty}\end{align*}
for $\operatorname{Im}\bigl({\bf b}^2\bigr)<0$, we should use
\[
\Phi_{{\bf b}^{-1}}\left(\frac{x}{2\pi {\bf b}}\right)=\frac{\bigl(-\tilde{q}^{-\frac{1}{2}}\mathrm{e}^{\frac{2\pi x}{\hbar}};\tilde{q}^{-1}\bigr)_\infty}{\bigl(-q^{-\frac{1}{2}}\mathrm{e}^{x};q^{-1}\bigr)_\infty},\qquad \operatorname{Im}({\bf b}^2)<0.\]
We define
\be \label{fadpoch}\Phi(x,\hbar)=\Phi_{\bf b}\left(\frac{x}{2\pi {\bf b}}\right). \ee
Hence
\begin{align}
\label{pdef}\Phi(x,\hbar)=\begin{cases}\displaystyle\frac{\bigl(-q^{\frac{1}{2}}\mathrm{e}^x;q\bigr)_\infty}{\bigl(-\tilde{q}^{\frac{1}{2}}\mathrm{e}^{\frac{2\pi x}{\hbar}};\tilde{q}\bigr)_{\infty}},&\operatorname{Im}(\hbar)>0,\vspace{1mm}\\
\displaystyle \frac{\bigl(-\tilde{q}^{-\frac{1}{2}}\mathrm{e}^{\frac{2\pi x}{\hbar}};\tilde{q}^{-1}\bigr)_\infty}{\bigl(-q^{-\frac{1}{2}}\mathrm{e}^{x};q^{-1}\bigr)_\infty},& \operatorname{Im}(\hbar)<0.\end{cases}.\end{align}
When $\arg(\hbar)=0$, we use \eqref{fadpoch} and the integral expression of the Faddeev quantum dilog \eqref{definePhixint}.

In the case of $\operatorname{Re}(\hbar)<0$, we define a similar function $\tilde{\Phi}(x,\hbar)$. $\tilde{\Phi}(x,\hbar)$ can be obtained by taking advantage of the symmetry of Borel transform \eqref{hadamardc3} under $\hbar\rightarrow -\hbar$. Since
\begin{align}
\mathcal{L}\mathcal{B}\phi(x,-\hbar)&= -\E^{\ri\vartheta}\int_{0}^\infty{\bf \mathcal{B}}\phi\bigl(x,-\xi\E^{\ri\vartheta}\bigr)\E^{-\frac{\xi}{|\hbar|}}\rd\xi\nonumber\\
&= -\E^{\ri\vartheta}\int_{0}^\infty{\bf \mathcal{B}}\phi\bigl(x,\xi\E^{\ri\vartheta}\bigr)\E^{-\frac{\xi}{|\hbar|}}\rd\xi=-\mathcal{L}\mathcal{B}\phi(x,\hbar),\label{symhmh} \end{align}
and
 \[\frac{1}{\ri(-\hbar)}\operatorname{Li}_2\bigl(-\E^x\bigr)=-\frac{1}{\ri\hbar}\operatorname{Li}_2\bigl(-\E^x\bigr),\]
 we get
 \be\label{symh}\tilde{\Phi}(x,\hbar)=\re^{s(\phi)(x,\hbar)}={1\over \re^{s(\phi)(x,-\hbar)}}=\frac{1}{\Phi(x,-\hbar)},\qquad \hbar\in\mathbb{R}_-.\ee

Hence,
\begin{align}\label{tpdef}\tilde{\Phi}(x,\hbar)=\begin{cases}\displaystyle\frac{\bigl(-q^{\frac{1}{2}}\mathrm{e}^{x};q\bigr)_\infty}{\bigl(-\tilde{q}^{\frac{1}{2}}\mathrm{e}^{-\frac{2\pi x}{\hbar}};\tilde{q}\bigr)_\infty}&\text{for $\operatorname{Im}(\hbar)>0$},\vspace{1mm}\\
\displaystyle\frac{\bigl(-\tilde{q}^{-\frac{1}{2}}\mathrm{e}^{-\frac{2\pi x}{\hbar}};\tilde{q}^{-1}\bigr)_\infty}{\bigl(-q^{-\frac{1}{2}}\mathrm{e}^{x};q^{-1}\bigr)_{\infty}}&\text{for $\operatorname{Im}(\hbar)<0$}.\end{cases}\end{align}

Some other useful identities of ${\rm \Phi}_{\bf b}$ are
\begin{align*}
&{\rm \Phi}_{\bf b}\left(x-{\ri {\bf b}\over 2}\right) = {\rm \Phi}_{\bf b}\left(x+{\ri {\bf b}\over 2}\right) \bigl(1+\re^{2\pi {\bf b} x}\bigr),\\
& {\rm \Phi}_{\bf b} \left(x+\ri {\bf b}^{-1}\right)= {\rm \Phi}_{\bf b} (x){1\over 1-{\tilde q}\re^{2\pi {\bf b}^{-1}(x-c_{\bf b})}}.\end{align*}
It follows that
\[
{\rm \Phi}_{\bf b} \bigl(x-\ri {\bf b}^{-1}\bigr)= {\rm \Phi}_{\bf b} (x){ \bigl(1-{\tilde q}^{-1}\re^{2\pi {\bf b}^{-1}(x-\ri {\bf b}^{-1}-c_{\bf b})}\bigr)}.\]
Hence we have
\[
 {\rm \Phi}_{\bf b}\left(x+ {\ri n \over {\bf b}}\right)={\rm \Phi}_{\bf b} (x){1\over \prod_{k=1}^n\bigl( 1-{\tilde q}^{-1}\re^{2\pi {\bf b}^{-1}(x+\ri {k-1\over {\bf b}}-c_{\bf b})}\bigr)}\]
as well as
\begin{align}
{\rm \Phi}_{\bf b}\left(x- {\ri n \over {\bf b}}\right)&={\rm \Phi}_{\bf b} (x){ \prod_{k=1}^n\bigl( 1-{\tilde q}^{-1}\re^{2\pi {\bf b}^{-1}(x-\ri {k\over {\bf b}}-c_{\bf b})}\bigr)} \nonumber\\
&= {\rm \Phi}_{\bf b} (x){ \prod_{k=1}^n\bigl( 1-{\tilde q}^{k-1}\re^{2\pi {\bf b}^{-1}(x-c_{\bf b})}\bigr)}. \label{sh2}
 \end{align}

\section{Calculations for the solutions in all other sectors}
\label{summaryls}
\subsection[Fourth quadrant of the Borel plane and Re(x)<0]{Fourth quadrant of the Borel plane and $\boldsymbol{\operatorname{Re}(x)<0}$}
\label{sec:4threxg0}

In this case the relevant sectors of the Borel plane are
\begin{align*}
 {\mathcal I}_{m}^{\tcircled{4}}= \left(\vartheta_{m}^-; \vartheta^-_{m-1}\right),\qquad m\geq 1.
 \end{align*}
We sum over the residue contributions, so that the jump from the positive real axis solution to the $m$th sector solution in the fourth quadrant is
\begin{align*}
2\pi\ri\sum_{k=0}^{m-1}\sum_{n=1}^\infty\frac{(-1)^n}{2\pi\ri n}\E^{\frac{2\pi n (x+2\pi\ri k+\ri\pi)}{\hbar}}=-\log\bigl(\bigl(-\tilde{q}^{-\frac{1}{2}}\E^{\frac{2\pi x}{\hbar}};\tilde{q}^{-1}\bigr)_m\bigr).
\end{align*}
Thus the solution in the $m$th sector is
\be\label{4thx<0} s(\phi)(x,\hbar)=\frac{\bigl(-\tilde{q}^{-\frac{1}{2}}\mathrm{e}^{\frac{2\pi x}{\hbar}};\tilde{q}^{-1}\bigr)_\infty}{\bigl(-q^{-\frac{1}{2}}\mathrm{e}^{x};q^{-1}\bigr)_\infty}\frac{1}{\bigl(-\tilde{q}^{-\frac{1}{2}}\mathrm{e}^{\frac{2\pi x}{\hbar}};\tilde{q}^{-1}\bigr)_m},\qquad m\geq 1. \ee
\eqref{4thx<0} can also be written as
\begin{align*}
\Phi(x+2\pi\ri m,\hbar)&=\frac{\prod_{i=0}^\infty\bigl(1+\tilde{q}^{-i-\frac{1}{2}-m}\E^{\frac{2\pi x}{\hbar}}\bigr)}{\prod_{i=0}^\infty\bigl(1+q^{-i-\frac{1}{2}}\bigr)}\\
&=\frac{\bigl(-\tilde{q}^{-\frac{1}{2}}\mathrm{e}^{\frac{2\pi x}{\hbar}};\tilde{q}^{-1}\bigr)_\infty}{\bigl(-q^{-\frac{1}{2}}\mathrm{e}^{x};q^{-1}\bigr)_\infty}\frac{1}{\bigl(-\tilde{q}^{-\frac{1}{2}}\mathrm{e}^{\frac{2\pi x}{\hbar}};\tilde{q}^{-1}\bigr)_m},\qquad m\geq 1.
\end{align*}
Hence we have
\[ \boxed{s(\phi)(x,\hbar)=\log \Phi(x+2\pi\ri m,\hbar), \qquad \arg(\hbar)\in \mathcal{I}_m^{\tcircled{4}}, \qquad m\geq 0.} \]
The solution along negative imaginary axis is obtained by taking $m\to \infty$:
\be\label{nirexg0}\boxed{s(\phi)(x,\hbar)=-\log\bigl({\bigl(-q^{-\frac{1}{2}}\mathrm{e}^{x};q^{-1}\bigr)_\infty}\bigr),\qquad \hbar\in \ri \IR_-,\qquad \operatorname{Re}(x)<0.}\ee

\subsection[4th quadrant of the Borel plane and Re(x)>0]{$\boldsymbol{4}$th quadrant of the Borel plane and $\boldsymbol{\operatorname{Re}(x)>0}$}

In this case
the relevant sector of the Borel plane is
\begin{align*}
 {\mathcal I}_{m}^{\tcircled{4}}= \left(\vartheta^+_{m-1}; \vartheta^+_{m}\right),\qquad m\leq -1.
 \end{align*}
We sum over the residue contributions, so that the jump from positive real axis solution to the $m$th sector solution in the fourth quadrant is
\begin{align*}
2\pi\ri\sum_{k=0}^{-m-1}\sum_{n=1}^\infty-\frac{(-1)^n}{2\pi\ri n}\E^{-\frac{2\pi n (x-2\pi\ri k-\ri\pi)}{\hbar}}=\log\bigl(\bigl(-\tilde{q}^{-\frac{1}{2}}\E^{-\frac{2\pi x}{\hbar}};\tilde{q}^{-1}\bigr)_{-m}\bigr), \qquad m\leq -1.
\end{align*}
So the solution in the $m$th sector is
\[
\frac{\bigl(-\tilde{q}^{-\frac{1}{2}}\mathrm{e}^{\frac{2\pi x}{\hbar}};\tilde{q}^{-1}\bigr)_\infty}{\bigl(-q^{-\frac{1}{2}}\mathrm{e}^{x};q^{-1}\bigr)_\infty}\bigl(-\tilde{q}^{-\frac{1}{2}}\mathrm{e}^{-\frac{2\pi x}{\hbar}};\tilde{q}^{-1}\bigr)_{-m},\qquad m\leq -1,\]
which can also be written as
\[ \boxed{s(\phi)(x,\hbar)=\log \Phi(x+2\pi\ri m,\hbar)+\frac{2\pi m(x+m\pi\mathrm{i})}{\hbar}, \qquad \arg(\hbar)\in \mathcal{I}_m^{\tcircled{4}}, \qquad m\leq -1.}\]
Taking the limit $m\to -\infty$, the solution along negative imaginary axis is
\begin{align*}
&\frac{\bigl(-\tilde{q}^{-\frac{1}{2}}\mathrm{e}^{\frac{2\pi x}{\hbar}};\tilde{q}^{-1}\bigr)_\infty}{\bigl(-q^{-\frac{1}{2}}\mathrm{e}^{x};q^{-1}\bigr)_\infty}\bigl(-\tilde{q}^{-\frac{1}{2}}\mathrm{e}^{-\frac{2\pi x}{\hbar}};\tilde{q}^{-1}\bigr)_{\infty}\\
& \qquad {} =\frac{\vartheta_3\bigl(-\frac{\mathrm{i}\pi x}{\hbar},\tilde{q}^{-\frac{1}{2}}\bigr)}{\tilde{q}^{\frac{1}{24}}\eta\bigl(\frac{2\pi}{\hbar}\bigr)}\frac{1}{\bigl(-q^{-\frac{1}{2}}\mathrm{e}^x;q^{-1}\bigr)}_\infty
=\frac{\E^{\frac{x^2\ri}{2\hbar}}q^{\frac{1}{24}}}{\tilde{q}^{\frac{1}{24}}}{\bigl(-q^{-\frac{1}{2}}\mathrm{e}^{-x};q^{-1}\bigr)}_\infty.
\end{align*}
The matching with Borel summation is
\[
\boxed{s(\phi)(x,\hbar)={\frac{x^2\ri}{2\hbar}}+\log\left(\frac{q^{\frac{1}{24}}}{\tilde{q}^{\frac{1}{24}}}{\bigl(-q^{-\frac{1}{2}}\mathrm{e}^{-x};q^{-1}\bigr)}_\infty\right),\qquad \hbar\in \ri \IR_-,\qquad \operatorname{Re}(x)<0.}\]

\subsection[2nd quadrant, Re(x)<0]{$\boldsymbol{2}$nd quadrant, $\boldsymbol{\operatorname{Re}(x)<0}$}
In this case
the relevant sector of the Borel plane is
\begin{align*}
 {\mathcal I}_{m}^{\tcircled{2}}= \left(\vartheta^+_{m}; \vartheta^+_{m-1}\right),\qquad m\geq 1.
 \end{align*}

We sum over the residue contributions, so that the jump from negative real axis solution to the $m$th sector solution in the second quadrant is
\begin{align}
\label{2ndrexl0m}2\pi\ri\sum_{k=0}^{m-1}\sum_{n=1}^\infty-\frac{(-1)^n}{2\pi\ri n}\E^{-\frac{2\pi n (x+2\pi\ri k+\ri\pi)}{\hbar}}=\log\bigl(\bigl(-\tilde{q}^{\frac{1}{2}}\E^{-\frac{2\pi x}{\hbar}};\tilde{q}\bigr)_{m}\bigr),\qquad m\geq 1.
\end{align}

Solution along the negative real axis can be obtained by dividing \eqref{pisol} by exponential of~\eqref{2ndrexl0m} in the limit $m\rightarrow \infty$
\be\label{niph}\tilde{\Phi}(x)\equiv\frac{\bigl(-q^{\frac{1}{2}}\mathrm{e}^x;q\bigr)_\infty}{\bigl(-\tilde{q}^{\frac{1}{2}}\E^{-\frac{2\pi x}{\hbar}};\tilde{q}\bigr)_\infty},\qquad \operatorname{Im}(\hbar)>0.\ee

Or we can directly using the symmetry \eqref{symh} to get \eqref{tpdef}.

So the solution in the $m$th sector is
\be\label{2ndrexl0}
\frac{\bigl(-q^{\frac{1}{2}}\mathrm{e}^x;q\bigr)_\infty}{\bigl(-\tilde{q}^{\frac{1}{2}}\mathrm{e}^{-\frac{2\pi x}{\hbar}};\tilde{q}\bigr)_\infty}\bigl(-\tilde{q}^{\frac{1}{2}}\mathrm{e}^{-\frac{2\pi x}{\hbar}};\tilde{q}\bigr)_{m},\qquad m\geq 1,\ee
which can also be written as
\[
\tilde{\Phi}(x+2 \pi\mathrm{i} m,\hbar)=\frac{\bigl(-q^{\frac{1}{2}}\mathrm{e}^x;q\bigr)_\infty}{\bigl(-\tilde{q}^{\frac{1}{2}}\E^{-\frac{2\pi x}{\hbar}};\tilde{q}\bigr)_m}=\eqref{2ndrexl0},\qquad m\geq 1.\]
Hence
\[ \boxed{s(\phi)(x,\hbar)=\log\tilde{\Phi}(x+2 \pi\mathrm{i} m,\hbar), \qquad \arg(\hbar)\in \mathcal{I}_m^{\tcircled{2}}, \qquad m\geq 1.} \]

\subsection[2nd quadrant, Re(x)>0]{$\boldsymbol{2}$nd quadrant, $\boldsymbol{\operatorname{Re}(x)>0}$}

In this case
the relevant sector of the Borel plane is
\begin{align*}
 {\mathcal I}_{m}^{\tcircled{2}}= \left(\vartheta^-_{m-1}; \vartheta^-_{m}\right),\qquad m\leq -1.
 \end{align*}

Summing over the residue contributions, the jump from negative real axis solution to the $m$th sector solution is
\begin{align*}
2\pi\ri\sum_{k=0}^{-m-1}\sum_{n=1}^\infty\frac{(-1)^n}{2\pi\ri n}\E^{\frac{2\pi n (x-2\pi\ri k-\ri\pi)}{\hbar}}=\log\bigg(\frac{1}{\bigl(-\tilde{q}^{\frac{1}{2}}\E^{\frac{2\pi x}{\hbar}};\tilde{q}\bigr)_m}\bigg).
\end{align*}
Therefore, the solution in the $m$th sector is
\be\label{sol2ndx>0}\frac{\bigl(-q^{\frac{1}{2}}\mathrm{e}^x;q\bigr)_\infty}{\bigl(-\tilde{q}^{\frac{1}{2}}\mathrm{e}^{-\frac{2\pi x}{\hbar}};\tilde{q}\bigr)_\infty}\frac{1}{\bigl(-\tilde{q}^{\frac{1}{2}}\mathrm{e}^{\frac{2\pi x}{\hbar}};\tilde{q}\bigr)_{-m}}, \qquad m\leq -1,\ee
which can also be written as
\[
\tilde{\Phi}(x+2\pi\mathrm{i} m,\hbar)\mathrm{e}^{\frac{2\pi m(x+m\pi\mathrm{i})}{\hbar}}=\tilde{\Phi}(x,\hbar)\frac{\E^{\frac{2\pi m(x+m\pi\ri)}{\hbar}}}{\bigl(-\E^{-\frac{2\pi x}{\hbar}}\tilde{q}^{-\frac{1}{2}};\tilde{q}^{-1}\bigr)_{-m}}=\eqref{sol2ndx>0}.
\]
 Hence
\[
 \boxed{s(\phi)(x,\hbar)=\log\tilde{\Phi}(x+2 \pi\mathrm{i} m,\hbar)+\frac{2\pi m(x+m\pi\mathrm{i})}{\hbar}, \qquad \arg(\hbar)\in \mathcal{I}_m^{\tcircled{2}}, \qquad m\leq -1.} \]

\subsection[3rd quadrant, Re(x)<0]{$\boldsymbol{3}$rd quadrant, $\boldsymbol{\operatorname{Re}(x)<0}$}

In this case
the relevant sector of the Borel plane is
\begin{align*}
 {\mathcal I}_{m}^{\tcircled{3}}= \left(\vartheta^+_{m}; \vartheta^+_{m-1}\right),\qquad m\leq -1.
 \end{align*}
We sum over the residue contributions and get the jump from negative real axis solution to the $m$th sector solution is
\begin{align}\label{3rdrexl0m}
-2\pi\ri\sum_{k=0}^{-m-1}\sum_{n=1}^\infty-\frac{(-1)^n}{2\pi\ri n}\E^{-\frac{2\pi n (x-2\pi\ri k-\ri\pi)}{\hbar}}=\log\biggl(\frac{1}{\bigl(-\tilde{q}^{-\frac{1}{2}}\E^{-\frac{2\pi x}{\hbar}};\tilde{q}^{-1}\bigr)_{-m}}\bigg),\qquad m\leq -1.
\end{align}
The solution along negative real axis can be obtained by
multiply \eqref{nirexg0} by the inverse of~$\E^\eqref{3rdrexl0m}$ in the limit $m\rightarrow -\infty$. We get
\[
\tilde{\Phi}(x,\hbar)=\frac{\bigl(-\tilde{q}^{-\frac{1}{2}}\E^{-\frac{2\pi x}{\hbar}};\tilde{q}^{-1}\bigr)_{\infty}}{\bigl(-q^{-\frac{1}{2}}\mathrm{e}^{x};q^{-1}\bigr)_\infty},\qquad\operatorname{Im}(\hbar)<0,
\]
which is valid for $\operatorname{Im}(\hbar)<0$, compared to \eqref{niph} which is valid for $\operatorname{Im}(\hbar)>0$. Alternatively, we can directly using the symmetry \eqref{symh} to get \eqref{tpdef}.

So the solution in the $m$th sector is
\[
\frac{\bigl(-\tilde{q}^{-\frac{1}{2}}\mathrm{e}^{-\frac{2\pi x}{\hbar}};\tilde{q}^{-1}\bigr)_\infty}{\bigl(-q^{-\frac{1}{2}}\mathrm{e}^{x};q^{-1}\bigr)_\infty}\frac{1}{\bigl(-\tilde{q}^{-\frac{1}{2}}\mathrm{e}^{-\frac{2\pi x}{\hbar}};\tilde{q}^{-1}\bigr)_{-m}}, \qquad m\leq -1,\]
which can also be written as
\[
\boxed{s(\phi)(x,\hbar)=\log\tilde{\Phi}(x+2\pi\ri m,\hbar), \qquad \arg(\hbar)\in \mathcal{I}_m^{\tcircled{3}}, \qquad m\leq -1.} \]

\subsection[3rd quadrant, Re(x)>0]{$\boldsymbol{3}$rd quadrant, $\boldsymbol{\operatorname{Re}(x)>0}$}

In this case
the relevant sector of the Borel plane is
\begin{align*}
 {\mathcal I}_{m}^{\tcircled{3}}= \left(\vartheta^-_{m-1}; \vartheta^-_{m}\right),\qquad m\geq 1.
 \end{align*}
We sum over the residue contributions, so that the jump from negative real axis solution to the $m$th sector solution in the fourth quadrant is
\begin{align*}
-2\pi\ri\sum_{k=0}^{m-1}\sum_{n=1}^\infty\frac{(-1)^n}{2\pi\ri n}\E^{\frac{2\pi n (x+2\pi\ri k+\ri\pi)}{\hbar}}=\log\bigl(\bigl(-\tilde{q}^{-\frac{1}{2}}\E^{\frac{2\pi x}{\hbar}};\tilde{q}^{-1}\bigr)_m\bigr), \qquad m\geq 1.
\end{align*}
Therefore the solution in the $m$th sector is
\[
\frac{\bigl(-\tilde{q}^{-\frac{1}{2}}\mathrm{e}^{-\frac{2\pi x}{\hbar}};\tilde{q}^{-1}\bigr)_\infty}{\bigl(-q^{-\frac{1}{2}}\mathrm{e}^{x};q^{-1}\bigr)_\infty}\bigl(-\tilde{q}^{-\frac{1}{2}}\mathrm{e}^{\frac{2\pi x}{\hbar}};\tilde{q}^{-1}\bigr)_m,\qquad m\geq 1,\]
which can also be written as
\[\boxed{s(\phi)(x,\hbar)=\log\tilde{\Phi}(x+2\pi\ri m,\hbar)+\frac{2\pi m(x+m\pi\mathrm{i})}{\hbar}, \qquad \arg(\hbar)\in \mathcal{I}_m^{\tcircled{3}}, \qquad m \geq 1.} \]

\subsection[Re(x)=0]{$\boldsymbol{\operatorname{Re}(x)=0}$}\label{appen:rex0}
 As discussed in Section~\ref{sec:rex0}, along the positive imaginary axis, the solution is
\begin{align*}
 s(\phi)(x,\hbar)={}&\frac{1}{2} \Bigg(\frac{\ri \left(12 x^2+\hbar ^2+4 \pi ^2\right)}{24 \hbar }\\
 &-\log\Bigg(\frac{\bigl(-\E^{-x+\frac{\ri \hbar }{2}};\E^{i \hbar }\bigr){}_{\infty }}{\bigl(-\E^{x+\frac{\ri \hbar }{2}};\E^{i \hbar }\bigr){}_{\infty }}\Bigg)\Bigg), \qquad \operatorname{Re}(x)=0, \qquad \hbar\in \ri \IR_+.
\end{align*}
Similarly, along the negative imaginary axis, we have
\begin{align*}
 s(\phi)(x,\hbar)={}&\frac{1}{2} \Bigg(\frac{\ri \left(12 x^2+\hbar ^2+4 \pi ^2\right)}{24 \hbar }\\
 &+\log\Bigg(\frac{\bigl(-\E^{-x-\frac{i \hbar }{2}};\E^{-i \hbar }\bigr){}_{\infty }}{\bigl(-\E^{x-\frac{i \hbar }{2}};\E^{-i \hbar }\bigr){}_{\infty }}\Bigg)\Bigg), \qquad \operatorname{Re}(x)=0, \qquad \hbar\in \ri \IR_-.
 \end{align*}
 For all other values of $\hbar$ with $\operatorname{Re}(\hbar)>0$, the Borel summation matches with $\Phi(x)$:
\[ s(\phi)(x,\hbar)= \log\Phi(x,\hbar), \qquad \operatorname{Re}(x)=0, \qquad
\operatorname{Re} (\hbar)> 0. \]
And for all other values of $\hbar$ with $\operatorname{Re}(\hbar)<0$, the Borel summation matches with $\tilde{\Phi}(x)$:
 \[s(\phi)(x,\hbar)= \log\tilde{\Phi}(x,\hbar), \qquad \operatorname{Re}(x)=0, \qquad
\operatorname{Re} (\hbar)< 0. \]

\section[Borel transform at generic alpha]{Borel transform at generic $\boldsymbol{\alpha}$ }\label{Ga}
In this appendix, we show the calculation for \eqref{gafin}.
It is convenient to write \eqref{Galpha} as
\begin{align*}
 G_\alpha(t,\xi)={}& -\sum_{k=0}^\infty {\bigl(1-2^{2k-1}\bigr)B_{2k}\over (2k)!}\\
 &\times\sum_{g=k}^\infty
\bigl(1-2^{2g-2k-1}\bigr)B_{2g-2k}(\ri/2)^{2g-2}{\operatorname{Li}_{3-2g}(Q)\over (2g-3)! }{(\alpha)^{2g-2k-1}\over (2g-2k)!}\xi^{2g-3},
\end{align*}
which we write as
\[ G_\alpha(t,\xi)=-\sum_{i=1}^3G_\alpha^{(i)}(t,\xi),
\]
where
 \begin{align*}
G_\alpha^{(1)}(t,\xi)={}&\sum_{g\geq 2}\frac{4^{-g} \bigl(4^g-2\bigr) B_{2 g} \alpha ^{2 g-1} (\ri \xi )^{2 g} \operatorname{Li}_{3-2 g}\bigl(\E^{-t}\bigr)}{\xi ^3 (2 g)! (2 g-3)!},\\
G_\alpha^{(2)}(t,\xi)={}&-\sum_{g\geq 2}\frac{2^{-2 g-3} \bigl(4^g-8\bigr) B_{2 g-2} \alpha ^{2 g-3} (i \xi )^{2 g} \operatorname{Li}_{3-2 g}\bigl(\E^{-t}\bigr)}{3 \xi ^3 (2 g-3)! (2 g-2)!},\\
G_\alpha^{(3)}(t,\xi)={}&\sum_{k=2}^\infty {\bigl(1-2^{2k-1}\bigr)B_{2k}\over (2k)!}\\
 &\times\sum_{g=k}^\infty
\bigl(1-2^{2g-2k-1}\bigr)B_{2g-2k}(\ri/2)^{2g-2}{\operatorname{Li}_{3-2g}(Q)\over (2g-3)! }{(\alpha)^{2g-2k-1}\over (2g-2k)!}\xi^{2g-3}.
\end{align*}
{{We begin with $G_\alpha^{(3)}(t,\xi)$}}.
 Let us first consider the second sum only. We have
\[ \sum_{g=k}^\infty
\bigl(1-2^{2g-2k-1}\bigr)B_{2g-2k}(\ri/2)^{2g-2}{\operatorname{Li}_{3-2g}(Q)\over (2g-3)! }{(\alpha)^{2g-2k-1}\over (2g-2k)!}\xi^{2g-3}=f_1^\alpha(\xi) \star f_2(\xi,t),\]
where
\begin{align*}
& f_1^{\alpha}(\xi)=\xi^{-1}\sum_{g\geq k }\bigl(1-2^{2g-2k-1}\bigr)B_{2g-2k}(\ri \xi/2)^{2g-2}{(\alpha)^{2g-2k-1}\over (2g-2k)!}
 =-\frac{4^{-k} (i \xi )^{2 k} \csc \bigl(\frac{\alpha \xi }{2}\bigr)}{\xi^2 },
\end{align*}
and
\begin{align*}
 f_2(\xi,t,k)&=\sum_{g\geq k} \xi^{2g-3}
{f^{2g-3}(t)\over (2g-3)!}={1\over 2}\left(f(t+\xi)-f(t-\xi)\right)-\sum_{g=2}^{k-1}~\xi^{2g-3}{f^{2g-3}(t)\over (2g-3)!}\\
&=\hat f_2(\xi, t)-\sum_{g=2}^{k-1}~\xi^{2g-3}{f^{2g-3}(t)\over (2g-3)!},\end{align*}
where
\begin{align*}
&f(t)=\frac{1}{1-\re^t},\qquad \hat f_2(\xi,t)= \frac{\sinh (\xi )}{2 \cosh (t)-2 \cosh (\xi )}.\end{align*}
Then we have
\begin{align*}
G^{(3)}_\alpha(t,\xi)&={1\over 2\pi \ri}\oint_\gamma f_1^\alpha(s) f_2\left({\xi\over s},t\right){\rd s \over s}\\
&= {1\over 2\pi \ri}\oint_\gamma f_1^\alpha(s) \hat f_2\left({\xi\over s},t\right){\rd s \over s}-
\sum_{g=2}^{k-1}{1\over 2\pi \ri}\oint_\gamma f_1^\alpha(s) \left({\left(\xi\over s\right)}^{2g-3}{f^{2g-3}(t)\over (2g-3)!}\right){\rd s \over s},
\end{align*}
where $\gamma$ only include poles of $\hat f_2$ at $s=\pm \frac{\xi }{t+2 \ri \pi n}$, $n \in \IZ$ with residue
\[
-\frac{\ri}{4\pi}\sum_{n\in \IZ}4^{-k} \xi ^{2 k-2} (2 \pi n-\ri t)^{1-2 k} \operatorname{csch}\left(\frac{\alpha \xi }{4 \pi n-2 \ri t}\right).\]
Hence we have
\begin{align*}
 G_\alpha^{(3)}(t,\xi)&= \sum_{k=2}^\infty {\bigl(1-2^{2k-1}\bigr)B_{2k}\over (2k)!}\sum_{n\in \IZ}4^{-k} \xi ^{2 k-2} (2 \pi n-\ri t)^{1-2 k} \operatorname{csch}\left(\frac{\alpha \xi }{4 \pi n-2 \ri t}\right)
\\
&=\sum_{n\in \IZ}\frac{\operatorname{csch}\big(\frac{\xi }{4 \pi n-2 \ri t}\big) \operatorname{csch}\big(\frac{\alpha \xi }{4 \pi n-2 \ri t}\big)}{4 \xi }+\frac{\bigl(\xi ^2-24 (2 \pi n-\ri t)^2\bigr) \operatorname{csch}\big(\frac{\alpha \xi }{4 \pi n-2 \ri t}\big)}{48 \xi ^2 (2 \pi n-\ri t)}.
\end{align*}
{We now look at $G_\alpha^{(1)}(t,\xi)$.}
We have
\[
 G_\alpha^{(1)}(t,\xi)={1\over 2 \pi \ri} \oint f_3^\alpha(s) \hat f_2({\xi/s},t) {\rd s \over s},\]
where $\hat f_2$ is defined above and
\begin{align*}
f_3^\alpha(\xi)=&\sum_{g\geq 2}\frac{4^{-g} \left(4^g-2\right) B_{2 g} \alpha ^{2 g-1} (\ri \xi )^{2 g} }{\xi ^3 (2 g)!}=\frac{\alpha \xi ^2-12 \xi \csc \left(\frac{\alpha \xi }{2}\right)+\frac{24}{\alpha }}{24 \xi ^3}.
\end{align*}
Looking at the residue we find
\[ G_\alpha^{(1)}(t,\xi)=\sum_{n\in\mathbb{Z}} \frac{\alpha ^2 \xi ^2-12 \alpha \xi t \csc \big(\frac{\alpha \xi }{2 (t+2 \ri \pi n)}\big)+24 \pi \alpha \xi n \operatorname{csch}\big(\frac{\alpha \xi }{4 \pi n-2 \ri t}\big)-24 (2 \pi n-\ri t)^2}{24 \alpha \xi ^3}\]
{We now look at $G_\alpha^{(2)}(t,\xi)$.}
We have
\[ G_\alpha^{(2)}(t,\xi)={1\over 2 \pi \ri} \oint f_4^\alpha(s) \hat f_2({\xi/s},t) {\rd s \over s},\]
where $\hat f_2$ is defined above and
\[ f_4^\alpha(\xi)=-\sum_{g\geq 2}\frac{2^{-2 g-3} \left(4^g-8\right) B_{2 g-2} \alpha ^{2 g-3} (i \xi )^{2 g}}{3 \xi ^3 (2 g-2)!}= \frac{1}{48} \left(\frac{2}{\alpha \xi }-\csc \left(\frac{\alpha \xi }{2}\right)\right).\]
Looking at the residue we find
\[ G_\alpha^{(2)}(t,\xi)=\sum_{n\in\mathbb{Z}}\frac{1}{48} \left(\frac{2}{\alpha \xi }+\frac{\operatorname{csch}\bigl(\frac{\alpha \xi }{4 \pi n-2 \ri t}\bigr)}{-2 \pi n+\ri t}\right).\]
Hence by combining all together, we find
\[ G_\alpha (t,\xi)=- \sum_{n\in \IZ}\frac{\bigl(\alpha ^2+1\bigr) \xi ^2+6 \alpha \xi ^2 \operatorname{csch}\bigl(\frac{\xi }{4 \pi n-2 \ri t}\bigr) \operatorname{csch}\bigl(\frac{\alpha \xi }{4 \pi n-2 \ri t}\bigr)-24 (2 \pi n-\ri t)^2}{24 \alpha \xi ^3}.\]

\subsection*{Acknowledgements}
We would like to thank Murad Alim, Mat Bullimore, Fabrizio Del Monte, Lotte Hollands, Yakov Kononov, Pietro Longhi, Marcos Mari\~no, Sebastian Schulz, Shamil Shakirov, Ivan Tulli and Daniel Zhang for helpful discussion. We also thank the referees for reviewing the manuscript. The work of AN is supported by National Science Foundation grant 2005312 (DMS). The work of AG is partially supported by the Fonds National Suisse, Grant No.~185723 and by the NCCR ``The Mathematics of Physics'' (SwissMAP).

\pdfbookmark[1]{References}{ref}
\LastPageEnding

\end{document}